\documentclass[12pt]{article}

\usepackage{amsmath,amssymb,amsthm,amscd,mathrsfs}
\usepackage{latexsym,amsmath,amssymb,graphicx}
\usepackage[all]{xy} 
\usepackage{multirow}
\usepackage{color}
\usepackage{epsf}
\usepackage{youngtab}
\xyoption{v2} \xyoption{2cell} \xyoption{dvips}
\CompileMatrices \numberwithin{equation}{section}

\numberwithin{equation}{section}

\newcommand{\be}{\begin{equation}}
\newcommand{\ee}{\end{equation}}

\newcommand{\IP}{\mathbb{P}}

\newcommand{\wC}{{\widetilde C}}
\newcommand{\wA}{{\tilde A}}
\newcommand{\wtI}{{\tilde I}}
\newcommand{\wJ}{{\tilde J}}
\newcommand{\wE}{{\tilde E}}

\newcommand{\wY}{{\widetilde Y}}
\newcommand\IZ{\mathbb {Z}}
\newcommand\ID{\mathbb {D}}
\newcommand\IV{\mathbb {V}}

\newcommand{\tgamma}{{\tilde \gamma}}
\newcommand\IQ{\mathbb {Q}}
\newcommand{\IC}{\mathbb{C}}
\newcommand{\IE}{\mathbb{E}}
\newcommand{\IR}{\mathbb{R}}
\newcommand{\CU}{{\mathcal U}}
\newcommand{\ba}{\begin{array}}
\newcommand{\ea}{\end{array}}

\newcommand{\wH}{{\widetilde H}}

\newcommand{\CV}{{\mathcal V}}

\newcommand{\wG}{{\widetilde G}}

\newcommand{\CS}{{\mathcal S}}
\newcommand{\CB}{{\mathcal B}}

\newcommand{\CN}{{\mathcal N}}

\newcommand{\IH}{{\mathbb H}}
\newcommand{\bal}{\begin{aligned}}
\newcommand{\eal}{\end{aligned}}

\newcommand{\CZ}{{\mathcal Z}}

\newcommand{\wX}{{\widetilde X}}

\newcommand{\rk}{{\rm{rk}}}

\newcommand{\wP}{{\widetilde P}}
\newcommand{\tS}{{\widetilde S}}

\newcommand{\ualpha}{{\underline \alpha}}

\newcommand{\wCH}{{\widetilde {\mathcal H}}}
\newcommand{\wCU}{{\widetilde {\mathcal U}}}

\newcommand{\longto}{\longrightarrow}

\newcommand{\ch}{{\mathrm{ch}}}

\newcommand{\CO}{{\mathcal O}}

\newcommand{\CE}{{\mathcal E}}

\newcommand{\CH}{{\mathcal H}}
\newcommand{\CF}{{\mathcal F}}

\newcommand{\CM}{{\mathcal M}}

\newcommand{\CC}{{\mathcal C}}
\newcommand{\CI}{{\mathcal I}}

\newcommand{\CQ}{{\mathcal Q}}

\newcommand{\um}{{\underline m}}
\newcommand{\un}{{\underline n}}

\newcommand{\wS}{{\widetilde S}}

\newcommand{\wtJ}{{\widetilde J}}

\newcommand{\ulambda}{{\underline \lambda}}
\newcommand{\CT}{{\mathcal T}}

\newcommand{\us}{{\underline s}}

\newcommand{\calP}{{\mathcal P}}

\newcommand{\IA}{{\mathbb A}}
\newcommand{\uxi}{{\underline \xi}}
\newcommand{\bmu}{{\boldsymbol{\mu}}} 
\CompileMatrices \LaTeXdiagrams \UseAllTwocells
\newdimen\tableauside\tableauside=1.0ex
\newdimen\tableaurule\tableaurule=0.4pt
\newdimen\tableaustep
\def\phantomhrule#1{\hbox{\vbox to0pt{\hrule height\tableaurule width#1\vss}}}
\def\phantomvrule#1{\vbox{\hbox to0pt{\vrule width\tableaurule height#1\hss}}}
\def\sqr{\vbox{%
  \phantomhrule\tableaustep
  \hbox{\phantomvrule\tableaustep\kern\tableaustep\phantomvrule\tableaustep}%
  \hbox{\vbox{\phantomhrule\tableauside}\kern-\tableaurule}}}
\def\squares#1{\hbox{\count0=#1\noindent\loop\sqr
  \advance\count0 by-1 \ifnum\count0>0\repeat}}
\def\tableau#1{\vcenter{\offinterlineskip
  \tableaustep=\tableauside\advance\tableaustep by-\tableaurule
  \kern\normallineskip\hbox
    {\kern\normallineskip\vbox
      {\gettableau#1 0 }%
     \kern\normallineskip\kern\tableaurule}%
  \kern\normallineskip\kern\tableaurule}}
\def\gettableau#1 {\ifnum#1=0\let\next=\null\else
  \squares{#1}\let\next=\gettableau\fi\next}
\begin{document}

\title{Parabolic refined invariants and Macdonald polynomials}
\author{Wu-yen Chuang, 
Duiliu-Emanuel 
Diaconescu,
Ron Donagi, Tony Pantev}
\date{}
\maketitle

\begin{abstract} 
A string theoretic derivation is given for the conjecture of Hausel,
Letellier and Rodriguez-Villegas on the cohomology of character
varieties with marked points. Their formula is identified with a
refined BPS expansion in the stable pair theory of a local root stack,
generalizing previous work of the first two authors in collaboration
with G. Pan. Haiman's geometric construction for Macdonald polynomials
is shown to emerge naturally in this context via geometric
engineering. In particular this yields a new conjectural relation
between Macdonald polynomials and refined local orbifold curve counting
invariants. The string theoretic approach also leads to a new 
spectral cover construction for parabolic Higgs bundles in terms 
of holomorphic symplectic orbifolds.
\end{abstract}

\tableofcontents

\section{Introduction}

The main goal of this paper is a string theoretic derivation of the
conjecture of Hausel, Letellier and Rodriguez-Villegas \cite{HLRV} on
the topology of character varieties of punctured Riemann surfaces.
Analogous results have been obtained in \cite{wallpairs,BPSPW} in the
absence of marked points, identifying the main conjecture of Hausel
and Rodriguez-Villegas \cite{HRV} with a refined Gopakumar-Vafa
expansion. The same framework yields a recursion relation for
Poincar\'e and Hodge polynomials of Higgs bundle moduli spaces using
the wallcrossing formula of Kontsevich and Soibleman
\cite{wallcrossing}.  A motivic version of this recursion relation is
derived by Mozgovoy in \cite{ADHMrecursion}, and proved to be in
agreement with the Hausel-Rodriguez-Villegas formula.  The string
theoretic construction also provides quantitative supporting evidence
\cite{BPSPW} for the $P=W$ conjecture formulated by de Cataldo, Hausel,
and Migliorini in \cite{hodgechar}, and proven in loc. cit. for rank two
Higgs bundles.  The present paper carries out a similar program for
character varieties with marked points, the starting point being the
main conjecture formulated in \cite{HLRV}, which is briefly reviewed
below.

\subsection{The Hausel-Letellier-Rodriguez-Villegas 
formula}\label{HLRVsection}
Let $C$ be a smooth complex projective curve of genus $g\geq 0$, and
$D=p_1+\cdots +p_k $ a divisor of distinct reduced marked points on
$C$. Let $\gamma_1,\ldots, \gamma_k$ denote the generators of the
fundamental group $\pi_1(C\setminus D)$ corresponding to the marked
points.  For any nonempty partition $\mu=(\mu^1, \ldots, \mu^l)$ of $r\geq 1$, let ${\sf C}_\mu$ be a semisimple  conjugacy class in $GL(r,\IC)$ such that the eigenvalues of 
any matrix in ${\sf C}_\mu$ have multiplicities 
$\{\mu^1,\ldots, \mu^l\}$.

Let $\bmu =(\mu_1,\ldots, \mu_k)$ be a collection of 
partitions of an integer $r\geq 1$. 
Then the character variety
$\CC(C,D;\bmu)$ is the moduli space of conjugacy classes
of representations
\[ 
f : \pi_1(C\setminus D) \to GL(r,\IC)
\]
such that $f(\gamma_i)\in {\sf C}_{\mu_i}$
for all $1\leq i\leq k$.
The character variety $\CC(C,D;\bmu)$ actually depends on 
the choice eigenvalues but we will suppress this dependence from the notation since the topological invariants we compute below are independent of this choice. 

According to \cite[Thm. 2.1.5]{HLRV}, for sufficiently generic 
conjugacy classes ${\sf C}_{\mu_i}$,
$\CC(C,D; \bmu)$ is either empty or 
a smooth quasi-projective variety of dimension $d_\bmu=
r^2(2g-2+k)-\sum_{i=1}^k \sum_{j=1}^{l_i} (\mu_{i}^j)^{2} + 2$, where $l_i$ is the length of the partition $\mu_i$, $1\leq i\leq k$, as above. 
The compactly supported cohomology $H^*_{cpt}(\CC(C,D;\bmu))$ carries
a weight filtration $W_\bullet$ and the mixed Poincar\'e polynomial is
defined by 
\be\label{eq:mixedPA} P_c(\CC(C,D;\bmu); u,t) =
\sum_{i,k\geq 0} {\rm dim}\,
\left(Gr^W_{i}H_{cpt}^k(\CC(C,D;\bmu))\right)\, u^{i/2} (-t)^k.
\ee 
A priori the right hand side of \eqref{eq:mixedPA} takes values in
$\IZ[u^{1/2},t]$, but it was conjectured in \cite{HLRV} that it is in
fact a polynomial in $(u,t)$.

In order to formulate the main conjecture of \cite{HLRV}, for any 
partition $\lambda$ let
\be\label{eq:defhook}
\CH^g_\lambda(z,w) = \prod_{\Box\in \lambda} 
{ (z^{2a(\Box)+1}-w^{2l(\Box)+1})^{2g} \over 
(z^{2a(\Box)+2}-w^{2l(\Box)})(z^{2a(\Box)} - w^{2l(\Box)+2})}.
\ee
where $a(\Box), l(\Box)$ denote the arm, respectively leg length of $\Box 
\in \lambda$. Moreover, for each marked point $p_i$, 
let ${\sf x}_i=\left({\sf x}_{i,1}, 
{\sf x}_{i,2}, \ldots\right)$ be an infinite collection of 
formal variables, $1\leq i\leq k$, 
and $\wH_\lambda(z^2,w^2;{\sf x}_i)$  be the
modified MacDonald ploynomial \cite{gradedMD,MDgeom}
labelled by $\lambda$.
Then \cite[Conjecture~1.2.1.(iii)]{HLRV} states that 
\be
\boxed{ \label{eq:HLRVformB}
Z_{HLRV}(z,w, {\sf x}_i) = {\rm exp} \left(\sum_{k=1}^\infty 
\sum_{\bmu}
{1\over
  k}{w^{-kd_\bmu}  
P_c(\CC(C,D;\bmu); z^{-2k}, -(zw)^k) \over
(1-z^{2k})(w^{2k}-1)} 
\prod_{i=1}^k m_{\mu_i}({\sf x}_i^k) 
\right)
}
\ee
where 
\[
Z_{HLRV}(z,w, {\sf x}_i) =
\sum_{\lambda} \CH^{g}_{\lambda}(z,w) 
\prod_{i=1}^k{\widetilde H}_\lambda(z^2,w^2;{\sf x}_i) 
\]
and
$m_{\mu_i}({\sf x}_i)$ are the monomial symmetric functions. 
For ease of exposition equation \eqref{eq:HLRVformB} 
 will be referred to as the HLRV formula.

Note also that the character variety $\CC(C,D;\bmu)$ is diffeomorphic
to a moduli space of strongly 
parabolic Higgs bundles on $C$. 
By analogy with the $P=W$ conjecture
formulated in \cite{hodgechar}, one expects the weight filtration on
the compactly supported cohomology on the character variety to be
identified with a perverse Leray filtration for the Hitchin map on the
moduli of parabolic Higgs bundles. This conjectural identification
plays an important role in this paper.

\subsection{The main conjecture}\label{mainconj}

In this paper we propose a program for verifying \eqref{eq:HLRVformB}
by following a sequence of string-theoretic and geometric dualities
providing identifications of various counting functions.  Our main
string theoretic construction relies on a conjectural identification
of the generating function $Z_{HLRV}(z,w,{\sf x})$ with the stable
pair theory of a Calabi-Yau orbifold $\wY$. This orbifold is
constructed in Section \ref{orbisection} using the results of
\cite{orb_higgs,HilbertHiggs}, which identify parabolic Higgs bundles
on $C$ with Higgs bundles on a root stack. The root stack is an
orbifold curve $\wC$ equipped with a natural projection to $C$, which
makes $C$ its coarse moduli space.  Its construction depends on the
discrete invariants of the parabolic structure and is reviewed in
detail in Section \ref{rootsection}.
In particular, note that the closed points of 
$\wC$ have generically trivial stabilizers, the orbifold points
being in one-to-one correspondence with the marked points on $C$.

Given
a line bundle $M$ on $C$, the three dimensional Calabi-Yau
orbifold $\wY_M$ is defined to be the total space of the rank two bundle 
$\wY_{M} := \text{tot}\left(\nu^*M^{-1}\oplus
\left(K_{\wC}\otimes_\wC\nu^*M\right)\right)$
on  ${\wC}$. 
In what follows
we will call such three dimensional Calabi-Yau orbifolds {\em\bfseries
  local orbifold curves}.
Initially we focus on
$\wY := \wY_{\CO} := \text{tot}\left(\CO_{\widetilde C}\oplus K_{\wC}\right)$.


By analogy with Pandharipande and Thomas \cite{stabpairsI}, the {\em\bfseries stable pair
  theory} of a local orbifold curve $\wY$ is defined in Section
\ref{orbisection} as a counting theory for pairs $(F,s)$ with $F$ a
pure dimension one sheaf on $\wY$, and $s:\CO_\wY \to F$ a generically
surjective section. The discrete invariants of $F$ are the Euler
character $n=\chi({\widetilde F})$ and a collection of integral
vectors $\um=(\um_i)_{1\leq i\leq k}$, $\um_i\in (\IZ_{\geq
  0})^{s_i}$, $s_i\geq 1$, encoding the $K$-theory class of $F$.  In
Section \ref{ADHMpar} we extend the analysis of \cite{modADHM}: given
a line bundle $M$ on $C$, we reformulate the stable pair theory of the
local orbifold curve $\wY_{M}$ in terms of parabolic
ADHM sheaves on the curve $C$.  This yields an explicit construction
of the perfect obstruction theory of the moduli space, and makes the
relation with parabolic Higgs bundles more transparent.

Assuming the foundational aspects of motivic Donaldson-Thomas 
theory from \cite{wallcrossing}, one obtains a series of refined
Pandharipande-Thomas (PT) invariants for the orbifold $\wY_{M}$:
\be\label{eq:orbstpairA} 
Z^{ref}_{{\widetilde Y}_{M}}(q,{\underline x}, y) = \sum_{n\in \IZ} \sum_{\um} 
PT({\widetilde Y}_{M}, n, \um; y) q^n \prod_{i=1}^k 
{\underline x}_i^{\um_i}
\ee
for some formal variables ${\underline x} = 
\big({\underline x}_1,\ldots,{\underline x}_k\big)$, 
${\underline x}_i = (x_{i,0}, \ldots, x_{i,s_i-1})$, 
$1\leq i\leq k$. 
Then the main conjecture in this paper is the following identity:

\

\smallskip

\noindent
{\bfseries Conjecture.} \ {\em After a change of variables, the counting
  function for the refined PT 
  invariants on the orbifold Calabi-Yau $\wY =
  \text{tot}(\mathcal{O}_{\wC}\oplus K_{\wC})$  is identified with the 
combinatorial HLRV partition function:
\be\label{eq:conjrelA} 
\boxed{
Z^{ref}_{\widetilde Y}(z^{-1}w,{\underline x},z^{-1}w^{-1}) = 
Z_{HLRV}(z,w,{\underline x}).
}
\ee
}

\

\smallskip

In Section \ref{parPW} we explain how
the relation \eqref{eq:conjrelA} and the parabolic $P=W$ 
conjecture imply that  the same change of variables converts 
the HLRV formula \eqref{eq:HLRVformB} 
into a refined Gopakumar-Vafa expansion.
Moreover, an application of the wallcrossing formula of Kontsevich and 
Soibelman yields a recursion relation for Poincar\'e  
polynomials analogous to \cite{wallpairs}. 
As shown in Section \ref{recursion}, the main arguments of
\cite{ADHMrecursion} apply to the present case as well, proving that
the  
solution of this recursion formula is in agreement with the predictions 
of the HLRV formula. 

A rigorous proof of  identity \eqref{eq:conjrelA} is one of the most 
important open problems emerging from this paper. 
Supporting evidence for this conjecture 
is provided in Sections \ref{geomeng} and \ref{conifold}, which are briefly 
summarized below. 

\subsection{Macdonald polynomials via geometric engineering}\label{MDgeom}
Geometric engineering is used in Section \ref{geomeng} to 
relate the stable pair generating function \eqref{eq:orbstpairA} 
to a D-brane quiver quantum mechanical partition function. 
Analogous results in the physics literature were obtained in 
\cite{geom_eng_qft,Lawrence:1997jr,EK-I,Nekrasov:2002qd,IKP-I,IKP-II,
EK-II,Hollowood:2003cv,Konishi-I,LLZ,Iqbal:2007ii} while a general mathematical theory of geometric engineering is currently 
being developed by Nekrasov and Okounkov in \cite{Mvert}. 
The treatment in Section \ref{geomeng} 
follows the usual approach in the physics literature 
via IIA/M-theory duality and D-brane dynamics. 
A detailed comparison with the formalism of \cite{Mvert} is left as an open 
problem, as briefly explained below.

For simplicity it is assumed that there is only one marked point 
on $C$. The local orbifold curve is taken of the form 
form $\wY_M = {\rm tot}\big(\nu^*M^{-1}\oplus K_{\wC}\otimes_\wC\nu^*M\big)$ where 
$M$ is a degree $p\geq 0$ line bundle on $C$. 
As shown in Section \ref{nested}, a two step chain of dualities relates the  
resulting stable pair theory to a series of equivariant 
$K$-theoretic invariants of nested Hilbert schemes of points in
$\IC^2$. The construction of the $K$-theoretic partition function 
is explained in Section \ref{Kpartfct}. The final formula recorded in 
equation \eqref{eq:parinstA} is a generating function of the form 
\be\label{eq:parinstX} 
Z_K(q_1,q_2; {\tilde y},{\tilde x}) = 
\sum_{\gamma} \chi_{\tilde y}^{\bf T}(\CV(\gamma)) m_{\mu(\gamma)}({\tilde x}) 
\ee
where $\chi_{\tilde y}^{\bf T}(\CV(\gamma))$ is the equivariant Hirzebruch 
genus of a vector bundle $\CV(\gamma)$ on the nested Hilbert scheme 
$\CN(\gamma)$. Here the sum is over all finite collections $\gamma=(\gamma_\iota)_{0\leq \iota\leq \ell}$ of positive integers 
labelling discrete invariants of flags of ideal sheaves on $\IC^2$, 
as explained in Section \ref{nested},  above equation \eqref{eq:idealflag}. 
For any $\gamma$, $\mu(\gamma)$ denotes the unordered partition of 
$|\gamma|=\sum_{\iota=0}^\ell \gamma_\iota$ determined by $\gamma$, and $m_{\mu}({\tilde x})$ are the monomial symmetric 
functions in an infinite set of variables $({\tilde x}_0, {\tilde x}_1, \ldots)$. 

Note that the formalism of \cite{Mvert} relates the above stable pair theory 
with the equivariant K-theoretic stable pair theory of the product 
$\wC\times \IC^2$. Then one expects the partition function 
\eqref{eq:parinstX} to follow from this theory by virtual localization computations. 
In particular the bundle of fermion zero modes derived in 
Appendix \ref{zeromodes} is expected to be naturally determined by the 
induced perfect obstruction theory on the fixed loci. This computation 
will be left as an open problem.

The main result of section \ref{geomeng} is identity \eqref{eq:parinstB} 
expressing the generating function \eqref{eq:parinstX}  in terms of 
modified Macdonald polynomials, 
\be\label{eq:parinstY} 
Z_K(q_1,q_2; {\tilde x}, {\tilde y}) = \sum_{\mu} \Omega_{\mu}^{g,p}(q_1,q_2,{\tilde y}) 
{\tilde H}_\mu(q_2,q_1,{\tilde x}).
\ee
The $\Omega_{\mu}^{g,p}(q_1,q_2,{\tilde y})$ are rational 
functions of the equivariant parameters $(q_1,q_2)$ and ${\tilde y}$ 
determined explicitely by a fixed point theorem, according to equation 
\eqref{eq:omegaformula}. 

Formula \eqref{eq:parinstY} is proven in Section \ref{nestedMD} 
using Haiman's geometric construction of Macdonald polynomials in 
terms of isospectral Hilbert scheme \cite{MDgeom,polygraphs}. 
The proof also requires some geometric comparison results between 
nested and isospectral Hilbert schemes established in Section 
\ref{isosection}. 

As supporting evidence for equation \eqref{eq:conjrelA}, 
it is shown in Section \ref{parPW}, equation \eqref{eq:reforbD}, 
that a simple change of variables relates
the right hand side of equation \eqref{eq:parinstY} 
to the HLRV generating function $Z_{HLRV}(z,w; {\sf x})$, 
\be\label{HLRVK} 
Z_{HLRV}(z,w; {\sf x})= Z_K\big(w^2, z^2; (zw)^{-1}, (-1)^{g-1}(zw)^g{\sf x}\big).
\ee
Further supporting evidence is provided in Section \ref{conifold}, which is 
briefly summarized below. 

\subsection{Parabolic conifold invariants and the equivariant index} 
A direct computational test of conjecture \eqref{eq:conjrelA} 
is carried out in Section \ref{conifold} using the formalism developed 
by Nekrasov and Okounkov in \cite{Mvert}. The computations are 
carried out for the special 
case where $C$ is the projective line with one marked point $p$, and the local threefold is 
${\widetilde Y}_{\CO_C(1)}$ i.e. the total space of the rank two bundle $\nu^*\CO_C(-1) 
\oplus K_{\widetilde C} \otimes_{\widetilde C} \nu^*\CO_C(1)$ on  
${\widetilde C}$. A conjectural relation between 
the equivariant index defined in  \cite{Mvert} and 
orbifold refined stable pair invariants is formulated in 
Section \ref{compsection}, equation \eqref{eq:conjrelA}.
This identification is checked by explicit virtual localization 
computations for low degree terms up to three box 
partitions in Section \ref{fixedsection}. 

An important outcome of the string theoretic derivation is a 
new geometric construction of spectral data for parabolic Higgs 
bundles which lays the ground for a generalization of the HLRV 
formula. This is carried out in Section \ref{specholsurf}, 
a brief outline being provided below.

\subsection{Outline of the program} \label{flowchart} 
For the convenience of the reader we now list all the ingredients in the physical
derivation of the HLRV conjecture 
\eqref{eq:HLRVformB}  in their logical sequence:
\begin{description}
\item[{\bfseries Step 1.}] Identify the combinatorial left hand side
  of the HLRV formula with the counting function for refined stable
  pair invariants on the three dimensional Calabi-Yau orbifold
  $\widetilde{Y}$. This identification is provided by the conjectural
  formula \eqref{eq:conjrelA}. 
The construction of the orbifold stable pair theory for this
step is
presented in Section \ref{orbisection}.
\item[{\bfseries Step 2.}] Identify the counting function for the
  refined stable pair invariants on $\widetilde{Y}$ with the
  generating function for the perverse Poincar\'{e} polynomials of the
  moduli of parabolic Higgs bundles.  This identification is a
  combination of two components: 
\begin{itemize}
\item[{\bfseries (i)}] A geometric isomorphism of the moduli of
  Bridgeland stable pure dimension one sheaves on $\widetilde{Y}$ and
  the product of the moduli space of parabolic Higgs bundles on $C$
  with the affine line. This identification is based on the spectral 
cover construction explained in Section \ref{rootsection}. 
\item[{\bfseries (ii)}] 
A conjectural refined Gopakumar-Vafa expansion of the stable pair theory of
  $\widetilde{Y}$ generalizing the unrefined conjecture 
formulated in \cite{stabpairsI}. 
Granting identity \eqref{eq:conjrelA}, the 
specialization of the HLRV formula to Poincar\'e 
polynomials follows recursively from the 
Kontsevich-Soibelman wall-crossing formula 
\cite{wallcrossing} for the variation of
  Bridgeland stabilities on the stable pair moduli
by analogy with \cite{wallpairs,ADHMrecursion}. 
The details are presented in Sections \ref{parPW}, 
\ref{recursion}. 
\end{itemize}
\item[{\bfseries Step 3.}]  Identify the generating function for the
  perverse  Poincar\'{e} polynomials of the
  moduli of parabolic Higgs bundles with the generating function for
  the weight-refined Poincar\'{e} polynomial of the character
  variety. This is a parabolic version of  the $P=W$ conjecture of Hausel, de Cataldo, and Migliorini. A brief discussion
is provided in Section \ref{parPW}. 
\end{description}

 Note that the refined 
Gopakumar-Vafa expansion 
needed here was conjectured for toric Calabi-Yau 
threefolds in \cite{Iqbal:2007ii} and also 
\cite{CKK} building on previous 
work of \cite{GVII,KKV}. This conjecture was extended 
to higher genus local curves in \cite{BPSPW}. Here it is 
further extended to local orbifold curves. 

In the mathematical literature, a weak form 
of the unrefined Gopakumar-Vafa 
conjecture 
stable pair theory was proven in
\cite{Hall_curve_counting}, \cite{generating_wallcrossing}, while the
full unrefined conjecture was proven in
\cite{stab_curve_counting}. These results prove the existence of a
suitable integral expansion, but do not provide a cohomological
intepretation of the resulting integral invariants. The latter is
also needed in the string theory derivation of the HLRV formula.

The geometric framework developed in this program 
admits a generalization to parabolic Higgs bundles with
nontrivial eigenvalues at the marked points. This yields in 
particular a new orbifold spectral cover presentation 
for such objects, generalizing the construction in Section 
\ref{rootsection}. This is carried out in Sections 
\ref{orbisurfaces}, \ref{equivalence} and \ref{proofoutline}, 
which are summarized below.

\subsection{Orbifold spectral data for nontrivial eigenvalues}
The orbifold spectral cover construction 
applies to a particular flavor of parabolic 
Higgs bundles introduced in Section 
\ref{diagonalres}. These are Higgs bundles with 
simple poles at the marked points, whose 
residues are $\xi$-parabolic maps. 
This condition requires each graded component of the Higgs field residue 
at a marked point with respect to the flag to be a specified multiple 
of the identity. Parabolic Higgs bundles satisfying this condition are called {\it diagonally parabolic}, or, more specifically, 
$\uxi$-{\it parabolic},  and form a closed substack of the moduli stack of 
semistable parabolic Higgs bundles. 

The unordered eigenvalues of parabolic Higgs bundles are parameterized 
by the quotient $Q$ of the Hitchin base defined in equation 
\eqref{eq:eigenvalproj}. For each point $q\in Q$, the construction in 
Section \ref{orbisurfaces} produces a holomorphic symplectic orbifold 
surface $\tS_\delta$. The moduli space of semistable $\uxi$-parabolic 
Higgs bundles is conjecturally  identified with a moduli space of semistable 
torsion sheaves on $\tS_\delta$ with fixed $K$-theory class.  
The precise statement of this conjecture is given 
in Section \ref{equivalence}, a brief outline of the 
proof being provided in Section \ref{proofoutline}. 

In this geometric framework string theory arguments predict a formula 
of the form \eqref{eq:HLRVformB}, where the left hand side is given 
by the refined stable pair theory $Z_{PT}^{ref}(\tS_\delta\times \IC)$ up to a 
change of variables. The right hand side will be a similar generating 
function for perverse Poincar\'e polynomials for moduli spaces of 
stable $\uxi$-parabolic Higgs bundles. 
As pointed out by Emmanuel Letellier and 
Tamas Hausel, the latter is expected to 
be identical with the right hand side of equation 
\eqref{eq:HLRVformB}, even away from the nilpotent locus. 
This leads to a rather surprising conjecture
stating that the refined stable pair theory 
$Z_{PT}^{ref}(\tS_\delta \times \IC)$ is independent 
on $\delta$. In particular, it should be identical 
with the stable pair theory of ${\wY}$ in equation
\eqref{eq:orbstpairA}.

\subsection{Open problems}\label{openproblems}
We conclude the introduction with a list of open problems 
emerging from this work. 
Several such questions have already been encountered above, including:

$(a)$
the proof of \eqref{eq:conjrelA},

$(b)$ 
the derivation of an analogous formula for the stable pair theory 
of the orbifolds $\tS_\delta \times \IC$, confirming 
deformation invariance, and 

$(c)$ an explicit comparison with the equivariant $K$-theoretic stable pair theory of $\wC \times \IC^2$ in the context of geometric engineering.

\noindent
Additional possible future directions  include:

$(d)$ Section \ref{parPW} presents quantitative evidence for a parabolic version of the $P=W$ conjecture formulated 
in \cite{hodgechar}. It would be very interesting if the parabolic 
$P=W$ 
can be proven by direct comparison methods in certain classes of examples.

$(e)$ Another problem is to 
prove the crepant resoltuion conjecture for stable 
pair invariants formulated in \cite[Conj. 4]{crepantcounting}, for the orbifolds 
$\tS_\delta\times \IC$. Similar results in Donaldson-Thomas theoriers have been proven in 
\cite{crepant_res, gen_DT_orb}.

$(f)$ Elaborating on the same topic, a further question is whether 
$\uxi$-parabolic Higgs bundles admit a spectral cover presentation 
in terms of torsion sheaves on the resolutions of the coarse moduli 
spaces. Again, the Fourier-Mukai transform should provide 
important input in finding the answer. 

$(g)$ Finally, a natural question is whether one can construct a
 TQFT formalism for 
(unrefined) curve counting invariants of local orbifold curves, 
by analogy with the results of Bryan and Pandharipande \cite{BP},
and Okounkov and Pandharipande \cite{OP}. 

\bigskip

\subsection{Notation and conventions} \label{notation}

\begin{description}
\item[$C$] - a smooth complex projective curve of genus $g\geq 0$.
\item[$D=p_1+\cdots +p_k $] - a divisor of distinct reduced marked points on
$C$.
\item[$\bmu =(\mu_1,\ldots, \mu_k)$] -  a collection of
partitions of an integer $r\geq 1$.
\item[$\CC(C,D;\bmu)$] - the  character variety, i.e. the moduli space of conjugacy classes
of representations of $\pi_{1}(C\setminus D)$ with values in fixed
conjugacy classes at the punctures. 
\item[$P_c(\CC(C,D;\bmu); u,t)$] -  the mixed Poincar\'e polynomial
  for the weight filtration on the compactly supported cohomology of
  the character variety.
\item[$\CH^g_\lambda(z,w)$] - the HLRV $(z,w)$-deformation of the
  $2g-2$ power of the hook polynomial given in 
equation \eqref{eq:defhook}. 
\item[${\widetilde H}_\mu(q_2,q_1,{\tilde x})$] - the modified
  MacDonald polynomial \cite{gradedMD,MDgeom}. 
\item[$Z_{HLRV}(z,w, {\sf x})$] - the combinatorial HLRV partition
  function appearing in the left hand side of the HLRV formula. 
\item[$\wC$] - an orbifold curve equipped with a morphism $\nu : \wC
  \to C$, which is an isomorphism outside $D$.
\item[$\wY$] - a three dimensional Calabi-Yau orbifold given as
$\wY = \text{tot}\left(\mathcal{O}_{\wC}\oplus K_{\wC}\right)$. 
\item[$\wY_{M}$] - a three dimensional Calabi-Yau orbifold given as
$\wY_{M} = \text{tot}\left(\nu^{*}M\oplus \left(K_{\wC}\otimes
  \nu^{*}M^{-1}\right)\right)$ for some line bundle $M$ on $C$. 
\item[$Z^{ref}_{{\widetilde Y}_{M}}(q,y,x)$] - the counting function
  of refined stable pair invariants on $\wY_{M}$. 
\item[$Z_K(q_1,q_2; {\tilde y},{\tilde x})$] - the counting function
  of equivariant $K$-theoretic invariants of nested Hilbert schemes of
  points in $\mathbb{C}^{2}$.
\item[$\CH_{\uxi}^{ss}(C,D; \um, e, \ualpha)$] - the moduli stack of
  semistable $\uxi$-parabolic Higgs bundles on $C$.
\item[$\widetilde{S}_{\delta}$] - a symplectic orbifold surface
  associated with a zero dimensional subscheme inside
  $\text{tot}(K_{C}(D))$. 
\end{description}

\bigskip

\noindent
{\it Acknowledgements.} We are very grateful to Vincent Bouchard, Ugo Bruzzo, 
Jim Bryan, Tamas
Hausel, Marcos Jardim, Sheldon Katz, Ludmil Katzarkov, Emmanuel Letellier, Davesh
Maulik, Sergey Mozgovoy, Alexei Oblomkov, Andrei Okounkov, Rahul
Pandharipande, Fernando Rodriguez-Villegas, Vivek Shende, Andras
Szenes, Richard Thomas and Zhiwei Yun for very helpful discussions. WYC and DED would like to thank Guang Pan 
for collaboration at an incipient stage of this project. 
WYC was supported by NSC grant 101-2628-M-002-003-MY4 and a fellowship 
from the Kenda Foundation. DED was partially supported by 
NSF grant PHY-0854757-2009. 
RD acknowledges partial support by NSF grants DMS 1304962 and RTG
0636606.
DED and TP also acknowledge partial support from NSF grants DMS 1107452,
1107263, 1107367 "RNMS: GEometric structures And Representation
varieties" (the GEAR Network) during the initial stages of this work.
TP was partially supported by NSF RTG grant DMS-0636606 and NSF grant
DMS-1302242.

\section{Parabolic Higgs bundles and spectral covers}\label{meromparHiggs} 
The goal of this section is to provide some background 
on parabolic Higgs bundles, summarizing the 
main results used throughout the paper.  
Let $C$ be a smooth projective curve over $\IC$ and 
$D=\sum_{i=1}^k p_i$ a reduced effective divisor on $C$.
A meromorphic Higgs bundle is a pair $(E,\Phi)$ with a $E$ a 
locally free sheaf and $\Phi:E\to E\otimes K_C(D)$ 
a morphism of sheaves on $C$. 
Parabolic Higgs bundles are a refinement of meromorphic ones defined by 
specifying a parabolic structure at each marked point, as 
discussed in detail below. 

\subsection{Parabolic structures}\label{background} 

In order to fix notation, let  $V$ be a finite dimensional vector space and 
$\um=(m_a)_{0\leq a\leq s-1}\in \IZ_{\geq 0}^s$
an ordered collection of non-negative integers
such that 
\be\label{eq:sumcond}
\sum_{a=0}^{s-1}m_a = {\rm dim}\, V.
\ee
A  flag of type $\um$ in $V$ is a filtration  
\[ 
0=V^s\subseteq V^{s-1} \cdots \subseteq V^{1}\subseteq V^{0}=V
\]
by vector subspaces such that 
\begin{equation} \label{oifd}
{\rm{dim}}\, (V^a/V^{a+1})=m_a,   \quad 0\leq a\leq s-1.
\end{equation}
 Note that degenerate flags are allowed 
i.e. the inclusions do not have to be strict,
but the length of the filtration is fixed. 

Suppose $V,W$ are finite dimensional vector spaces equipped 
with filtrations $V^\bullet$, $W^\bullet$ of the same length $s$. 
A linear map $f:V\to W$ will be called parabolic if 
$f(V^a) \subseteq W^a$ for all $0\leq a\leq s$. The map 
$f$ will be called 
strongly parabolic if $f(V^a)\subseteq W^{a+1}$ for all $0\leq a\leq s-1$. 

A more refined compatibility condition can be defined when 
$W=V\otimes L$, with $L$ a one dimensional vector space, 
and $W^\bullet$ is the natural filtration determined by  $V^\bullet$.
Given a collection of linear maps $\xi=\big(\xi_a:\IC\to L\big)_{0\leq a\leq s-1}$,
a linear map $f:V\to W$ will be called $\xi$-parabolic 
if $f$ is parabolic and the induced maps 
$f_a: V^a/V^{a+1} \to  V^a/V^{a+1}\otimes L$ are of the form 
\[
f_a = {\bf 1}_{V^a/V^{a+1}} \otimes \xi_a.
\] 

Following the notation introduced 
in \cite{rankthreepar}, let 
${\rm PHom}(V,W)$, ${\rm SPHom}(V,W)$ denote the 
linear space of parabolic, respectively strongly parabolic 
linear maps. Let also
\[
{\rm APHom}(V,W) = {\rm Hom}(V,W) /{\rm PHom}(V,W) 
\]
be the vector space parameterizing equivalence classes of 
morphisms not preserving the filtrations.

For any  exact sequence 
\[
0\to V' \to V \to V'' \to 0,
\] 
 a flag $V^\bullet$ in $V$ of length $s$ induces canonical 
flags $V'^\bullet$, $V''^\bullet$ of $V',V''$ of the same length. 
In fact vector spaces equipped with flags of fixed length $s$ form 
an abelian category.

Given a reduced effective divisor $D=\sum_{i=1}^k p_i$ on 
the curve $C$, for each $i\in \{1,\ldots, k\}$ let $\um_i=(m_{i,a})$, $0\leq a\leq s_i$, $s_i\geq 1$, 
be an ordered collection of integers of length $s_i\geq 1$ 
 satisfying conditions \eqref{eq:sumcond}.
A quasi-parabolic structure on a vector bundle $E$ on $C$ 
is a collection 
$(E_i^\bullet)_{1\leq i\leq k}$ 
of flags of type $\um_i$ in the fiber $E_{p_i}$ for each 
$i\in \{1,\ldots,k\}$. For ease of exposition, such a quasi-parabolic 
vector bundle will be denoted by $E^\bullet$, and its numerical 
type by $\um=(\um_i)_{1\leq i\leq k}$. 

For any exact sequence of vector bundles 
\[ 
0\to F\to E \to G \to 0,
\] 
a quasi-parabolic structure on $E$ at $D$ of type $\um^E$ 
induces quasi-parabolic structures of types $\um^F$, $\um^G$ 
on $F,G$ such that $\um^F+\um^G= \um^E$. 
Moreover, as explained in \cite[Sect 2.2]{rankthreepar},
for any two parabolic bundles $E^\bullet$, $F^\bullet$ there is a 
sheaf of parabolic morphisms $PHom_C(E^\bullet, F^\bullet)$ which fits in an exact sequence
\be\label{eq:parhomA}  
0 \to PHom_C(E^\bullet, F^\bullet) \to Hom_C(E,F)
\to \oplus_{i=1}^k 
{\rm APHom}(E^\bullet_{p_i},F^\bullet_{p_i}) \otimes 
\CO_{p_i}\to 0
\ee
\be\label{eq:parhomB}  
0 \to Hom_C(E,F(-D)) \to PHom_C(E^\bullet, F^\bullet) 
\to \oplus_{i=1}^k 
{\rm PHom}(E^\bullet_{p_i},F^\bullet_{p_i}) \otimes 
\CO_{p_i}\to 0
\ee
Similarly, there is a sheaf of local strongly parabolic morphisms 
$SPHom_C(E^\bullet, F^\bullet)$ which fits in analogous 
exact sequences. Note also that there is a natural duality 
relation \cite[Prop. 2.3.i]{rankthreepar}
\be\label{eq:pardual}
PHom_C(E^\bullet, F^\bullet)^\vee \simeq SPHom_C(F^\bullet, E^\bullet)
\otimes_C \CO_C(D).
\ee

A parabolic bundle on $C$ is a quasi-parabolic bundle $E^\bullet$ 
equipped in addition with collections of weights 
$\ualpha_i = (\ualpha_{i,a})_{0 \leq a\leq s_i-1}\in \IR^{s_i}$ 
for each $i\in \{1,\ldots, l\}$ such that 
\be\label{eq:parweights} 
0\leq \alpha_{i,0} < \cdots < \alpha_{i,s_i-1} < 1. 
\ee
The data $(\ualpha_i)_{1\leq i\leq k}$ will be denoted 
by $\ualpha$ and parabolic bundles will be denoted by $(E^\bullet, 
\ualpha)$.

There is a natural stability condition for parabolic bundles formulated in terms of 
parabolic slopes. The parabolic degree of $(E^\bullet, \ualpha)$ is defined as 
\be\label{eq:pardegA} 
{\rm deg}(E^\bullet, \ualpha)= {\rm deg}(E) + \sum_{i=1}^k 
\sum_{a=0}^{s_i-1} m_{i,a} \alpha_{i,a},
\ee
and the parabolic slope is given by 
\be\label{eq:parslope}
\mu(E^\bullet, \ualpha) = {\chi(E^\bullet, \ualpha)
\over {\rm rk}(E)}.
\ee
with
\be\label{eq:parchi}
\chi(E^\bullet,\ualpha) = {\rm deg}(E^\bullet, \ualpha) - 
{\rm rk}(E)(g-1) = \chi(E) + \sum_{i=1}^k \sum_{a=0}^{s_i-1} 
m_{i,a}\alpha_{i,a}.
\ee
Any nontrivial proper saturated subsheaf $0\subset E'\subset E$ inherits an induced parabolic structure $({E'}^\bullet,\ualpha')$ 
on $E'$. The parabolic bundle $(E^\bullet, \ualpha)$ is 
(semi)stable if any such subsheaf satisfies the parabolic 
slope condition
\be\label{eq:parslopeineq}
\mu({E'}^\bullet, \ualpha')\ (\leq)\ \mu(E^\bullet, \ualpha).
\ee
As shown in \cite{modpar}, this stability condition yields  projective moduli spaces of $S$-equivalence classes of semistable objects. Moreover, for sufficiently generic weights 
these moduli spaces are smooth. 

\subsection{Higgs fields, spectral covers, and foliations}\label{higgsleaves}
A quasi-parabolic Higgs bundle on $C$ is a quasi-parabolic 
vector bundle $E^\bullet$ equipped with a Higgs field $\Phi:E\to 
E\otimes_C K_C(D)$ such that the 
residue ${\rm Res}_{p_i}(\Phi) : E_{p_i}\to E_{p_i}$
is a parabolic map for each marked point. 
A parabolic Higgs bundle is defined by specifying in addition a collection 
of weights $\ualpha$ as in the previous section. 

There is a natural notion of stability for parabolic 
Higgs bundles, defined by 
imposing the parabolic slope inequality \eqref{eq:parslopeineq}
for all proper saturated 
subsheaves preserved by the Higgs field. The results of \cite{cptparmod}, 
imply that semistable parabolic 
Higgs bundles with fixed numerical invariants $\um$, $\deg(E)=e$ form 
 an algebraic stack of finite type 
 $\CH^{ss}_{\rm \small par}(C,D;\um,e,\ualpha)$. 
 The stable ones form an open substack $\CH^{s}_{\rm \small par}(C,D;\um,e,\ualpha)$. Moreover there is a  coarse moduli space 
 $H^{ss}_{\rm \small par}(C,D;\um,e,\ualpha)$ parameterizing 
 $S$-equivalence classes of semistable objects which contains 
 an open subspace $H^s_{\rm \small par}(C,D;\um,e,\ualpha)$ parameterizing 
 isomorphism classes of stable objects. According to \cite{infparHiggs}, 
 $H^{ss}_{\rm \small par}(C,D;\um,e,\ualpha)$ is a normal quasi-projective
 variety while the stable open subspace is smooth. 
 Note also that any semistable object must be stable for sufficiently generic weights and primitive numerical invariants.

Similar considerations apply to strongly parabolic Higgs 
bundles, in which case the moduli stacks/spaces will be labelled 
with a subscript {\it s-par} instead of {\it par}. 
In addition, one can construct similarly moduli spaces 
of parabolic and strongly parabolic 
Higgs bundles where the Higgs field takes 
values in an arbitrary coefficient line bundle $M$, that is 
$\Phi: E \to E\otimes_C M(D)$. In this case, the line bundle 
$M$ will be specified in the notation of the moduli space
e.g. $H^{ss}_{\rm par}(C,D,M; \um,e,\ualpha)$.

Taking polynomial invariants of the Higgs field
yields the Hitchin map
\be\label{eq:Hitchinmap} 
{h}:H^{ss}_{\rm \small par}(C,D;\um,e,\ualpha) \to B(C,D;r),\qquad
B(C,D;r)= \oplus_{l=1}^r H^0(C, (K_C(D))^l).
\ee
This is a surjective proper morphism, its generic fibers being 
disjoint unions of abelian varieties.
As observed in \cite{spectral_int, parabolic_poisson}, the unordered eigenvalues 
of the Higgs field at the marked points are parameterized by the quotient 
$B/B_0$ where $B_0\subset B$ is the linear subspace 
\[
B_0(C,D;r)= \oplus_{l=1}^r H^0(C, (K_C(D))^l\otimes \CO_C(-D)) \subset 
B(C,D;r).
\]
The moduli space is foliated by the fibers of the resulting projection, 
\be\label{eq:eigenvalproj}
p: H^{ss}_{\rm \small par}(C,D;\um,e,\ualpha) \to B(C,D;r)/B_0(C,D;R). 
\ee

Parabolic Higgs bundles admit a spectral cover presentation 
 as parabolic pure dimension one sheaves on the total space 
 $P$ of  $K_C(D)$. Let $\pi:P\to C$ denote the canonical projection,
 $P_i=\pi^{-1}(p_i)$ the fiber at the marked point $p_i$, and  
 $D_P= \sum_{i=1}^k P_i$. A quasi-parabolic structure $F^\bullet$
 on a pure dimension 
 one sheaf $F$ on $P$ is defined by a sequence of surjective morphisms 
 \be\label{eq:torsionqpar}
 F\otimes_P P_i \twoheadrightarrow F_i^{s_i-1}\twoheadrightarrow\cdots 
 \twoheadrightarrow F_i^{1}
 \ee
 where $F_i^a$ are sheaves on $P_i$ for all $1\leq i\leq k$. 
 Moreover $F$ is required to have compact support, which implies 
 that the sheaves $F\otimes_P P_i$ are zero dimensional. In this 
 case $\ch_1(F) = d\sigma$ with $\sigma$ the class of the zero 
 section. 
  A parabolic structure is defined by specifying in addition parabolic 
 weights $\ualpha=(\alpha_i^a)$ as above. 
 
  Any saturated sub sheaf $F'\subset F$ inherits a natural induced 
 parabolic structure. Then one defines a stability condition using the 
 parabolic slope 
\[
\mu(F^\bullet,\ualpha) = {1\over d}\bigg(\chi(F) + 
\sum_{i=1}^k \sum_{a=0}^{s-1} \alpha_{i,a} (\chi(F_i^{a+1})-\chi(F_i^{a}))\bigg).
\]
This yields an algebraic moduli stack of semistable objects which is isomorphic 
to the moduli stack of semistable parabolic Higgs bundles on $C$ 
with numerical invariants 
\[
m_i^a = d-\chi(F_i^a), \qquad e = \chi(F) + d(g-1).
\]
This isomorphism assigns to any sheaf $F$ the bundle $E=\pi_*F$, 
the flags being determined by 
\[
E_i^a = {\rm Ker}( E_{p_i} \twoheadrightarrow \pi_*F_i^a).
\]
The Higgs field $\Phi:E\to E\otimes_C K_C(D)$ is the pushforward
$\Phi=\pi_*y$ of the multiplication map $F\to F\otimes_P\pi^*K_C(D)$ 
by the tautological section $y\in H^0(P, \pi^*K_C(D))$.
 
Again, a similar spectral construction
applies to parabolic Higgs 
bundles with coefficients in a line bundle $M$, as 
defined above \eqref{eq:Hitchinmap}. In that case, $P$ will be 
the total space of the line bundle $M(D)$. 
 
 \subsection{Diagonally parabolic Higgs bundles}\label{diagonalres}
 For further reference it will be convenient to note here that the moduli 
 stack of semistable Higgs bundles contains a closed substack where 
 the Higgs field has $\xi_i$-parabolic residues at each marked point $p_i$, 
 where $\xi_i$ is a collection 
  $\xi_i=(\xi_i^0,\ldots, \xi_i^{s_i-1})\in K_C(D)_{p_i}^{\oplus s_i}$. 
  Using the notion introduced in Section \ref{background}, 
  this means that $\Phi|_{p_i}:E_{p_i}\to 
  E_{p_i}\otimes K_C(D)_{p_i}$ is parabolic, and the induced maps 
  $$
 E^a_{p_i}/E^{a+1}_{p_i}\to E^a_{p_i}/E^{a+1}_{p_i}\otimes K_C(D)_{p_i}
 $$
 are of the form ${\bf 1}\otimes \xi_i^a$. Such objects will be called
 $\uxi$-parabolic, where $\uxi=(\xi_i^a)$, $1\leq i\leq k$, 
 $0\leq a\leq s_i$. The closed substack of such objects 
 will be denoted by $\CH^{ss}_{\uxi-{\rm \small par}}(C,D; \um,e,\ualpha)$. For $\xi_i=(0,\ldots, 0)$, $1\leq i\leq k$, 
one recovers the moduli stack of strongly parabolic 
Higgs bundles. 

 It will be shown in Section \ref{specholsurf} that moduli spaces of $\uxi$-parabolic Higgs bundles occur naturally in string theory.

\section{Spectral data via holomorphic symplectic 
orbifolds}\label{specholsurf}

The goal of this section is to formulate a variant of the spectral
cover construction for parabolic Higgs bundles. In this variant the
spectral data are torsion sheaves on holomorphic symplectic orbifold
surfaces. This construction is different from the standard spectral
construction from Section \ref{higgsleaves} in which the spectral data
are parabolic dimension one sheaves on the total space $P$ of the line
bundle $K_C(D)$.  The main motivation for this alternative approach
resides in string theory, where parabolic structures must arise
naturally from D-brane moduli problems rather than being specified as
additional data.

For a brief outline, suppose $C$ is a smooth projective curve equipped
with a reduced divisor $D=\sum_{i=1}^k p_i$ of marked points. We want
to describe orbifold spectral data for parabolic Higgs bundles on
$(C,D)$. Consider the moduli stack $\CH^{ss}_{\uxi-{\rm \small
    par}}(C,D; \um,e,\ualpha)$ of diagonally parabolic Higgs bundles
introduced in Section~\ref{diagonalres}. In this section we will show
that any Higgs bundle $(E,\Phi)$ in $\CH^{ss}_{\uxi-{\rm \small
    par}}(C,D; \um,e,\ualpha)$ can be represented by a spectral datum
$\wG$ which is a Bridgeland semistable pure dimension one coherent
sheaf on a certain orbifold symplectic surface
$\widetilde{S}_{\delta}$. 

The surface $\widetilde{S}_{\delta}$ depends on $C$, the divisor $D$,
and a zero dimensional subscheme $\delta$ inside
$\text{tot}(K_{C}(D))$. To describe it let $P$ denote the total space of
$K_C(D)$, and let $P_i$ be the fiber over $p_i$, $1\leq i\leq k$.  Let
$s_i\in \IZ_{\geq 1}$, $1\leq i\leq k$ be fixed positive integers and
$\delta = (\delta_i)_{1\leq i\leq k}$ be a fixed collection of degree
$s_i$ divisors
\[
\delta_{i} = \sum_{j=1}^{\ell_i} s_{i,j} \wp_{i,j}, \qquad 
 \ell_{i}\geq 1, s_{i,j} \geq 1, \ 1\leq j\leq \ell_i,  
\]
on $P_i$ for each $1\leq i\leq k$. By convention, set $s_{i,0}=0$ for
each $1\leq i\leq k$.  In Section \ref{orbisurfaces} we check that the
weighted blowup of $P$ along $\delta$ produces a holomorphic
symplectic orbifold surface ${\widetilde S}_\delta$.

The particular $\delta$ needed for the spectral description of the
Higgs bundles in $\CH^{ss}_{\uxi-{\rm \small
    par}}(C,D; \um,e,\ualpha)$ is constructed out of the eigenvalues
$\uxi$ of the residues of the Higgs fields, and the  flag types
$\um$.  Specifically we take  
$s_{i}$ to  be the number of
steps in the parabolic filtration at the point $p_{i}$ and
$\ell_{i}$
to be the number of distinct entries in the vector $\xi_{i} =
\left( \xi_{i}^{0}, \ldots, \xi_{i}^{s_{i}-1}\right) \in
K_{C}(D)^{\oplus s_{i}}_{p_{i}}$. We label the
distinct entries of $\xi_{i}$ by $\wp_{i,1},
\ldots, \wp_{i,\ell_{i}}$, and we write $s_{i,j}$ for the multiplicity
with which $\wp_{i,j}$ is repeated as a coordinate inside $\xi_{i}$. 
In other words we choose a function $\jmath : \{0, \ldots, s_{i}-1\} \to
  \{1, \ldots, \ell_{i}\}$  so that
\be\label{eq:deltaxirelation} 
\bal 
\xi_i^a = \wp_{i,\jmath(a)}, \qquad 0\leq a\leq s_i-1. 
\eal 
\ee
These choices define a zero dimensional subscheme $\delta \subset P$
and an orbifold symplectic surface $\widetilde{S}_{\delta}$. 

Then the main result of this section is the existence of  an isomorphism 
\begin{equation} \label{eq:iso.moduli}
\boxed{
\CH^{ss}_{\uxi-{\rm \small
    par}}(C,D; \um,e,\ualpha) \cong
\mathcal{M}^{ss}_{\beta}\left(\wS_{\delta}, \boldsymbol{d}\right) 
}
\end{equation}
of the moduli
stack $\CH^{ss}_{\uxi-{\rm \small
    par}}(C,D; \um,e,\ualpha)$
between semistable $\uxi$-parabolic bundles on $C$ with the moduli
stack $\mathcal{M}^{ss}_{\beta}\left(\wS_{\delta}, \boldsymbol{d}\right)$ of 
Bridgeland $\beta$-semistable pure dimension one sheaves on
$\tS_\delta$ with $K$-theory class $\boldsymbol{d} \in K_{c}^{0}(\wS_{\delta})$.

To set up this  isomorphism we first construct an identification of discrete
invariants 
\[
(r,\um,e) \longleftrightarrow \boldsymbol{d}
\]
and an identification
\[
\ualpha \longleftrightarrow \beta
\]
of the parabolic weights on the Higgs side with the Bridgeland
stability parameters $\beta$ on the spectral data side. These
identifications are based on an explicit computation of the compactly
supported $K$-theory of $\tS_\delta$.  A precise statement is formulated
in  Section \ref{equivalence}.

The simplest instance of this construction is 
$\uxi=0$, in which case 
$\uxi$-parabolic bundles are the same as strongly parabolic bundles. 
In this case the construction of ${\widetilde S}_0$ 
follows from standard root 
stack constructions in the literature, as explained below. 
\subsection{Root stacks and orbifold spectral covers}\label{rootsection}
Using the construction of \cite{parabolic_orbifold,parabolic_root}, 
parabolic Higgs bundles have been 
identified with ordinary Higgs bundles on an orbicurve in \cite{orb_higgs,HilbertHiggs}. 
This section reviews the basics of this construction following 
the algebraic approach of \cite{HilbertHiggs}.

Given the curve $C$ with marked points  $p_i$, $1\leq i\leq k$ 
one first constructs an orbicurve ${\widetilde C}$ as follows. Let 
$U=C \setminus \{p_1,\ldots, p_k\}$. 
For any point $p_i$, let ${\mathbb D}_{p_i}$, denote the 
formal disc centered at $p_i$ and ${\mathbb D}_{p_i}^\circ= {\mathbb D}_{p_i}\times_C U$ the punctured formal disc. Let 
${\varphi}_i:{\widetilde \ID}_{p_i}\to {\mathbb D}_{p_i}$ be the 
$s_i:1$ cover 
given by $z_i \mapsto z_i^{s_i}$.
There is a natural $\mu_{s_i}$-action on 
${\widetilde \ID}_{p_i}$ sending $z_i \mapsto \omega_i z_i$, where 
$\omega_i = {\rm exp}(2\pi\sqrt{-1}/s_i)$. 
The quotient stacks $[{\widetilde \ID}_{p_i}/\mu_{s_i}]$ are then glued to $U$ 
using the morphisms $\varphi_i$ to identify the open substacks 
$[{\widetilde \ID}^\circ_{p_i}/\mu_{s_i}]$ with the punctured 
disks $\ID^\circ_{p_i}$. In characteristic zero this yields a smooth Deligne-Mumford stack ${\widetilde C}$ equipped with a map 
$\nu:{\widetilde C}\to C$ which identifies $C$ with its  
coarse moduli space. 

Following \cite[Sect 2.4]{HilbertHiggs}, a Higgs bundle on
${\widetilde C}$ is a vector bundle ${\tilde E}$ equipped with a 
Higgs field ${\tilde \Phi}: {\tilde E}\to {\tilde E}\otimes_{\widetilde C}
K_{\widetilde C}$. This data determines a parabolic Higgs 
bundle on $C$ as follows. For each point $p_i$ there is a line bundle ${\tilde L}_i$ on 
${\widetilde C}$ such that ${\tilde L}_i^{s_i} = \nu^*\CO_C(p_i)$. 
Locally, $\nu^*\CO_{C}(p_i)$ corresponds to the rank one free 
$\IC[[z_i]]$-module generated by $z_i^{-s_i}$ 
while ${\tilde L}_i$ corresponds to the $\IC[[z_i]]$-module generated by $z_i^{-1}$. Now let 
\[
E = \nu_*\, {\tilde E},
\]
and
\be\label{eq:orbfiltrationA} 
F_{i}^a = \nu_*\, \big({\tilde E}\otimes_{\widetilde C} 
{\tilde L}_i^{-a}\big)
\ee
for each $1\leq i\leq k$, $0\leq a\leq s_i-1$. By the base change theorem, all direct images are 
locally free and the sheaves $F_i^a$, $0\leq a\leq s_i-1$, form a 
filtration 
\be\label{eq:orbfiltrationB}
E(-p_i) \subseteq F_i^{s_i-1} \subseteq \cdots \subseteq F_i^0 =E
\ee
for each $1\leq i\leq k$. 

For concreteness, note that any 
locally free sheaf ${\tilde E}$ is locally 
isomorphic to a sum of line bundles of the form $\oplus_{j=1}^r {\tilde L}_i^{n_{i,j}}$, corresponding to the $\IC[[z_i]]$-module 
$\bigoplus_{i=1}^r z_i^{-n_{i,j}}\IC[[z_i]]$. The morphism $\nu:\wC \to C$ is locally of the form $t_i=z_i^{s_i}$, where $t_i$ is a local coordinate 
on $C$ centered at $p_i$. The direct image $E=\nu_*{\tilde E}$ 
corresponds locally to the $\IC[[t_i]]$-module obtained by taking 
the $\mu_{s_i}$-fixed part of 
$\bigoplus_{i=1}^r z_i^{-n_{i,j}}\IC[[z_i]]$. The subsheaves $F_i^a$ are 
obtained similarly by taking the $\mu_{s_i}$-fixed part of 
$\bigoplus_{i=1}^r z_i^{-n_{i,j}+a}\IC[[z_i]]$, $0\leq a\leq s_i-1$.

The filtration \eqref{eq:orbfiltrationB} determines a flag 
\be\label{eq:orbflagA}
E_i^a = {\rm Ker}(E_{p_i} \otimes \CO_{p_i} \twoheadrightarrow 
E/F_i^a) 
\ee
in the fiber $E_{p_i}$, hence one obtains a quasi-parabolic 
bundle $E^\bullet$ on $C$.  
Note also that the snake lemma yields an
isomorphism 
\be\label{eq:orbflagB} 
E_i^a \otimes \CO_{p_i} \simeq F_i^a / E(-p_i) 
\ee
for all $0\leq a\leq s_i$, $1\leq i\leq k$. 

According to \cite[Prop. 2.16]{HilbertHiggs} 
assigning the quasi-parabolic bundle $E^\bullet$ to
 ${\tilde E}$ 
yields an equivalence of groupoids. Moreover, the degree of ${\tilde E}$ 
as an orbibundle equals the parabolic degree of $E^\bullet$ with 
weights $\alpha_a = {a\over s_i}$.

Next consider a Higgs field ${\tilde \Phi}:{\tilde E} \to {\tilde E}\otimes_{\widetilde C}K_{\widetilde C}$ on the stack and note the 
isomorphisms
\[
K_{\widetilde C} \simeq \otimes_{i=1}^k {\tilde L}_i^{(s_i-1)} 
\otimes_{\widetilde C} \nu^*K_C 
\simeq \otimes_{i=1}^k {\tilde L}_i^{-1} 
\otimes_{\widetilde C} \nu^*K_C(D).
 \] 
Then $\Phi=\nu_*{\tilde \Phi}: E \to E\otimes_C \Omega^1_C(D)$ 
is a Higgs field on $E$ and  equations \eqref{eq:orbfiltrationA} imply that 
$$\Phi(F_i^a) \subseteq F_i^{a+1}\otimes_C \Omega^1_C(D)$$
for all $0\leq a\leq s_i-1$, $1\leq i\leq k$. 
Using isomorphisms \eqref{eq:orbflagB}, this 
implies that the residue ${\rm Res}_{p_i}(\Phi)$ is strongly parabolic with respect to the flag \eqref{eq:orbflagA} 
for each $i\in \{1,\ldots, k\}$.

According to \cite[Prop. 20]{HilbertHiggs},
 the above construction yields 
a one-to-one correspondence between Higgs fields ${\tilde 
\Phi}$ on the orbifold ${\widetilde C}$ and Higgs fields $\Phi$ on 
$C$ with strongly parabolic residues with respect to the flag $E_i^\bullet$ 
at each $p_i$, $1\leq i\leq k$. 
Furthermore, the degree of the orbifold bundle ${\tilde E}$ 
equals the parabolic degree ${\rm deg}(E,\ualpha)$ for 
special values of the weights 
\be\label{eq:orbiweights} 
\alpha_{i,a} = {a\over s_i}, \qquad 0\leq a\leq s_i-1, \qquad 
1\leq i\leq k.
\ee
Based on this observation, it is straightforward to check that 
this correspondence maps (semi)stable orbibundles to 
(semi)stable parabolic Higgs bundles with weights \eqref{eq:orbiweights}. Since it works for flat families as well, 
it yields isomorphisms of moduli stacks. 

For completeness, note that the fiber ${\tilde E}_{i,0}$ of 
${\tilde E}$ at the closed point $0\in 
[{\widetilde \ID}_{p_i}/\mu_{s_i}]$ carries a natural action 
of the stabilizer group $\mu_{s_i}$. Hence it decomposes 
into irreducible representations, 
\be\label{eq:fractdecomp}
{\tilde E}_{i,0} \simeq \bigoplus_{a=0}^{s_i-1} R_{i,a}^{\oplus n_{i,a}},
\ee
where $R_{i,a}$ denotes the one dimensional representation of $\mu_{s_i}$ with character $\omega_i^a$. 
Then 
{\cite[Lemma 2.19]{HilbertHiggs} 
proves that the discrete invariants $m_{i,a}$ 
of the flag \eqref{eq:orbflagA} are given by $m_{i,a}=n_{i,a}$,
for $0\leq a\leq s_i-1$.
In string theoretic language this means that the flag type encodes 
the fractional  charges with respect to the twisted sector 
Ramond-Ramond fields.

Using the standard spectral cover construction, 
an orbifold Higgs bundle $({\tilde E}, {\tilde \Phi})$ corresponds 
to a pure dimension one sheaf ${\tilde F}$ on the total 
space ${\widetilde S}$ of  
$K_{\widetilde C}$, finite with respect to the natural projection 
$\pi : {\widetilde S} \to {\widetilde C}$. The bundle ${\tilde E}$ is
obtained by push-forward, ${\tilde E}=\pi_* {\tilde F}$, and 
the Higgs field ${\tilde \Phi}$ is the push-forward of the 
multiplication map  ${\tilde F} \to {\tilde F} \otimes \pi^*K_\wC$ by the tautological section. This is a one-to-one 
correspondence which holds for  flat families as well, preserving 
(semi)stability. Therefore it induces again isomorphisms of moduli stacks of semistable objects. As shown in detail in the next subsection, 
one can in fact obtain arbitrary values of the parabolic weights 
using Bridgeland stability conditions for pure dimension one sheaves on $\tS$. 

Finally, all above statements generalize immediately to Higgs fields 
with values in an arbitrary line bundle $M$ on $C$. Namely, let 
\[
{\widetilde M} = \otimes_{i=1}^k {\tilde L}_i ^{s_i-1}
\otimes_{\widetilde C} 
\nu^* M.
\]
Then there is again a one-to-one correspondence between 
strongly parabolic Higgs bundles $(E,\Phi)$, with $\Phi:E \to E\otimes_C 
M(D)$ and Higgs orbibundles $({\tilde E},{\tilde \Phi})$ 
with ${\tilde \Phi}:{\tilde E} \to {\tilde E}\otimes_{\widetilde C} 
{\widetilde M}$. The latter admit again a spectral cover presentation, 
as pure dimension one sheaves on the total space of ${\widetilde M}$. 

At the same time analogous considerations 
hold for parabolic Higgs bundles $(E^\bullet,\Phi)$ on $C$ 
with $\Phi:E\to E\otimes_C M$ a Higgs field preserving the flag 
$E_i^\bullet$ at each point $p_i$, $1\leq i\leq k$. Note that in this case $\Phi$ is regular at $p_i$ and $\Phi|_{p_i}$ is not necessarily strongly parabolic. Repeating 
the above arguments, such objects are in one-to-one correspondence 
with Higgs orbibundles $({\tilde E}, {\tilde \Phi})$ on ${\widetilde C}$, 
with ${\tilde \Phi}: {\tilde E}\to {\tilde E} \otimes_{\widetilde C}
\nu^*M$.

\subsection{Orbifold spectral data for diagonally  
parabolic Higgs bundles}\label{orbisurfaces}

As in the second paragraph of Section 
\ref{specholsurf}, 
let $\delta=(\delta_i)_{1\leq i\leq k}$ be a collection 
of degree $s_i$ divisors 
\[
\delta_i=\sum_{j=1}^{\ell_i} s_{i,j} \wp_{i,j}, \qquad 
 \ell_{i}\geq 1, s_{i,j} \geq 1, \ 1\leq j\leq \ell_i,  
\]
on $P_i$ for each $1\leq i\leq k$. 
Note that one can choose an affine chart $(x_i,y_{i,j})$ 
on $P$ centered at each point $\wp_{i,j}$, where $x_i$ is an affine 
coordinate on $C$ centered at $p_i$, and $y_{i,j}$ a linear coordinate 
on the fibers of $P$ centered at $\wp_{i,j}$.  
Let $V_{i,j}={\rm Spec}\, \IC[x_i,y_{i,j}]\subset P$ 
denote the domain of this affine coordinate chart.
Then any  collection $\delta=(\delta_i)_{1\leq i\leq k}$, 
determines a smooth orbifold surface ${\widetilde P}_{\delta}$, 
the stack theoretic weighted projective blow-up of $P$ 
at the points $\wp_{i,j}$ with weights $(s_{i,j}, 1)$ with respect 
to the affine chart $(x_i,y_{i,j})$. 

This is a standard construction in the orbifold literature employed for 
example in \cite{orbiflips,weightedblowup}. 
Following
\cite[Sect. 2.1]{weightedblowup}, the weighted projective 
blow-up of $P$ at $\wp_{i,j}$ is a quotient stack 
$[X_{i,j}/\IC^\times]$, where $X_{i,j}$ is a scheme obtained by 
gluing $(\IC^2\setminus \{0\}) \times \IC$ to $(P\setminus \{\wp_{i,j}\})\times 
\IC^\times$ along $(V_{i,j}\setminus \{0\}) \times \IC^\times$.  
In terms of linear coordinates $(u,v,t)$ on $\IC^2\times \IC$, 
the gluing map reads 
\be\label{eq:weightedblowupA} 
x_i =t^{s_{i,j}} u, \qquad y_{i,j} = t v, \qquad z= t^{-1}
\ee
where $z$ is a linear coordinate on the target $\IC^\times$. 
The $\IC^\times$-action on $X$ is induced by the $\IC^\times$-action 
\[ 
(\zeta, (u,v,t)) \mapsto (\zeta^{s_{i,j}} u, \zeta v, \zeta^{-1} t)
\]
on $(\IC^2\setminus \{0\}) \times \IC$ and the scaling action on the 
second factor of $(P\setminus \{\wp_{i,j}\})\times 
\IC^\times$.

Proceeding this way for each 
 point $\wp_{i,j}$, one obtains a 
smooth Deligne-Mumford stack ${\widetilde P}_\delta$ equipped with a natural projection $\eta : {\widetilde P}_\delta \to P$. In terms of the above affine coordinates, the projection map is given by 
\be\label{eq:locblowup}
x_i = u t^{s_{i,j}}, \qquad y_{i,j}= v t. 
\ee
For each pair $(i,j)$
the inverse image $\eta^{-1}(V_{i,j})$ is an 
open substack of $\wP_\delta$ isomorphic to 
the smooth toric stack ${\mathcal X}_{i,j}=[{\rm Spec}\, \IC[u,v,t]/\IC^\times]$ with 
weights $(s_{i,j},1,1)$. According to 
\cite[Prop. 4.5]{orbiflips}, the open substack of 
${\mathcal X}_{i,j}$ where $u\neq 0$ is isomorphic to a 
quotient stack $[{\rm Spec}\, \IC[{\tilde x}, {\tilde y}]/\mu_{s_{i,j}}]$, where 
\[
{\tilde x}^{s_{i,j}} = ut^{s_{i,j}}, \qquad {\tilde y}^{s_{i,j}} = u^{-1} v^{s_{i,j}}.
\] 
This shows that there is an orbifold point 
${\widetilde \wp}_{i,j}$ with stabilizer $\mu_{s_{i,j}}$ mapping to 
each blow-up center $\wp_{i,j}$. All other closed points of 
${\widetilde P}_{\delta}$ have trivial stabilizers. The reduced exceptional 
divisor ${\Xi}_{i,j}$ corresponding to $\wp_{i,j}$ 
is  isomorphic to a 
weighted projective line $\IP[s_{i,j}, 1]$ passing through the orbifold 
point. Moreover, the total transform of the fiber $P_i$ is 
\be\label{eq:fiberpullbackA}
\eta^* P_i = P_i' + \sum_{j=1}^{\ell_i} s_{i,j} {\Xi}_{i,j}
\ee
where $P_i'$ is a line on ${\widetilde P}_\delta$ disjoint from the orbifold 
points. 

In terms of the local coordinates $({\tilde x}, {\tilde y})$ 
exceptional divisor is given by ${\tilde x} =0$, and the 
projection map \eqref{eq:locblowup} is given by 
\be\label{eq:locblowupB} 
x_i = {\tilde x}^{s_{i,j}}, \qquad y_{i,j} = {\tilde x}{\tilde y}.
\ee
Then 
\[ 
\eta^* (dx_i\wedge dy_{i,j}) = s_{i,j} {\tilde x}^{s_{i,j}} 
d{\tilde x} \wedge d{\tilde y},
\] 
which implies that the canonical class $K_{\wP_\delta}$, is given by 
\[
K_{\wP_\delta} =\eta^*K_P + \sum_{i=1}^k \sum_{j=1}^{\ell_i} s_{i,j} \Xi_{i,j}. 
\]
Since $P$ has canonical class 
\[ 
K_P = -\sum_{i=1}^k P_i,
\]
using equation \eqref{eq:fiberpullbackA},
the canonical class of $\wP_\delta$ is 
\be\label{eq:Kblowup} 
K_{{\widetilde P}_\delta} = -\sum_{i=1}^k P_i'.
\ee
Therefore the complement \[
\tS_\delta = \wP_\delta \setminus \cup_{i=1}^k P_i'
\]
is a holomorphic symplectic orbifold surface. 
This holomorphic symplectic surface will be used 
below in the construction of spectral data for 
diagonally parabolic Higgs bundles. 

{The first task is to classify the 
discrete invariants of compactly supported torsion sheaves on $\tS_\delta$. 
Consider the natural bilinear pairing
\be\label{eq:Kpairing}
(\ ,\ ) : K^0(\wP_\delta) \times K^0_{cpt}(\wP_\delta) \to 
\IZ
\ee
defined by  
\[ 
([{\widetilde F}], [{\widetilde G}]) = 
\sum_{i\in \IZ} (-1)^i {\rm dim} {\rm Ext}^i({\widetilde F}, {\widetilde G}) 
\]
for any two coherent sheaves ${\widetilde F}, {\widetilde G}$ on $\wP_\delta$, where ${\widetilde G}$ has 
compact support. 
The discrete invariants of compactly supported 
coherent sheaves on $\wP_\delta$ will be numerical 
equivalence classes in 
\be\label{eq:Gammaquot} 
\Gamma_{cpt}(\wP_\delta) = K^0_{cpt}(\wP_\delta)/
K^0(\wP_\delta)^{\perp}, 
\ee
where 
\[ 
K^0(\wP_\delta)^{\perp} = \big\{ \kappa\in K^0_{cpt}(\wP_\delta)\, |\, ( \kappa, \gamma) =0,\ 
\forall \gamma \in K^0(\wP_\delta)\, \big\}.
\]
For future reference, let 
\[
\Gamma(\wP_\delta) = 
K^0(\wP_\delta)/K^0_{cpt}(\wP_\delta)^\perp
\]
be defined analogously. Then 
 \eqref{eq:Kpairing} descends to a nondegenerate 
bilinear pairing
\be\label{eq:Gpairing}
\chi : \Gamma(\wP_\delta) \times 
\Gamma_{cpt}(\wP_\delta) \to \IZ.
\ee
Note that similar definitions apply equally well to $P$, 
resulting in a nondegenerate bilinear pairing 
\[ 
\Gamma(P)\times \Gamma_{cpt}(P)\to \IZ.
\]
Since $\pi:P\to C$ is the total space of a line bundle 
over $C$, $\Gamma(P)$ is generated by the line bundle 
classes 
\[
[\CO_P], \quad [\CO_P(f)] 
\] 
where $f=\pi^{-1}(p)$ is a fiber of $\pi$ over a generic 
point $p\in C \setminus D$. The compactly supported 
lattice $\Gamma_{cpt}(P)$ is generated by the 
sheaf classes 
\[
[\CO_{\sigma}], \quad [\CO_\wp]
\]
where $\wp \in P$ is a generic point, $\pi(\wp) 
\in C\setminus D$. 

Then an
 explicit presentation of $\Gamma(\wP_\delta)$, 
$\Gamma_{cpt}(\wP_\delta)$ 
follows from  
\cite[Thm. 2]{weightedblowup},  which proves a structure 
result for 
the derived category 
$D^b(\wP_\delta)$. According to loc. cit. $D^b(\wP_\delta)$ admits a semiorthogonal decomposition 
\be\label{eq:blowupdercat}
D^b(\wP_\delta) = \langle \eta^*(D^b(P)), \CT_{i,j}^l\rangle 
\ee
where $\CT_{i,j}^l$ are extensions by zero of standard line bundles 
on the exceptional divisors $\Xi_{i,j} \simeq \IP[s_{i,j},1]$: 
\[
\CT_{i,j}^l = \CO_{\Xi_{i,j}}(l), \qquad 0\leq l \leq s_{i,j}-1. 
\] 
Here $\CO_{\Xi_{i,j}}(l)$ denotes the $l$-th power of the
 line bundle $\CO_{\Xi{i,j}}(1)$ on the weighted 
projective line $\IP[s_{i,j},1]$. In the present context,
$\CO_{\Xi_{i,j}}(-1)$ is the restriction:
\be\label{eq:exdivrestrict} 
\CO_{\Xi_{i,j}}(-1)\simeq \CO_{\wP_\delta}(\Xi_{i,j})\big|_{\Xi_{i,j}}
\ee 
on $\Xi_{i,j}$, as shown in the 
proof of 
\cite[Prop. 3]{weightedblowup}.

In the following it will be more convenient to work with the 
alternative $K$-theory generators 
\be\label{eq:altgenA}
\CQ_{i,j}^l = \CT_{i,j}^{s_{i,j}-l-1}\otimes_{\wP_\delta} 
\CO_{\wP_\delta} (K_{\wP_\delta}).
\ee
Using \eqref{eq:fiberpullbackA}, \eqref{eq:Kblowup}
and 
\eqref{eq:exdivrestrict}, one has an isomorphism 
\be\label{eq:altgenB}
\CQ_{i,j}^l \simeq \CO_{\Xi_{i,j}}(-l-1) 
\ee
for all $(i,j)$ and all $0\leq l\leq s_{i,j}-1$,
since $\eta^*P_i\big|_{\Xi_{i,j}}=0$. 
The semiorthogonal decomposition \eqref{eq:blowupdercat}, implies  
that $\Gamma(\wP_\delta)$ is generated by the numerical equivalence classes of the sheaves 
\[ 
\eta^*\CO_P =\CO_{\wP_\delta}, \quad 
\eta^*\CO_P(f), \quad \CQ_{i,j}^l.
\]

Now suppose  $\wG$ is a (nonzero) pure dimension one sheaf  on ${\widetilde P}_\delta$ 
with compact support. 
Then the discrete invariants of $\wG$ will be defined as 
\be\label{eq:discrinvA} 
n(\wG) = \chi(\eta^*\CO_P, \wG), \quad r(\wG) = 
n(\wG)-\chi(\eta^*\CO_P(f),\wG),\quad 
d_{i,j}^l(\wG) = -\chi(\CQ^l_{i,j}, \wG)
\ee
with $1\leq i\leq k$, $1\leq j\leq \ell_i$, 
 $0\leq l\leq s_{i,j}-1$. 
Note that $\eta_*\wG$ is a compactly supported torsion 
sheaf on $P$, which is pure dimension one on the 
complement $P\setminus \cup_{i=1}^k P_i$. Then 
it is straightforward to check that 
$\ch_1(\eta_*\wG)= r(\wG)\sigma$, and 
\be\label{eq:rankrel}
r(\wG) = \chi(\CO_P, \eta_*\wG \otimes_P \CO_f)  >0, 
\ee
for a generic fiber $f$ disjoint from $P_i$, $1\leq i\leq k$.

Next suppose in addition that $\wG$ has compact support 
in $\tS_\delta \subset \wP_\delta$, therefore disjoint 
from the strict transforms 
$P_i'$, $1\leq i\leq k$ of the marked fibers. 
Then, as shown below,  there is a relation of the form
\be\label{eq:discrinvC} 
r(\wG) = \sum_{j=1}^{\ell_i} \sum_{l=0}^{s_{i,j}-1} 
d_{i,j}^l
\ee
for each $1\leq i\leq k$. In conclusion the discrete 
invariants of such sheaves can be labelled by $(n(\wG),d_{i,j}^l(\wG))$, keeping in mind that they satisfy 
relations \eqref{eq:discrinvC}. 

Using the isomorphism 
$\CO_{\wP_\delta}(\Xi_{i,j})|_{\Xi_{i,j}}\simeq \CO_{\Xi_{i,j}}(-1)$ and equations \eqref{eq:altgenB},
each $\CQ_{i,j}^l$ has a locally free resolution 
\[ 
0\to \CO_{\wP_\delta}(l\Xi_{i,j}) \to 
\CO_{\wP_\delta}((l+1)\Xi_{i,j}) \to \CQ_{i,j}^l\to 0.
\]
This yields 
\[
d_{i,j}^l = \chi(\CO_{\wP_\delta}(l\Xi_{i,j}),\wG)-
\chi(\CO_{\wP_\delta}((l+1)\Xi_{i,j}) ,\wG).
\]
Summing the above relations from $l=0$ to $l=s_{i,j}-1$ 
yields 
\[ 
\sum_{l=0}^{s_{i,j}-1} 
d_{i,j}^l = \chi(\CO_{\wP_\delta},\wG)-\chi( \CO_{\wP_\delta}(s_{i,j}\Xi_{i,j}), \wG) =
\chi( \CO_{\wP_\delta}, \wG\otimes_{\wP_\delta} \CO_{s_{i,j}\Xi_{i,j}})
\]
Since the support of $G$ is disjoint from
the strict transforms $P_i'$ 
for each $1\leq i\leq k$, summing the above 
relation from $j=1$ to $j=\ell_i$ and using 
relation \eqref{eq:fiberpullbackA}, one obtains 
\[ 
\sum_{j=1}^{\ell_i} 
\sum_{l=0}^{s_{i,j}-1} d_{i,j}^l = 
\chi(\CO_{\wP_\delta}, \wG\otimes_{\wP_\delta} \CO_{\eta^*P_i}) = 
\chi(\CO_{\wP_\delta}, \wG\otimes_{\wP_\delta} \CO_{\eta^*f}).
\]
Since $\eta^*P_i$ is linearly equivalent with the $\eta^*f$, 
for $f$ an arbitrary fiber of $P$ over $C\setminus D$, 
it follows that 
\[ 
\sum_{j=1}^{\ell_i} 
\sum_{l=0}^{s_{i,j}-1} d_{i,j}^l = \chi(\CO_{\wP_\delta}, \wG\otimes_{\wP_\delta} \CO_{\eta^*f}).
\]
Using relations \eqref{eq:rankrel}, this implies relation \eqref{eq:discrinvC}. }

For the spectral cover construction one also needs a suitable 
stability condition for the torsion  sheaves $\wG$. 
This will be a twisted stability condition depending on a 
function $\beta: K^0_{cpt}(\wS_\delta)\to \IR$,
which should be regarded as the real part of a 
Bridgeland stability function on the $K$-theory of $\wP_\delta$. 
The resulting twisted 
stability condition should  be  regarded as the specialization 
of a Bridgeland stability condition to dimension one sheaves. 
In string theory, $\beta$ is the expectation value of the orbifold 
flat $B$-field on $\tS_\delta$. 

For any nontrivial pure dimension sheaf $\wG$ as above define the 
twisted slope 
\[
\mu_\beta(\wG) = {\chi(\wG) + \beta([\wG])\over d(\wG)}.
\]
Such a sheaf will be called $(\omega, \beta)$-(semi)stable if 
\[
\mu_{(\omega,\beta)}(\wG') \ (\leq)\ \mu_{(\omega,\beta)}(\wG) 
\]
for any proper nontrivial subsheaf $0\subset \wG'\subset \wG$. 

In order to conclude this section, note that
the above construction is especially simple 
in the case 
where all $s_{i,j}=1$. Then the surface $\wP_q$ is just the standard scheme theoretic  
blow-up of $P$ at the points $\wp_{i,j}$.

\subsection{Equivalence of moduli stacks}\label{equivalence}
We now summarize the constructions made up to this point and state the main claim.

\subsubsection{The parabolic moduli stack}\label{parstack} 
The stack $\CH_{\uxi}^{ss}(C,D; \um, e, \ualpha)$
was constructed in section \ref{diagonalres}. 
It is the moduli stack of semistable 
$\uxi$-parabolic Higgs bundles on $C$ 
with poles on the divisor $D =\sum_{i=1}^k p_i$ and
with discrete 
invariants $\um = (m_{i,a})$, weights 
$\ualpha =(\alpha_{i,a})$, $0\leq a\leq s_i-1$, $1\leq i\leq k$ and degree $e$.
Here $s_i$ is the number of steps in the parabolic filtration at point $p_i$, 
and the $\um = (m_{i,a})$ specify the type of the flag, as in section \ref{background}:
the dimension of the flag spaces $V_i^a$ are determined by equation \eqref{oifd}, namely:
${\rm{dim}}\, (V_i^a/V_i^{a+1})=m_{i,a},   \quad 0\leq a\leq s_i-1.$
Finally, $\uxi = (\xi_i) = (\xi_i^a)$ specifies the residues of the Higgs fields 
on the subquotients $V_i^a/V_i^{a+1}$, with  
$\xi_{i} = \left( \xi_{i}^{0}, \ldots, \xi_{i}^{s_{i}-1}\right) \in
K_{C}(D)^{\oplus s_{i}}_{p_{i}}$.

\subsubsection{Some combinatorics}\label{combinatorics}
For each $i$, $1 \leq i \leq k,$  let 
$\{ \wp_{i,j}$, $j \in J_i \}$
be the set consisting of  the $\xi_i^a$ (ignoring multiplicities), 
and let $\ell_i$ be the cardinality of the index set $J_i$. 
The relation 
$  \wp_{i,\jmath(a)} = \xi_i^a$
determines a natural map $\jmath: \{ 0, \dots, s_i -1 \} \to J_i$.
Let $s_{i,j}$ be the number of $a$'s that map to a given $j$,
and denote the set of such $a$'s, in increasing order: 
$  \{a_{ij}^0, \dots, a_{ij}^{s_{ij} -1}\} = \{a_{i,j}^l \ | \  l  \in S_{i,j} \}$, where
$S_{i,j} = \{0, \dots, s_{i,j} -1\}$.
There is a natural bijection
${\sf a}:  \sqcup_{j \in J_i} S_{i,j}  \to  \{ 0, \dots, s_i -1 \} $
sending $(j,l) \mapsto a_{ij}^l$.
In particular $ \sum_{j \in J_i} s_{i,j} =s_i$.
The composition $\jmath \circ {\sf a}$ sends $S_{i,j}$ to $j$.

\subsubsection{The symplectic orbifold 
surface}\label{symporbsurf}

The orbi-surface $\tS_\delta$ was
constructed in the beginning of section \ref{orbisurfaces}. 
We start with the total space $P$ of the line bundle $K_C(D)$ on $C$, 
and in the fiber $P_i$ above each marked point $p_i \in C$ we fix a divisor 
$\delta_i=\sum_{j=1}^{\ell_i} s_{i,j} \wp_{i,j}$
consisting of
$ \ell_{i}\geq 1$ 
points 
$ \wp_{i,j},   \ 1\leq j\leq \ell_i,$ 
with assigned multiplicities
$s_{i,j} \geq 1.$
We then consider the stack theoretic weighted projective blowup 
$\wP_\delta$
of $P$ at the $ \wp_{i,j}$ with weights $(s_{ij},1)$. 
It is a smooth orbifold surface
containing a unique orbifold point above each $\wp_{i,j}$. 
Our symplectic orbifold surface  $\tS_\delta$ 
is then obtained from this as the complement  
$\tS_\delta = \wP_\delta \setminus \cup_{i=1}^k P_i'$, 
where $P_i'$ is the proper transform of the fiber $P_i$.

\subsubsection{The orbifold spectral stack}\label{specstack} 

The stack $\CM^{ss}_\beta(\tS_\delta,\{d_{i,j}^l\}, n)$ 
was also constructed in section \ref{orbisurfaces}. 
It is the moduli stack of compactly supported 
$\beta$-semistable pure dimension one sheaves 
$\wG$ on 
$\tS_\delta$ with 
{discrete invariants
\[ 
n(\wG)=\chi(\CO_{\tS_\delta},\wG)=n,\quad 
 d_{i,j}^l(\wG)= 
\chi(\CO_{i,j}^l, \wG)=d_{i,j}^l
\]
for $1\leq i\leq k$, $1\leq j\leq \ell_i$, $0\leq l\leq s_{i,j}$. 
For the purpose of the spectral construction, the 
 twisted stability condition will be  specified by a 
function $\beta: \Gamma_{cpt}(\wP_\delta)\to \IR$, 
\be\label{eq:stabfctA}
\beta(\gamma) = 
\sum_{i,j,l} \beta_{i,j}^l \chi(\CQ_{i,j}^l,\, \gamma)
\ee
with $\beta_{i,j}^l\in \IR$. }

\subsubsection{Equivalence of moduli stacks.}\label{modulieq} 

Our main result is:

\noindent There is an isomorphism of stacks 
\be\label{eq:modulisom}
\CH_{\uxi}^{ss}(C,D; \um, e, \ualpha) \simeq \CM^{ss}_\beta(\tS_\delta,\{d_{i,j}^l\}, n),
\ee
where the labels on the right hand side are determined by those on the left:
\begin{itemize}
\item The $J_i$, the $s_{i,j}$, and the map $a:  \sqcup_{j \in J_i} S_{i,j}  \to  \{ 0, \dots, s_i -1 \} $ are determined by the combinatorics of the $\xi_i$ as above.
\item $\beta$ is determined by: 
{ $\beta_{i,j}^l:= \alpha_{i,{\sf a}(j,l)}$}
\item $\delta = (\delta_i)$, where $\delta_i = \sum_{j \in J_i} s_{i,j} \wp_{i,j} $
\item $d_{i,j}^l = m_{i,{\sf a}(j,l)}$
\item $n=e+r(g-1)$.
\end{itemize}

\subsection{Outline of the proof}\label{proofoutline} 
The construction of a natural morphism
from pure dimension 
sheaves on $\tS_\delta$ to $\uxi$-parabolic 
Higgs bundles is sketched below.   
An inverse morphism can be constructed in principle using 
methods analogous to \cite{parabolic_root}. The details
of this construction and a complete proof of the above stack isomorphism will appear elsewhere. 

{
Let $\beta:\Gamma_{cpt}(\tS_\delta) \to \IR$ be  a 
stability function of the form \eqref{eq:stabfctA}. 
Suppose $\{ \beta_{i,j}^l\, |\, 1\leq j\leq \ell_i,\ 0\leq l\leq s_{i,j}-1\} $ are  pairwise 
distinct for each $1\leq i\leq k$, 
and satisfy the inequalities
\be\label{eq:betachamber} 
0<\beta_{i,j}^0 < \beta_{i,j}^1 < \cdots < \beta_{i,j}^{s_{i,j}-1}<1
\ee
independently, for 
 each fixed values $1\leq i\leq k$, $1\leq j\leq \ell_i$.

Such a function $\beta$  determines a collection
of combinatorial data as defined in Section \ref{combinatorics}. 
Since the $\beta_{i,j}^l$ are assumed pairwise distinct for each $i$, there is a bijection 
${\sf a}: \sqcup_{j\in J_i} S_{ij} \to \{0, \ldots, s_i-1\}$ defined by 
\[
{\sf a}(j,l) = |\{ \beta_{i,j'}^{l'} < \beta_{i,j}^l\}|.
\]
Here $|A|$ denotes the cardinality of the finite set $A$. 
One also has the function 
\[
j:={\sf p}\circ {\sf a}^{-1} : \{0, \ldots, s_i-1\} 
\to J_i
\]
where 
\[ 
{\sf p}: \sqcup_{j\in J_i} S_{ij}\to J_i
\]
is the natural projection. 

Suppose $\wG$ is a $\beta$-stable 
pure dimension one sheaf on $\wP_\delta$ 
with compact support in $\tS_\delta$. 
Using the  
spectral correspondence in Section \ref{higgsleaves},  
it suffices to construct a stable parabolic 
pure dimension sheaf $F$ 
on $P$ with parabolic structure along the marked fibers $P_i$, $1\leq i\leq k$. By pushforward along $\pi:P\to C$, 
one will then obtain a stable $\uxi$-parabolic bundle 
on $C$ with combinatorial data ${\sf a}, 
\jmath$ at each marked point.

Let $F=\eta_*\wG$, which is a 
pure dimension one sheaf with compact support
on $P$. 
 For each $1\leq j\leq \ell_i$ there is a canonical filtration 
\be\label{eq:GfiltrationB}
0\subset \wG(-s_{i,j}\Xi_{i,j}) \subset \cdots \subset \wG(-\Xi_{i,j}) \subset \wG
\ee
of $\CO_{\wP_\delta}$-modules. Moreover, relations \eqref{eq:fiberpullbackA}
yield an isomorphism 
\[
F(-P_i) \simeq \eta_*(\wG \otimes_{\wP_\delta} \eta^*\CO_P(-P_i))
\simeq \eta_* \wG(-\Xi_i) 
\]
where ${\Xi}_i =\sum_{j=1}^{\ell_i} s_{i,j} {\Xi}_{i,j}$. 
Therefore for fixed $(i,j)$ the filtration \eqref{eq:GfiltrationB} 
yields a filtration 
\be\label{eq:FfiltrationB}
0\subset F(-P_i) \subset \eta_* \wG(-(r_{i,j}-1)\Xi_{i,j}) \subset 
\cdots \subset \eta_* \wG(-\Xi_{i,j}) \subset F.
\ee
Taking quotients one obtains a sequence of surjective morphisms 
\be\label{eq:parblocks} 
F|_{P_i} \twoheadrightarrow F_{i,j}^{s_{i,j}-1} \twoheadrightarrow \cdots 
\twoheadrightarrow  F_{i,j}^{1} 
\ee
for each pair $(i,j)$. 

The sequences \eqref{eq:parblocks} are then assembled into 
the following quasi-parabolic structure along $P_i$: 
\be\label{eq:quasiparA}
F|_{P_i}\twoheadrightarrow F_{i}^{s_i-1} \twoheadrightarrow \cdots \twoheadrightarrow F_i^1 \twoheadrightarrow 
F_i^0=0 
\ee
where 
\[
F_i^a =\bigoplus_{\substack{j\in J_i,\ 0\leq l\leq s_{ij}-1\\
{\sf a}(j,l)\leq a}} F_{i,j}^l.
\]
The epimorphisms in \eqref{eq:quasiparA} are canonically determined by those in \eqref{eq:parblocks}. 
Using the weights $\alpha_{i,a} = \beta_{i,j}^l$ with 
 ${\sf a}(j,l) = a$, one obtains a parabolic pure dimension 
one sheaf $F$ on $P$, as claimed above. 

As stated above, compatibility with stability conditions and 
the construction of the inverse morphism will be left 
for future work. 
}

\section{Orbifold stable pairs and parabolic ADHM sheaves}\label{orbisection}

This section introduces stable pair invariants of local orbifold curves and 
explains their relation with parabolic ADHM sheaves on ordinary curves. 

Let $C$ be a smooth projective curve over $\IC$, 
$D=\sum_{i=1}^k p_i$ a reduced effective divisor on $C$.
Let ${\widetilde C}$ be a root stack as in Section 
\ref{rootsection} with stabilizers $\mu_{s_i}$, $1\leq i\leq k$ at 
each marked point. 
Let $\wX$ be the total space 
of a 
rank two bundle ${\widetilde M}_1\oplus {\widetilde M}_2$ 
on ${\widetilde C}$, where ${\widetilde M}_1,{\widetilde M}_2$ are 
line bundles such that ${\widetilde M}_1\otimes_{\widetilde C} {\widetilde M}_2 
\simeq K_{\widetilde C}$. Hence ${\widetilde X}$ is a smooth Calabi-Yau three dimensional Deligne-Mumford stack with generically trivial stabilizers. 

A stable pair on ${\widetilde X}$ will be defined as a pair $({\tilde F}, 
{\tilde s})$ where 
\begin{itemize}
\item 
${\tilde F}$ is a pure dimension one 
sheaf with proper support, finite-to-one over ${\widetilde C}$, and 
\item ${\tilde s}: \CO_Y \to {\tilde F}$ is a section with zero 
dimensional cokernel. 
\end{itemize}

The string theoretic derivation of the HLRV formula is based on a 
relation between orbifold stable pairs and parabolic Higgs bundles
which generalizes the similar relation found in \cite{wallpairs,BPSPW} 
for ordinary Higgs bundles. In order to understand this relation in detail, 
note that stable pairs on ${\widetilde X}$ admit a presentation in 
terms of ADHM sheaves on the orbicurve ${\widetilde 
C}$ by analogy with \cite{modADHM}. 

An ADHM sheaf on a curve $C$ with coefficient line bundles 
$M_1,M_2$ was defined 
\cite{modADHM} as a collection 
$(E,\Phi_1,\Phi_2, \phi,\psi)$ where  $E$ is a vector bundle on $C$
and $\Phi_{j}:E\to E\otimes M_j$, $j=1,2$ , 
$\phi:E \to M_1\otimes_C M_2$, 
$ \psi:\CO_C\to E$ are morphisms of sheaves 
satisfying the ADHM relation 
\be\label{eq:ADHMrelation}
(\Phi_1\otimes 1_{M_2})\circ(\Phi_2) - (\Phi_2\otimes 1_{M_1})\circ \Phi_1
+ (\psi\otimes 1_{M_1\otimes_C M_2}) \circ \phi =0.
\ee
Such an object $\CE=(E,\Phi_1,\Phi_2, \phi,\psi)$ is called asymptotically stable 
if the restriction $\CE|_x$ is a cyclic ADHM quiver representation for all but finitely 
many closed points $x\in C$. Equivalently, the subsheaf 
${\rm Im}(\psi)\subset E$ generates $\CE$ as a quiver sheaf at all closed points 
$x\in C$ except a finite set.

The above definitions admit a straightforward generalization 
to an orbicurve ${\widetilde C}$ 
equipped with two line bundles ${\widetilde M}_1, {\widetilde M}_2$. 
By analogy with \cite[Sect. 7]{modADHM} one obtains an isomorphism between the moduli space of stable pairs on ${\widetilde X}$ 
and the moduli space of asymptotically stable orbifold ADHM data 
on ${\widetilde C}$.  
Moreover, these moduli spaces carry natural perfect obstruction theories, 
which are also identified by this isomorphism. 
The details are not essential for the main goals of this paper and will be omitted. 
Instead it will be helpful to note that for certain choices of the line 
bundles ${\widetilde M}_1, {\widetilde M}_2$ orbifold stable pairs are further 
identified with parabolic ADHM sheaves on the ordinary curve $C$. This second identification also provides an efficient construction of the perfect 
obstruction theory, as shown below. 

\subsection{ADHM parabolic structure}\label{ADHMpar}

Suppose ${\widetilde M}_1=\nu^*M^{-1}$, ${\widetilde M}_2= 
K_{\widetilde C}\otimes_{\widetilde C}\nu^*M$ for a line bundle $M$ 
on $C$. In this case $\wX$ will be denoted by $\wY_M$ as 
in Section \ref{mainconj}. 

Using the correspondence in Section \ref{rootsection}, 
an ADHM sheaf ${\widetilde \CE}$ on ${\widetilde C}$ 
 with coefficients $({\widetilde M}_1, {\widetilde M}_2)$ 
corresponds to an ADHM sheaf $\CE$ on $C$ with 
coefficient line bundles $M_1=M^{-1}$, $M_2=K_C\otimes_C M (D)$. 
Moreover, one obtains a flag $E_{i}^\bullet$ in the fiber of $E$ at each 
marked point satisfying natural compatibility conditions with 
the two Higgs fields $\Phi_1:E\to E\otimes_{C} M^{-1}$, 
$\Phi_2:E\to E\otimes_C K_C\otimes_C M (D)$. 
This yields the following definition for 
(quasi-)parabolic ADHM sheaves. 

A quasi-parabolic ADHM sheaf on $C$ of type 
$\um$ is a collection $(E^\bullet, \Phi_1, \Phi_2, 
\phi, \psi)$, where 
\begin{itemize} 
\item[(a)] $E^\bullet$ is a  quasi-parabolic bundle 
on $C$ of type $\us$.
\item[(b)] $\Phi_1: E\to E\otimes M_1$ and 
$\Phi_2:  E\to E\otimes M_2(D)$ are  morphisms of sheaves 
such that $\Phi_1|_{p_i}: E_{p_i} \to E_{p_i}$ is 
parabolic and $\Phi_2|_{p_i}:E(D)|_{p_i} \to E\otimes_C M_2(D)|_{p_i}$ 
is strongly parabolic for all $1\leq i \leq k$.
\item[(c)] $\phi:E \to M_1\otimes_C M_2$, $\psi:\CO_C 
\to E$ are morphisms of sheaves such that the following relation
is satisfied 
\be\label{eq:parADHMc}
(\Phi_1\otimes 1_{M_2(D)})\circ(\Phi_2) - (\Phi_2\otimes 1_{M_1})\circ \Phi_1 + ((s_D\circ \psi) \otimes 1_{M_1\otimes_C M_2}) \circ \phi =0.
\ee
\end{itemize}
A collection of data satisfying the above conditions will be denoted 
$\CE^\bullet$. 

By analogy with the Higgs bundle case, a parabolic ADHM sheaf 
will be defined as a quasi-parabolic object $\CE^\bullet$ 
equipped with weights $\ualpha = (\ualpha_i)_{1\leq i\leq k}$
satisfying conditions \eqref{eq:parweights}. 

Generalizing the results of \cite{chamberI}, parabolic ADHM sheaves 
admit natural stability conditions depending on a stability 
parameter $\delta \in \IR$. For a nonzero parabolic 
bundle $(E^\bullet,\ualpha)$, define the $\delta$-parabolic 
slope by 
\[
\mu_\delta(E^\bullet, \ualpha) =  { \delta + 
\chi(E^\bullet, \ualpha)\over \rk(E)},
\]
where $\chi(E^\bullet, \ualpha)$ is given in \eqref{eq:parchi}. 
Recall that the parabolic slope $\mu(E^\bullet, \ualpha)$  was defined in \eqref{eq:parslope}.

Then a parabolic 
ADHM sheaf  $(\CE^\bullet,\ualpha)$ is $\delta$-(semi)stable 
if the following conditions hold 
\begin{itemize}
\item[$(i)$] Any proper nontrivial saturated subsheaf $0\subset E'\subset E$ 
preserved by $\Phi_1,\Phi_2$ and contained in ${\rm Ker}(\phi)$ 
satisfies 
\[\mu({E'}^\bullet, \ualpha)\ \leq\  { \delta + \chi(E^\bullet, \ualpha)\over \rk(E)}
\]
\item[$(ii)$] Any proper nontrivial saturated subsheaf $0\subset E'\subset E$ 
preserved by $\Phi_1,\Phi_2$ containing ${\rm Im}(\psi)$ 
satisfies 
\[
{ \delta + \chi({E'}^\bullet, \ualpha)\over \rk(E')}\
(\leq)\ 
{ \delta +\chi(E^\bullet, \ualpha)\over \rk(E)}
\]
\end{itemize}

There is also a natural duality transofrmation for 
parabolic ADHM sheaves, generalizing the one introduced in \cite{chamberI} for ADHM sheaves. 
Given a parabolic ADHM sheaf 
$(\CE^\bullet,\ualpha)$ let $\CE$ denote its underlying 
ADHM sheaf, forgetting the parabolic structure. 
Let 
${\check \CE}=({\check E},{\check\Phi_1}, {\check\Phi_2}, 
{\check\phi}, {\check\psi})$
denote the dual of $\CE$ defined in \cite[Sect 2.1]{chamberI}. 
Note that 
${\check E} = E^\vee \otimes_C M_1\otimes_C M_2$ and
the morphism data $({\check\Phi_1}, {\check\Phi_2}, 
{\check\phi}, {\check\psi})$ is naturally determined by 
$(\Phi_1,\Phi_2, \phi,\psi)$. 

Next note that given a flag $E_i^\bullet$ in the fiber $E_{p_i}$, 
one can define a dual flag in the fiber $E^{\vee}_{p_i}$. 
Each subspace $E_i^a\subset E_{p_i}$ determines a locally free 
sheaf $F_i^a= {\rm Ker}(E\twoheadrightarrow E_{p_i}/E_i^a\otimes \CO_{p_i})$ 
on $C$, which yields a filtration 
\[
0\subset E(-p_i) = F^{s}_i \subseteq F^{s-1}_i \subseteq \cdots 
\subseteq F^{1}_i \subseteq F^0_i =E.
\]
The dual filtration yields a flag 
\[
0 = ({E^\vee})_i^0 \subseteq ({E^\vee})_i^{1} \subseteq \cdots \subseteq ({E^\vee})_i^{s-1}
\subseteq ({E^\vee})_i^{s}=({E^\vee})_{p_i},
\]
hence also a flag 
${\check E}_i^\bullet$ on ${\check E}_{p_i}$ 
by taking tensor product with $M_1\otimes_C M_2$. 
It is straightforward to check that successive 
quotients of the dual flag have dimensions
\[ 
{\check m}_{i,a} = m_{i,s_i-a-1}
\]
for $0\leq a \leq s_i-1$. Moreover the morphisms ${\check \Phi}_1$, ${\check \Phi}_2$
and the filtrations ${\check E}_i^\bullet$ satisfy naturally 
condition $(b)$ in Section \ref{ADHMpar}. 

In conclusion 
the data $({\check E}^\bullet, {\check \Phi}_1, {\check \Phi}_2, 
{\check \phi}, {\check \psi})$ determines a quasi-parabolic ADHM sheaf 
with numerical invariants 
$({\check \um},\, -e+r({\rm deg}(M_1)+{\rm 
deg}(M_2))$. In addition, let 
${\check \alpha}_{i,a} = - \alpha_{i,s_i-1-a}$ 
for $1\leq i\leq k$, 
$0\leq a \leq s_i-1$. Then it easily follows that the parabolic 
ADHM sheaf $({\check \CE}^\bullet, {\check \ualpha})$ is 
$\delta$-(semi)stable if and only if $(\CE^\bullet, \ualpha)$ 
is $(-\delta)$-(semi)stable. Note that the dual 
parabolic weights ${\check \ualpha}$ defined here differ from 
the usual conventions in the literature, where they are  
defined as $1-\alpha_{i, s_i-1-a}$ in order to bring them 
in the range $[0, \ 1)$. 

\subsection{Moduli spaces and counting invariants}\label{modulisect}
Moduli spaces of $\delta$-semistable parabolic ADHM sheaves 
are constructed by analogy with \cite{modADHM,chamberI}, 
using the boundedness results for parabolic sheaves 
proven in \cite{modpar}. Repeating the arguments of 
\cite{modADHM,chamberI} in the parabolic framework, 
one obtains a moduli 
an algebraic moduli stack 
${\mathcal {PM}}^{ss}_{\delta}(C,M_1,M_2,D; \um, e, \ualpha)$ 
of $\delta$-semistable ADHM sheaves 
with discrete invariants $(\um,e)$  and parabolic weights 
$\ualpha$.

Varying $\delta\in 
\IR$ for fixed $\ualpha$ yields a finite chamber structure 
on the real axis analogous 
to the one studied in \cite{chamberI}. 
For sufficiently 
large stability parameter, $\delta$-stability reduces to asymptotic 
stability \cite[Def. 4.5]{chamberI} of the underlying ADHM 
sheaf, forgetting the parabolic structure. 
This condition 
 is completely independent of the parabolic weights, 
hence the  moduli space in the asymptotic 
chamber will be denoted by ${\mathcal {PM}}_\infty(X,M_1,M_2,D; 
\um,e)$. 

Let also $\CM_\infty(X,M_1,M_2(D); r,e)$ denote the moduli space of 
space of asymptotically stable ADHM sheaves without 
parabolic structure and with fixed 
numerical invariants $(r,e)$. Then note that there is a proper morphism 
\be\label{eq:forgetflags}
{\mathcal {PM}}_\infty(X,M_1,M_2,D; \um, e) \to \CM_\infty(X,M_1,M_2(D); r,e) 
\ee
forgetting the flags at the marked points. Properness 
follows from the fact that for fixed morphisms $(\Phi_1,\Phi_2,\phi,\psi)$ the moduli space of collections of flags $(E_{p_i}^\bullet)_{1\leq i\leq k}$ compatible with the ADHM data is proper. 

Note also that the stability condition for parabolic ADHM 
takes a special form in the asymptotic chamber $\delta<<0$ as
well.  Keeping the numerical invariants $(\um,e)$ 
parabolic weights $\ualpha$ fixed a 
sheaf $(\CE^\bullet,\ualpha)$ is $\delta$-semistable for 
$\delta <<0$ 
if and only its dual $({\check \CE}^\bullet,{\check\ualpha})$ 
is assymptotically stable. 

In order to define parabolic virtual invariants, the moduli space must 
be equipped with a perfect obstruction theory. This is carried out 
in analogy with \cite[Sect. 5]{modADHM} using known results on the deformation theory of parabolic Higgs bundles
\cite{varmod,infparHiggs}.  A concise summary of such results is provided in 
\cite[Sect. 2.2]{rankthreepar}. 

By analogy with the non-parabolic case, 
\cite[Sect. 4.1]{modADHM}, the deformation 
complex of a parabolic ADHM sheaf $\CE^\bullet$ 
is the three 
term complex of amplitude $[0,\ 2]$ 
\be\label{eq:defcpxA} 
0\to \CC^0(\CE^\bullet) \to \CC^1(\CE^\bullet) \to 
\CC^2(\CE^\bullet)\to 0
\ee
where 
\[
\CC^0(\CE^\bullet) = PEnd_C(E^\bullet) 
\]
\[
\bal 
\CC^1(\CE^\bullet) = & PEnd_C(E^\bullet)\otimes M_1 \oplus 
SPEnd_C(E^\bullet)\otimes M_2(D) \\
& \oplus Hom_C(\CO_C,E) \oplus 
Hom_C(E,M_1\otimes_C M_2) \\
\eal 
\]
\[
\CC^2(\CE^\bullet) = SPEnd_C(E^\bullet)\otimes M_1\otimes_C M_2(D).
\]
The differentials are the same as in \cite[Def. 4.3]{modADHM}, but 
their explicit form will not be needed in the following. The main technical result needed in the construction of a perfect obstruction 
theory requires the hypercohomology groups $\IH^i(\CC(\CE^\bullet))$ to vanish for $i\leq 0$ as well as $i\geq 3$. This follows 
by analogy with \cite[Lemma 4.10]{modADHM} 
using the duality relation \eqref{eq:pardual}. Then the existence of a 
perfect obstruction theory follows by the same formal arguments 
as in \cite[Sect. 5]{modADHM}.
Furthermore, it is important to note that the resulting 
perfect obstruction theory is symmetric provided that 
$M_1\otimes_C M_2\simeq K_C$. 

Finally, since the moduli spaces are noncompact, parabolic invariants 
will be defined by equivariant virtual integration with respect to a torus action with compact fixed locus. For asymptotically stable 
parabolic ADHM sheaves, such an action is again obtained by analogy with \cite[Sect 3]{modADHM}. Namely, ${\bf T}=\IC^\times \times 
\IC^\times$ acts by 
\be\label{eq:toractA}
(t_1,t_2) \times (E^\bullet,\Phi_1,\Phi_2,\phi,\psi)\mapsto 
(E^\bullet,t_1\Phi_1, t_2\Phi_2, t_1t_2\phi, \psi).
\ee
Compactness of the fixed locus follows from 
\cite[Prop. 3.1]{modADHM} and the observation 
that the proper forgetful morphism \eqref{eq:forgetflags} 
is equivariant. 

Motivic and refined invariants will be defined using the theory 
of Kontsevich and Soibelman \cite{wallcrossing}
and assuming all the 
required foundational results. The refined parabolic ADHM invariants 
will be denoted by $A_\delta(\um,e, \ualpha;y)$. The asymptotic 
ones will be denoted by $A_{\pm\infty}(\um,e;y)$. 
The duality transformation introduced 
at the end of Section \ref{ADHMpar} yields 
relations of the form 
\be\label{eq:dualinv} 
A_{-\delta}(\um,e,\ualpha; y) = 
A_{\delta}({\check \um},{\check e},{\check \ualpha}; y)
\ee
where 
\[ 
{\check m}_{i,a} = m_{i,s_i-1-a}, \qquad {\check \alpha}_{i,a} 
= - \alpha_{i, s_i-1-a}, \qquad 
{\check e} = -e + 2r(g-1).
\] 
for all $1\leq i\leq k$, $0\leq a \leq s_i-1$. 

In conclusion, one obtains a series of asymptotic refined invariants 
\be\label{eq:orbstpairAB} 
Z^{ref}_{ADHM}(q,{\underline x},y) = 
\sum_{n\in \IZ} \sum_{\um} A_\infty(\um,e;y)
q^{e-r(g-1)} \prod_{i=1}^k 
{\underline x}_i^{\um_i}
\ee
for some formal variables ${\underline x} = 
\big({\underline x}_1,\ldots,{\underline x}_k\big)$, 
${\underline x}_i = (x_{i,0}, \ldots, x_{i,s_i-1})$, 
$1\leq i\leq k$. As explained in the paragraph preceeding 
Section \ref{ADHMpar}, generalizing the results of 
\cite{modADHM}, the above series is identical with the left hand side $Z^{ref}_{\wY_M}(q,{\underline x},y)$
 of equation \eqref{eq:orbstpairA}.

\section{Geometric engineering, Hilbert schemes and Macdonald polynomials }\label{geomeng}

A conjectural formula for the refined stable pair theory of the orbifold ${\widetilde Y}$ 
is derived in this section by geometric engineering.
Using IIA/M-theory duality, 
stable pair invariants are related to degeneracies of BPS wavefunction
in D-brane 
quiver quantum mechanics, which are counted by equivariant 
$K$-theoretic invariants as in \cite{Nekrasov:2002qd}.  Similar results have been obtained in 
\cite{geom_eng_qft,Lawrence:1997jr,EK-I,Nekrasov:2002qd,IKP-I,IKP-II,EK-II,Hollowood:2003cv,Konishi-I,LLZ,Iqbal:2007ii},
where the resulting quantum mechanical system describes 
instanton particles in five dimensional gauge theories. 
As shown in Section \ref{nested}, in the present case 
one obtains a moduli space of parabolic ADHM quiver representations
which can be identified with a nested Hilbert scheme of points in the complex plane. The quantum mechanical partition function 
is the generating function for equivariant $K$-theoretic invariants 
given in equation \eqref{eq:parinstA}. The main result of this section 
is formula \eqref{eq:parinstB} expressing this partition function 
in terms of Macdonald polynomials.

\subsection{Orbifold stable pairs in string theory}

In this section, ${\widetilde Y}$ will be the total space of 
a rank two bundle $\nu^*M^{-1}\oplus K_{\wC}\otimes 
\mu^*M$ as in Sections \ref{mainconj}, \ref{ADHMpar}. 
The subscript $M$ used in Sections \ref{mainconj}, 
and \ref{ADHMpar} will be suppressed for brevity. 

From a string theory perspective, the refined stable pair invariants of ${\widetilde Y}$
are identified with BPS degeneracies of D6-D2-D0 
bound states in the IIA vacuum ${\widetilde Y}\times \IR^{1,3}$.
This theory will be called ${\rm IIA}^{(1)}$. 
As explained below, a chain of string duality transformations relates such BPS  
states with D6-D2-D0 bound states in a different type IIA background 
of the form $T^*{\widetilde C} \times TN_1\times \IR^{1,1}$, where $TN_1$ 
is the one center Taub-NUT manifold. This theory will be called ${\rm IIA}^{(2)}$.

Applying a Wick rotation in the ${\rm IIA}^{(1)}$ theory
and making the time 
direction periodic  results in a background geometry 
${\widetilde Y}\times \IR^3 \times S^1_T$. 
For configurations with a single D6-brane on $\wY$,
 lifting ${\rm IIA}^{(1)}$ to M-theory
produces 
the eleven dimensional vacuum ${\widetilde Y}\times TN_1 \times S^1_T$. Note that 
there is a fiberwise 
circle action $S_M^1\times {\widetilde Y}\to {\widetilde Y}$ 
which scales the two line bundles ${\tilde L}_1, {\tilde L}_2$ 
with opposite weights leaving the 0-section pointwise fixed. 
Reducing the theory along the orbits of this action
yields the theory with underlying geometry 
 $T^*{\widetilde C}\times TN_1 \times \IR \times S^1_T$. 
Since the fixed locus of $S^1_M$-action on ${\widetilde Y}$ is 
the 0-section, in the new duality frame 
there is a D6-brane supported on the submanifold 
${\widetilde C} \times TN_1 \times \{0\}\times S^1_T$, where 
${\widetilde C}$ is embedded in $T^*{\widetilde C}$ as the 0-section. 
Note also that the ${\rm IIA}^{(2)}$ geometric background does not depend 
on the line bundle $M$ used in the construction of $\wY$.
By analogy with previous examples studied in the literature
\cite{IKP-I,IKP-II,EK-II,LLZ,Tachikawa:2004ur,Iqbal:2007ii} this dependence is 
expected to be encoded in the level $n$ of the Chern-Simons 
coupling in the efffective five dimensional 
 gauge theory on the D6-brane wrapped on $\wC \times TN_1 
\times S^1_T$, which is related by duality to the level 
of the five dimensional space-time Chern-Simons coupling 
in M-theory. In the ${\rm IIA}^{(2)}$ string theory, $n$ is determined by the 
Ramond-Ramond flux $\int_\wC G_2 =n$ via 
standard D-brane couplings. 
Inspecting the examples in loc. cit., for smooth local 
genus zero curves, the Chern-Simons level is 
given by $n={\rm deg}(M)-1$. This formula was extended to 
$n = g-1 + {\rm deg}(M)$ for local genus $g$ curves in 
\cite{wallpairs,BPSPW}. Here one needs a further generalization 
for local orbifold curves, which have never been studied in this 
context before. The solution to this string duality puzzle 
follows from the observation that $n$ is invariant under 
continuous deformations of the $M$-theory vacuum 
$\wY\times TN^1\times S^1_T$. 
Such deformations connect the present orbifold vacuum 
to a smooth geometric vacuum obtained by resolving the 
quotient singularities of the coarse moduli space $Y$ 
of $\wY$. The resolution ${\widehat Y}\to Y$ 
contains a local genus $g$ curve isomorphic to $C$ in addition with exceptional $(0,-2)$ rational curves which play no role 
in this argument. This leads to the conclusion that 
that the value of $n$ in the present orbifold vacuum 
must be the same 
as that found in \cite{wallpairs,BPSPW} for local genus $g$ 
curves i.e. 
\be\label{eq:CSlevel}
n = g-1 + {\rm deg}(M). 
\ee
The above derivation is not rigorous, hence it should be regarded as a conjecture. The explicit computations in Section \ref{conifold} provide ample evidence for this formula. 

As in \cite{Mtop}, D6-D2-D0 configurations in 
${\rm IIA}^{(1)}$ 
theory lift to 
spinning M2-branes in the M-theory. Since such states carry 
no momentum along the orbits of the fiberwise $S^1_M$-action, 
they reduce to D2-brane supported on the zero section $\wC$ 
in the the ${\rm IIA}^{(2)}$ vacuum with zero  
D0-brane charge. 
In conclusion one is left with a D6-D2 configuration in the IIA 
background $T^*{\widetilde C} \times TN_1\times \IR \times S^1_T$ 
consisting of one non-compact D6-brane on ${\widetilde C}\times 
TN_1\times S^1_N$ and a stack of D2-branes with some multiplicity  
$r\geq 1$ supported on ${\widetilde C}\times S^1_T$. The Chan-Paton bundle 
on the D2-branes is  topological trivial since these configurations 
carry no D0-brane 
charge. 

This chain of duality transformations relates orbifold stable pair invariants in the original ${\rm IIA}^{(1)}$ theory with D6-D2 supersymmetric bound states in the new IIA vacuum. 
The latter are counted by the partition function of the
 D2-brane  low energy theory, which is a topological 
gauge theory on $\wC\times S^1_T$. 
 The topological twist is determined by the normal bundle of the zero section ${\widetilde C}$ 
in $T^*{\widetilde C}$, which is isomorphic to $K_{\widetilde C}$.
Using the topological symmetry, the background manifold 
can be changed from $T^*{\widetilde C} \times TN_1 \times S^1_T$ 
to $T^*{\widetilde C} \times \IC^2 \times S^1_T$ leaving the BPS degeneracies 
invariant. 
Moreover, in order to detect the spin quantum numbers in the $M$-theory 
framework,  the D6-D2 theory must be placed in an $\Omega$-background \cite{Nekrasov:2002qd}
determined by the natural 
 $\IC^\times\times \IC^\times$ scaling action on  $\IC^2$.

\subsection{From D-branes to nested Hilbert 
schemes}\label{nested}

The next goal is to count BPS states of the above D6-D2 system on 
$T^*{\widetilde C} \times \IC^2 \times \IR$.  As explained above, 
the D2-brane low energy effective 
theory is a topologically twisted $\CN=4$ three 
dimensional gauge theory on $\wC \times S^1_T$.
By analogy with the five dimensional situation, \cite{Nekrasov:2002qd} 
the topological partition function counts BPS states of vortex particles 
on $\wC$. Mathematically, these are holomorphic sections of a certain 
bundle of fermionic zero modes on the vortex moduli space. Hence,
taking into account the $\Omega$-background along $\IC^2$, 
the partition function will be generating function for equivariant $K$-theoretic 
invariants of the moduli space. 

The field content of the D2-brane low energy theory consists of 
two adjoint chiral multiplets $\wA_1,\wA_2$ corresponding to 
fluctuations in the untwisted normal directions, and an adjoint 
valued one-form $\wA_3$ on $\wC$ corresponding to fluctuations 
in the normal directions to $\wC$ in $T^*\wC$. 
In addition, the D2-D6 open string sector yields two extra chiral 
multiplets $\wtI,\wJ$ in the fundamental, and anti-fundamental representation of the gauge group. 
BPS vortex solutions are field configurations in this gauge theory 
satisfying $F$ and $D$-flatness constraints.

Using Hitchin-Kobayashi 
correspondence for quiver bundles
\cite{HKquivers},  gauge
 equivalence classes of BPS vortex solutions are 
in one-to-one correspondence with
 isomorphism classes of holomorphic data 
$(\wE,\wA_1,\wA_2,\wA_3,\wtI,\wtJ)$ 
satisfying the $F$-term equations  and a 
stability condition determined by the $D$-term 
constraints. Namely,  
$E$ is a holomorphic vector bundle on $\wC$ and 
\[
\wA_1,\wA_2: \wE\to \wE, \qquad \wA_3:\wE\to \wE\otimes_{\wC} K_{\wC},\qquad \wtI: \CO_{\wC} \to \wE, \qquad 
\wJ : \wE \to \CO_{\wC}
\] 
are morphisms of sheaves on $\wC$. 
The $F$ term equations yield conditions of the form 
\be\label{eq:FtermeqA} 
[\wA_1,\wA_2] + \wtI\wJ =0, \qquad [\wA_3,\wA_1]=[\wA_3,\wA_2] =0, 
\qquad \wA_3 \wtI=0, \qquad \wJ \wA_3 =0.
\ee
Note that the data ${\widetilde \CE}=(\wE,\wA_1,\wA_2, \wtI,\wJ)$ determines 
an ADHM sheaf with trivial coefficient line bundles on the orbifold $\wC$, 
while $\wA_3$ is an extra field. 
The $D$ term constraints yield a stability condition for $({\widetilde \CE},\wA_3)$ via Hitchin-Kobayashi correspondence, which depends on a Fayet-Iliopoulos 
parameter. Then a straightforward computation shows that for suitable 
values of this parameter the stability condition requires ${\widetilde \CE}$ 
to be an asymptotically stable ADHM sheaf on $\wC$ as defined in 
Section \ref{orbisection}. In this case, it is also straightforward to prove that 
any field $\wA_3$ satisfying the $F$ term equations \eqref{eq:FtermeqA} 
must be identically 0. The details will be omitted for brevity.  

In conclusion, with a suitable choice of FI parameters, the vortex moduli 
spaces in the quiver gauge theory  on D2 branes are isomorphic to 
moduli spaces of asymptotically stable ADHM sheaves on the orbifold 
$\wC$ with trivial coefficient line bundles. 
According  Section \ref{orbisection}, such
orbifold ADHM sheaves are identified with  stable 
ADHM sheaves $\CE^\bullet=(E,A_1,A_2, I,J)$ on $C$
where $E$ is equipped with a parabolic structure $E^\bullet$ 
along $D$. Since the coefficient line bundles are trivial, 
the Higgs fields $A_1,A_2:E\to E$
are required to preserve the flag at each point $p_i$, 
$1\leq i\leq k$. Moreover the ADHM stability condition implies that 
$J$ is identically zero. In addition, as explained in the previous section, 
the bundle $E$ must be topologically trivial. 

In appendix
\ref{flatADHM} it is shown that 
a degree 0 asymptotically stable ADHM 
sheaf $\CE$ of arbitrary rank $r\geq 1$ on $C$ must have
 underlying bundle $E\simeq \CO_C^{\oplus r}$. 
This implies that all morphisms $A_1,A_2,I$ are constant maps, 
hence the moduli space of such sheaves 
is isomorphic to the moduli space of stable ADHM data 
\be\label{eq:stabADHM}
(A_1,A_2,I,0) \in \mathrm{End}(\IC^r)^{\oplus 2} 
\oplus \mathrm{Hom}(\IC,\IC^r).
\ee
The ADHM stability condition forbids the existence of 
linear subspaces 
$0\subsetneq V\subsetneq \IC^r$ 
preserved by $A_1,A_2$ and containing the image of $I$. 
Two such data are equivalent if they are related by the natural 
action $GL(r,\IC)$ action. 

Since asymptotic stability for parabolic ADHM sheaves reduces to 
asymptotic stability of the underlying ADHM sheaf, adding the 
parabolic structure yields 
a moduli space of data $(A_1,A_2,I,0; V_i^\bullet)_{1\leq i\leq k}$ where 
\begin{itemize} 
\item $(A_1,A_2,I,0)$ is a stable ADHM data 
\item $V_i^\bullet$ is a flag in $\IC^r$ of type $\um_i$ 
preserved by $A_1,A_2$ for all $1\leq i\leq k$. 
\end{itemize} 
Again two such data are equivalent if they are related by the 
natural $GL(r,\IC)$ action. 

For the remaining part of this section suppose there is a single marked point $p\in C$, and the flag at $p$ 
 is of the form 
 \be\label{eq:oneflagA}
 0=E^r \subseteq E^{r-1} \subseteq \cdots \subseteq E^0=E_p,
 \ee
where $r=\rk(E)$. Such 
flags are not neccessarily full  since the inclusions are not 
required to be strict. However 
let $a_\imath \in{0,\ldots, r-1}$, $0\leq \imath \leq \ell$ be the 
values of $0\leq a\leq r-1$ such that $m_{a_\imath}>0$. 
Then 
one can canonically associate 
a flag 
\[
0\subsetneq  E^{\ell} \subsetneq \cdots \subsetneq E^{1} 
\subsetneq E^0=E_p
\] 
to the flag \eqref{eq:oneflagA} such that all 
inclusions are strict and the discrete invariants are 
${\rm dim}\, E^\imath/E^{\imath+1} = m_{a_\imath}$ for 
$0\leq \imath \leq \ell$. 
This will be called the minimal flag associated to 
$E^\bullet$.
It is clear that the moduli space of asymptotically 
stable parabolic ADHM data depends only on the 
ordered sequence $\gamma=(m_{a_\imath})_{0\leq \imath 
\leq \ell}$, hence it will be denoted by $\CM(\gamma)$.
 The entries of $\gamma$ will be denoted by $\gamma_\imath= 
m_{a_{\imath}}$, $0\leq \imath\leq \ell$, in the following. 

The moduli space $\CM(\gamma)$ is identified in 
\cite{nested_quivers} 
with a nested Hilbert scheme of points in $\IC^2$
using the ADHM construction.  
Given an ordered sequence  
$\gamma=(m_{a_\imath})_{0\leq \imath \leq \ell}$ of positive integers as above, let 
$\CN(\gamma)$  denote the Hilbert scheme parameterizing 
flags of ideal sheaves 
\be\label{eq:idealflag}
0 \subset \CI_{\ell} \subset \cdots \subset \CI_0
\ee
of zero dimensional subschemes  $Z_\imath\subset \IC^2$ with 
\[ 
\chi(\CO_{Z_{\imath}})=\sum_{\jmath = 0}^{\imath} \gamma_{\jmath}
\]
for each $0\leq\imath \leq \ell$. 
Then, according to \cite{nested_quivers}, there is an isomorphism 
of moduli spaces ${\CM(\gamma)} \simeq \CN(\gamma)$.

\subsection{$K$-theoretic partition function}\label{Kpartfct} 
By analogy with 
\cite{wallpairs,BPSPW}, 
the $K$-theoretic partition function for fixed numerical 
invariants $\gamma$ 
will be the equivariant Hirzebruch genus of a bundle 
$\CV(\gamma)$ on  $\CN(\gamma)$ with respect to the 
 ${\bf T}=\IC^\times \times \IC^\times$ action 
induced by the scaling action on $\IC^2$. 
On general grounds, $\CV(\gamma)$ is 
the bundle of fermion zero modes on the moduli space
twisted by a line bundle determined by the space-time Chern-Simons coupling in M-theory \cite{Tachikawa:2004ur}. 
As shown in Appendix \ref{zeromodes}, the bundle of zero modes is simply the pullback 
$\eta^*\big(
{T^*\CH^r}\big)^{\oplus g}$ 
via the natural projection 
$\eta: \CN(\gamma)\to \CH^r$ to the Hilbert scheme of $r$ points in $\IC^2$. According to \cite{Tachikawa:2004ur}, 
the twisting line bundle is 
 $\eta^*
{\rm det}(\IV)^n$,
where $\IV$ is the tautological bundle on the Hilbert scheme 
and $n$ is the Chern-Simons level 
given in equation \eqref{eq:CSlevel}.
For completeness note that $\IV$ is 
 the pushforward of the structure sheaf 
of the universal subscheme $\CZ \subset \CH^r \times \IC^2$
to $\CH^r$. 
In conclusion, 
\[ 
\CV(\gamma) \simeq \eta^*\CV_{g,p}, \qquad \CV_{g,p}
=\big({T^*\CH^r}\big)^{\oplus g}
\otimes {\rm det}(\IV)^{g-1+p},
\]
and the $K$-theoretic partition function is given by
\be\label{eq:parinstA} 
 \CZ_K^{(r)}(q_1,q_2; {\tilde y}, {\tilde x}) =  \sum_{\substack 
 {\um=(m_0,\ldots, m_{r-1}) \in \IZ_{\geq 0}, 
 \\ m_0+\cdots+m_{r-1}=r}} 
\chi^{\bf T}_{\tilde y}( \CV(\gamma(\um))) \prod_{a=0}^{r-1} 
 {\tilde x}_{a}^{m_a} 
 \ee
 where $\chi^{\bf T}_{\tilde y}$ is the {\bf T}-equivariant 
 Hirzebruch genus. For any collection $\um=(m_0,\ldots, m_{r-1})$ 
of nonnegative integers, $\gamma(\um)$ denotes the sequence 
of distinct values in $\um$ defined below equation 
\eqref{eq:oneflagA}.

The above partition function will be expressed in terms of Macdonald 
polynomials in Section \ref{nestedMD}.  The next subsection summarizes the geometric results needed in that computation.

\subsection{Nested and isospectral Hilbert schemes}\label{isosection}
 This section explains the relation between the nested Hilbert scheme
$\CN(\gamma)$ and the isospectral Hilbert scheme employed 
in the work of Haiman \cite{MDgeom,polygraphs} on Macdonald polynomials. 
The main results are the pushforward formulas \eqref{eq:pushfwdB} 
and \eqref{eq:pushfwdD}. 

First consider the case of full flags of ideal sheaves, where 
$\ell=r-1$ and 
 $\gamma=\big(\, \underbrace{1,\ldots, 1}_r\, \big)$. Then it will be shown below that there exists a
surjective birational 
projection with connected fibers mapping 
$\CN(1,\ldots,1)$ to the isospectral Hilbert scheme. 

Each flag of ideal sheaves \eqref{eq:idealflag} 
determines a collection of nested zero-dimensional subschemes 
$Z_0\subset Z_1 \subset \cdots \subset Z_{r-1}$ where $Z_a$ has 
length $a+1$ for $0\leq a\leq r-1$. Hence there are exact sequences 
of sheaves on $\IC^2$ 
\be\label{eq:Zexseq}
0\to \CO_{p_{a+1}}\to \CO_{Z_{a+1}} \to \CO_{Z_{a}} \to 0 
\ee
for all $0\leq a\leq r-2$, where $p_a$ are closed points in $\IC^2$. 
Moreover, $Z_0$ must also be a closed point $p_0$ since it has 
length 1. This determines a morphism $\CN(1,\ldots,1) 
\to (\IC^2)^r$.  
Using also the natural projection onto the Hilbert scheme 
$\CH^r$, one obtains a commutative diagram 
\be\label{eq:hilbdiagA} 
\xymatrix{ 
\CN(1,\ldots,1) \ar[drr] \ar[ddr]_-{\eta} \ar@{.>}[dr]_-{\rho}
& & \\ 
& {\wCH}^r \ar[r] \ar[d]^{\pi}  & (\IC^2)^r \ar[d] \\
& \CH^r\ar[r] & S^r(\IC^2) \\}
\ee
where the square is Cartesian and the bottom horizontal arrow 
is the Hilbert-Chow morphism. The upper left corner of the square 
is by definition the isospectral Hilbert scheme ${\wCH}^r$. 
The above diagram determines a morphism 
$\rho:\CN(1,\ldots,1) \to {\wCH}^r$
such that $\eta=\pi\circ\rho$. Since both $\pi$ and $\eta$ are 
surjective and proper, so is $\rho$. 
Moreover, according to Appendix \ref{reducednested},  
$\CN(1,\ldots,1)$ is reduced, hence $\rho$ factors through 
a morphism $\rho_{\sf red}: \CN(1,\ldots, 1)\to 
{\wCH}^r_{\sf red}$, which is surjective and proper as well. 

The next step is to show that  
\be\label{eq:pushfwdA} 
{\rho_{\sf red}}_*\CO_{\CN(1,\ldots,1)} 
\simeq \CO_{{\wCH}^r_{\sf red}}.
\ee
The proof will rely on \cite[Prop. 3.3.2]{polygraphs} and 
\cite[Thm. 3.1]{polygraphs}, which prove that the 
reduced scheme ${\wCH}^r_{\sf red}$ is irreducible and  normal.
Then, 
by Stein factorization, $\rho_{\sf red}$ factors as
$$\CN(\gamma) {\buildrel f\over \longto} {\wCH}' 
{\buildrel g\over \longto} {\wCH}^r_{\sf red},$$ where 
${\wCH}'={\rm Spec}_{{\wCH}^r_{\sf red}}{\rho_{\sf red}}_*\CO_{\CN(\gamma)}$ and $g$ is a finite morphism. 
In particular 
\be\label{eq:pushfwd0}
f_*\CO_{\CN(1,\ldots,1)} \simeq \CO_{{\tilde \CH}'}. 
\ee
Since $\CN(1,\ldots,1)$ is reduced according to Appendix 
\ref{reducednested},  ${\tilde \CH}'$ is 
reduced as well. 

Next note that there is an open subset 
$\wCU\subset \wCH^r$ such that the restriction of $\rho$ 
to $\rho^{-1}({\wCU})$ is an  isomorphism onto  
to ${\wCU}$. This open subset is the inverse image 
$\pi^{-1}(\CU)$, where  $\CU\subset \CH^r$ is the open subset parameterizing 
subschemes $Z\subset \IC^2$ consisting of $r$ distinct 
 points in $\IC^2$. By construction, the restriction of $g$ to $g^{-1}(\wCU)$ is an isomorphism as well.

In order to conclude the proof of \eqref{eq:pushfwdA}, 
it suffices to show that $\rho_{\sf red}$ 
has connected  fibers. This implies that $g$ is one-to-one on closed points, which further implies that ${\wCH}'$ is irreducible since 
${\wCH}^r_{\sf red}$ is irreducible. Therefore ${\wCH}'$ is 
reduced and irreducible. Since $g$ is an isomorphism 
over the open subset $g^{-1}({\wCU})$, it follows 
that $g$ is birational, hence an isomorphism. 
Therefore \eqref{eq:pushfwdA} follows from \eqref{eq:pushfwd0}.

 To prove that the fibers of $\rho_{\sf red}$ are 
connected,  let $(p_0,\ldots, p_{r-1}; Z_{r-1})$ be a 
closed point of ${\wCH}^r_{\sf red}$, with 
$(p_0,\ldots, p_{r-1}) \in (\IC^2)^r$, and $Z_{r-1} \subset 
\IC^2$ a closed zero-dimesional subscheme of length $r$. 
In particular $(p_0,\ldots,p_{r-1})\in (\IC^2)^r$ and 
$Z_{r-1}$ are mapped to the same point in $S^r(\IC^2)$ 
in diagram \eqref{eq:hilbdiagA}. 
The fiber $N(p_0,\ldots, p_{r-1}; Z_{r-1})={\rho_{\sf red}}^{-1}( p_0,\ldots, p_{r-1}; Z_{r-1})$ 
parametrizes collections of length $a+1$ zero-dimensional 
subschemes $Z_a\subset \IC^2$, $0\leq a\leq r-2$,
such that their structure sheaves fit in exact sequences of
the form \eqref{eq:Zexseq}. The inductive argument given below 
shows 
that this fiber is connected. 

Since $Z_{r-1}$ and $p_{r-1}$ are fixed, the moduli space of 
subschemes $Z_{r-2}$ such that $\CO_{Z_{r-2}}$ fits in an 
exact sequence 
\[ 
0\to \CO_{p_{r-1}} \to \CO_{Z_{r-1}}\to \CO_{Z_{r-2}}\to 0
\]
is isomorphic to the projective space $\IP{\rm Hom}(\CO_{p_{r-1}}, 
\CO_{Z_{r-1}})$. Therefore there is a natural projection
\[ 
\pi_{r-1}:
N(p_0,\ldots, p_{r-1}; Z_{r-1}) \to \IP{\rm Hom}(\CO_{p_{r-1}}, 
\CO_{Z_{r-1}}).
\] 
The fiber of $\pi_{r-1}$ over a point parameterized by a 
subscheme $Z_{r-2}$ is isomorphic to $N(p_0,\ldots, p_{r-2}; 
Z_{r-2})$.  If all fibers $N(p_1,\ldots, p_{r-2}; 
Z_{r-2})$ are connected, it follows that the total space 
$N(p_0,\ldots, p_{r-1}; Z_{r-1})$ 
is also connected. Therefore it suffices to prove connectedness 
for $r=2$. In that case the moduli space $N(p_0,p_1;Z_1)$ 
is a single point for any choice of $(p_0,p_1; Z_1)$ as above, 
hence the 
claim follows. 

Now recall that 
the projection $\pi_{\sf red}: {\wCH_{\sf red}}\to \CH^r$ 
is flat, according to  \cite[Thm 3.1]{polygraphs},
hence the pushforward 
$\pi_{{\sf red}*}\CO_{\wCH_{\sf red}}$ 
is a rank $r!$ vector bundle ${\mathcal P}$ on the Hilbert 
scheme. Equation \eqref{eq:pushfwdA} 
implies that 
\be\label{eq:pushfwdB}
\eta_*\CO_{\CN(1,\ldots,1)} \simeq \calP.
\ee

Next consider the case of arbitrary 
discrete invariants $\gamma = (\gamma_{\imath})_{0\leq \imath\leq \ell}$. Let  $\CS_\gamma = \CS_{\gamma_\ell} 
\times \cdots \times \CS_{\gamma_0} \subset \CS_r$ be the stabilizer of the ordered partition $\gamma$. 
  The group action $\CS_r \times 
{\wCH}^r \to \wCH^r$
yields by restriction an action of $\CS_\gamma\times 
{\wCH}^r \to \wCH^r$. 
Let $\wCH^\gamma$ denote the quotient of $\wCH^r$ by $\CS_\gamma$, 
which is a quasi-projective scheme. Since 
$\wCH^r_{\sf red}$ is normal and reduced and irreducible, 
so is $\wCH^{\gamma}_{\sf red}$. 
 Similarly, the quotient 
$S^\gamma(\IC^2)=(\IC^2)^r/\CS_\gamma$ is a 
quasi-projective variety and  there is a commutative diagram 
\be\label{eq:hilbdiagB}
\xymatrix{ 
\wCH^r \ar[r] \ar[d]^-{\kappa}
 \ar@/_2pc/[dd]_-{\pi} & (\IC^2)^r \ar[d] \\
\wCH^\gamma \ar[r] \ar[d]^-{\pi^\gamma} & 
S^\gamma(\IC^2) \ar[d] \\
\CH^r \ar[r] & S^r(\IC^2)\\}
\ee
where both squares are Cartesian.

Next note that there is a Hilbert-Chow 
morphism $\CN(\gamma)\to S^\gamma(\IC^2)$ defined as follows. Given a 
flag of zero dimensional subschemes
\be\label{eq:flagschemes}
Z_{0}\subset \cdots \subset Z_{\ell}\subset \IC^2
\ee
there are exact sequences 
\[
0\to K_{\imath} \to \CO_{Z_{\imath}} \to \CO_{Z_{\imath-1}}
\to 0
\]
with $K_{\imath}$ a zero dimensional sheaf on $\IC^2$, for 
$1\leq \imath \leq \ell$. The morphism $\CN(\gamma)\to S^\gamma(\IC^2)$ sends a flag of subschemes of the form 
\eqref{eq:flagschemes} to the cycle classes associated to the 
zero dimensional sheaves $(\CO_{Z_0}, K_1, \ldots, K_{\ell})$
via the Hilbert-Chow morphism. 
Then the bottom Cartesian square in \eqref{eq:hilbdiagB} 
yields a morphism $\rho^\gamma: \CN(\gamma) \to 
\wCH^\gamma$ which factors through a 
morphism $\rho^\gamma_{\sf red}: \CN(\gamma) \to 
\wCH^\gamma_{\sf red}$. The following generalization 
of \eqref{eq:pushfwdA} will be proven below:
\be\label{eq:pushfwdC} 
\rho^\gamma_{{\sf red}*}\CO_{\CN(\gamma)} 
= \CO_{\wCH^\gamma_{\sf red}}.
\ee

Let $\CU^\gamma = (\pi^{\gamma})^{-1}\CU$ be the 
inverse image of the open subset parameterizing subschemes 
of $\IC^2$ supported at $r$ distinct closed points.
Then $\eta^\gamma$ is an isomorphism 
over the open subset $(\eta^{\gamma})^{-1}\CU^\gamma$. 
Then equation \eqref{eq:pushfwdC} follows from the Zariski 
Main Theorem provided one can prove that $\CN(\gamma)$ is
connected. This is shown in Appendix \ref{reducednested}.

For future reference, note that by construction the bundle 
$\calP=\pi_{{\sf red}*}\CO_{\wCH^r_{\sf red}}$ 
is equipped with a fiberwise action of the symmetric group $\CS_r$ 
such that its fiber over any point $[\CI]\in \CH^r$ 
is isomorphic to the regular representation. 
By construction, 
\[
\CO_{\wCH^\gamma_{\sf red}} \simeq 
\big({\kappa}_{{\sf red}*}\CO_{\wCH^r_{\sf red}}\big)^{\CS_\gamma}
\]
where $\kappa_{red}: \wCH^r_{\sf red}\to
 \wCH^\gamma_{\sf red}$ is the morphism of reduced 
 schemes determined by $\kappa$ in diagram \eqref{eq:hilbdiagB}. 
 Pushing forward this identity to $\CH^r$ via $\pi^\gamma_{\sf red}$, one 
 learns that 
 \be\label{eq:pushfwdD}
 \pi^{\gamma}_{{\sf red}*}\CO_{\wCH^\gamma_{\sf red}}
 \simeq \big(\pi_{{\sf red}*} \CO_{\wCH^r_{\sf red}}\big)^{\CS_\gamma} = \calP^{\CS_\gamma}.
 \ee
 In particular, since $\calP$ is locally free, so is $\calP^\gamma = 
  \pi^{\gamma}_{{\sf red}*}\CO_{\wCH^\gamma_{\sf red}}$. Moreover $\calP^\gamma$ is equipped with a fiberwise action 
  of  $\CS_r$ such that its fiber 
  at any closed point $[I]\in \CH^r$ is isomorphic to the permutation 
   representation $M_\gamma$ of $\CS_r$ 
   with stabilizer $\CS_\gamma = \prod_{\imath=0}^\ell 
   \CS_{\gamma_\imath} \subset \CS_r$.

\subsection{Nested partition function and Macdonald 
polynomials}\label{nestedMD}
This section concludes the computation of the partition function
\eqref{eq:parinstA} 
using the results of \cite{polygraphs} and the previous subsection.

As a preliminary remark, note that $\chi^{\bf T}_{\tilde y} (\CV_{g,p})$ 
on the Hilbert scheme $\CH^r$ can be easily computed by a fixed point theorem. 
The fixed points of the ${\bf T}$-action on $\CH^r$ are
monomial ideals $[I_\mu]\in \CH^r$ in one-to-one correspondence 
with partitions $\mu$ of $r$. For any equivariant bundle $\CF$ 
on $\CH^r$, let $\CF_\mu$ 
denote the fiber of $\CF$ at $[I_\mu]$. An exception will be made 
for the cotangent bundle $\CT^*\CH^r$, in which case the fiber 
at $[I_\mu]$ will be denoted by $\CT^*_\mu \CH^r$. 
Then equivariant localization yields 
\be\label{eq:chigenusidO} 
\chi^{\bf T}_{\tilde y} (\CV_{g,p})
= \sum_\mu  \Omega_\mu^{g,p}(q_1,q_2,{\tilde y}),  
 \ee
 where 
 \be\label{eq:omegaformula}
  \Omega_\mu^{g,p}(q_1,q_2,{\tilde y}) = {\ch_{\bf T}({\rm det}\IV^{g-1+p})\, 
 \ch_{\bf T}\Lambda_{\tilde y}({\CT_\mu^*\CH^r}^{\oplus g}_\mu)\over 
 \ch_{\bf T}\Lambda_{-1}(\CT_\mu^*\CH^r)}.
 \ee

Then the main formula proven in this section reads 
 \be\label{eq:parinstB} 
 \CZ_K^{(r)}(q_1,q_2, {\tilde y}, {\tilde x}) =  
 \sum_{\mu}  \Omega_\mu^{g,p}(q_1,q_2,{\tilde y}) 
 {\widetilde H}_\mu (q_2,q_1,{\tilde x})
 \ee
 where ${\widetilde H}_\mu(q_2,q_1,{\tilde x})$ are 
 the modified MacDonald polynomials. 

First note that the pushforward formulas \eqref{eq:pushfwdC},
\eqref{eq:pushfwdD}  
 are valid  in ${\bf T}$-equivariant  setting, hence 
 one obtains an identity
  \be\label{eq:chigenusidA} 
 \chi^{\bf T}_{\tilde y}( \CN(\gamma(\um)), \eta^{\gamma*}\CV_{g,p}) 
 =\chi^{\bf T}_{\tilde y}(\CH^r,({\calP}^{\CS_\gamma}\otimes_{\CH^r} 
 \CV_{g,p})). 
 \ee
 The right hand side of equation \eqref{eq:chigenusidA} can be 
evaluated again by equivariant localization:
\be\label{eq:chigenusidC} 
\chi^{\bf T}_{\tilde y}(\CH^r,({\calP}^{\CS_\gamma}\otimes_{\CH^r} 
 \CV_{g,p})) = \sum_\mu  \Omega_\mu^{g,p}(q_1,q_2,{\tilde y})
 \ch_{\bf T}(\calP^\gamma_\mu). 
 \ee

Now let ${\tgamma}$ denote the unordered partition of 
$r$ determined by the sequence $\gamma=(\gamma_0, \ldots, 
\gamma_\ell)$. Then following formula 
 \be\label{eq:chargammaD}
\ch_{\bf T}(\calP^\gamma_\mu) = \sum_\lambda K_{\lambda, \tgamma} {\widetilde K}_{\lambda, \mu}(q_1,q_2).
\ee 
will be proven bellow, where the sum is over all partitions $\lambda$ of $r$, 
$K_{\lambda, \tgamma}$ are the Kostka numbers, and 
${\widetilde K}_{\lambda, \mu}(q_2,q_1)$ are the modified 
Kostka--Macdonald coefficients.

 Since the fiberwise $\CS_r$-action on $\calP$
is compatible with the ${\bf T}$-equivariant structure, 
there is a direct sum decomposition
\be\label{eq:regdecomp}
\calP_\mu \simeq \bigoplus_{\lambda} V_{\mu,\lambda}
\otimes R_\lambda  
\ee
$R_\lambda$ is the 
irreducible $\CS_r$-representation 
labelled by the partition 
$\lambda$ and  $V_{\mu,\lambda}$ are finite dimensional 
representations of ${\bf T}$. 
According to 
\cite[Thm. 3.1, Prop. 3.7.3, Thm. 3.2]{polygraphs}, the ${\bf T}$-character of 
$V_{\lambda, \mu}$ 
is given by the modified Kostka-MacDonald coefficients,
\be\label{eq:KMDnumbers} 
\ch_{\bf T} V_{\lambda, \mu} = 
{\widetilde K}_{\lambda,\mu}(q_2,q_1). 
\ee
The pushforward formula \eqref{eq:pushfwdD} 
shows that the fiber $\calP^\gamma_\mu$ is the $\CS_\gamma$-fixed subspace of $\calP_\mu$. This yields 
\be\label{eq:gammacharA}
\ch_T \calP_\mu^\gamma = {1\over |\CS_\gamma|} 
\sum_{g=(g_0,\ldots, g_\ell) \in \CS_\gamma} \sum_\lambda
 \chi_{R_\lambda}(g)\,
 \ch_{\bf T} V_{\mu,\lambda}.
 \ee
 
Now recall the branching rule for representations of the symmetric 
group. Given a subgroup $\CS_{r_1}\times \CS_{r_2} \subset \CS_r$, with $r_1+r_2=r$, the irreducible $\CS_r$-representation $R_\lambda$ has a direct sum decomposition 
\be\label{eq:branchingA} 
R_\lambda \simeq \bigoplus_{\nu_1,\nu_2}
 N_{\nu_1,\nu_2, \lambda}\, \big( R_{\nu_1}\boxtimes R_{\nu_2}\big) 
\ee
where $\nu_1,\nu_2$ are partitions of $r_1,r_2$ respectively, 
and $N_{\nu_1, \nu_2, \lambda}$ are the Littlewood-Richardson 
coefficients. 
Applying the rule \eqref{eq:branchingA} recursively one finds 
\be\label{eq:branchingB} 
R_\lambda \simeq \bigoplus_{\nu_0,\ldots, \nu_\ell} 
N_{\nu_0,\ldots, \nu_\ell,\lambda} \, \big(R_{\nu_0}\boxtimes \cdots \boxtimes 
R_{\nu_\ell}\big)
\ee
where $\nu_\imath$ is a partition of $\gamma_\imath$ for 
$0\leq \imath \leq \ell$. Substitution in \eqref{eq:gammacharA}
yields
\be\label{eq:gammacharB}
\ch_T \calP_\mu^\gamma =  \sum_\lambda
\sum_{\nu_0,\ldots, \nu_\ell} 
N_{\nu_0,\ldots, \nu_\ell,\lambda} \prod_{\imath=0}^\ell 
\bigg({1\over |\CS_{\gamma_\imath}|} 
\sum_{g_\imath \in \CS_{\gamma_\imath}} 
\chi_{R_{\nu_\imath}}(g_\imath)\bigg)\, \ch_{\bf T}V_{\mu,\lambda}
\ee
Next note that 
\[ 
{1\over |\CS_{\gamma_\imath}|} 
\sum_{g_\imath \in \CS_{\gamma_\imath}} 
\chi_{R_{\nu_\imath}}(g_\imath)  = {\rm dim}\, 
R_{\nu_\imath}^{\CS_{\gamma_\imath}}
\]
is the dimension of the $\CS_{\gamma_\imath}$-fixed 
subspace of $R_{\nu_\imath}$. Since $R_{\nu_\imath}$ is an 
irreducible $\CS_{\gamma_\imath}$-representation, 
\[
{\rm dim}\, 
R_{\nu_\imath}^{\CS_{\gamma_\imath}}= 0
\]
unless $R_{\nu_\imath}$ is the trivial representation corresponding 
to the length one partition 
$\nu_\imath = (\gamma_\imath)$. In the latter case, 
\[
{\rm dim}\, 
R_{(\gamma_\imath)}^{\CS_{\gamma_\imath}}= 1.
\]
Then equation \eqref{eq:gammacharB} reduces to
\[
\ch_T \calP_\mu^\gamma =  \sum_\lambda
N_{(\gamma_0),\ldots, (\gamma_\ell),\lambda}\, \ch_{\bf T}V_{\mu,\lambda}.
\]
formula \eqref{eq:chargammaD} 
Using equation \eqref{eq:KMDnumbers}, formula \eqref{eq:chargammaD} follows from the identity 
\be\label{eq:KNidentity}
N_{(\gamma_0),\ldots, (\gamma_\ell),\lambda}=
K_{\lambda, \tgamma}. 
\ee
The latter is proven in \cite[Appendix 9]{reptheory}. 
More precisely, 
as shown in loc. cit., the Littlewood-Richardson coefficients 
occur in the decomposition of the product of two Schur 
functions: 
\be\label{eq:schurproductA}
s_{\nu_1}(x) s_{\nu_2}(x) = \sum_{\lambda} N_{\nu_1,\nu_2, \lambda} s_\lambda(x).
\ee
Furthermore, applying formula \eqref{eq:schurproductA} recursively 
as in \cite[Eqn. (A.9) pp. 456]{reptheory} yields 
\be\label{eq:schurproductB} 
s_{(r_1)}(x) \cdots s_{(r_k)}(x) = 
\sum_{\lambda} K_{\lambda, \rho} s_\lambda(x) 
\ee
where $\rho$ is the partition of $r$ 
determined by $(r_1,\ldots, r_k)$, and 
$K_{\lambda, \rho}$ are the Kostka numbers. This implies 
identity \eqref{eq:KNidentity}. 

Using equations \eqref{eq:chigenusidC}, 
\eqref{eq:chargammaD} the contribution of a 
fixed point $[I_\mu]$ to the partition function \eqref{eq:parinstA}
reduces to: 
\[ 
\bal
& \Omega_\mu^{(g,p)}(q_1,q_2,{\tilde y}) \sum_\nu  \sum_\lambda {\widetilde K}_{\lambda,\mu}(q_2,q_1) 
K_{\lambda,\nu}m_\nu({\tilde x}) =  \\
&  \Omega_\mu^{(g,p)}(q_1,q_2,{\tilde y})\sum_\lambda {\widetilde K}_{\lambda,\mu}(q_2,q_1) s_\lambda({\tilde x})  = 
 \Omega_\mu^{(g,p)}(q_1,q_2,{\tilde y})
 {\widetilde H}_\mu (q_2,q_1; {\tilde x}). \\
\eal
\]
where $m_\nu({\tilde x})$ are the monomial symmetric functions, 
and $s_\lambda({\tilde x})$ the Schur functions.  
This concludes the proof of equation \eqref{eq:parinstB}.

\section{BPS expansion and a parabolic 
$P=W$ conjecture}\label{parPW}

Collecting the results of the previous two sections, here it is  shown that geometric engineering yields a conjectural expression 
for the refined stable pair partition 
function \eqref{eq:orbstpairA}, which agrees with the 
left hand side of the HLRV formula \eqref{eq:HLRVformB} 
by a change of variables. Furthermore, it will be checked that the 
same change of variables relates the right hand side of 
equation \eqref{eq:HLRVformB} with a refined Gopakumar-Vafa 
expansion, completing the physical derivation of the HLRV formula.

As in Section \ref{isosection} 
it will be assumed that there is a single marked point on $C$. 
The root stack $\wC$ be the root stack has stabilizer $\mu_s$ 
at the unique orbifold point, for some $s\geq 1$. 
The  local threefold ${\wY}_M$ is the total space of the rank two 
bundle $\nu^*M^{-1}\oplus K_{\wC}\otimes_{\wC}\nu^*M$ 
on the root stack $\wC$, with $M$ a degree $p$ line bundle on $C$. Let $Z_{\wY_M}^{ref}(q,x,y)$ be the refined stable pair 
partition function of $Y$, where $x=(x_0,\ldots, x_{s-1}, 0, \ldots)$ are the formal counting variables associated to the marked point. Physically, these are chemical potentials for 
twisted sector Ramond-Ramond charges at the orbifold point. 
By analogy with 
\cite{wallpairs,BPSPW}, the geometric engineering 
conjecture reads: 
\be\label{eq:reforbB} 
\CZ^{ref}_{\widetilde Y} (q,x,y) = 
 1+ \sum_{r\geq 1} \CZ^{(r)}_K (qy^{-1}, q^{-1}y^{-1}, y,  (-1)^{(g-1+p)}y^{-g}x).
\ee
The terms in 
the right hand side are given by \eqref{eq:parinstB}. 
Equation \eqref{eq:omegaformula} yields 
\[
\begin{aligned} 
& \Omega_\mu^{g,p}(q_1,q_2,{\widetilde y}) =
\prod_{\Box\in \mu} (q_1^{l(\Box)} q_2^{a(\Box)})^{g-1+p} 
{(1-{\widetilde y} q_1^{-l(\Box)} q_2^{a(\Box)+1})^g 
 (1-{\widetilde y} q_1^{l(\Box)+1} q_2^{-a(\Box)})^g 
 \over (1- q_1^{-l(\Box)} q_2^{a(\Box)+1})
 (1- q_1^{l(\Box)+1} q_2^{-a(\Box)})}
\end{aligned} 
\]
where $a(\Box), l(\Box)$ are the arm and leg length of a box 
$\Box \in \mu$. Making the change of variables in equation 
\eqref{eq:reforbB} yields 
\be\label{eq:reforbC}
\CZ^{ref}_{\widetilde Y} (q,y,x) = 1+
\sum_{\mu\neq\emptyset} 
Z_\mu^{g,p}(q,y) {\widetilde H}_\mu(q^{-1}y^{-1},qy^{-1},x),
\ee
where
\[ 
\bal 
& Z_\mu^{g,p}(q,y) = (-1)^{p|\mu|}\prod_{\Box\in \mu} 
{ \big(q^{l(\Box)-a(\Box)} y^{1-h(\Box)}\big)^p
(qy^{-1})^{(2l(\Box)+1)(g-1)} 
 (1-y^{l(\Box)-a(\Box)}q^{-h(\Box)})^{2g}\over 
(1-y^{l(\Box)-a(\Box)-1}q^{-h(\Box)})
(1-y^{l(\Box)-a(\Box)+1}q^{-h(\Box)})}
\eal
\]
with $h(\Box)=a(\Box)+l(\Box)+1$, and the sum is over all 
Young diagrams $\mu$. This formula can be also written as 
\be\label{eq:transpformA}
\bal 
& Z_\mu^{g,p}(q,y) = (-1)^{p|\mu|} \prod_{\Box\in \mu} 
{ \big(q^{l(\Box)-a(\Box)} y^{1-h(\Box)}\big)^p
(qy)^{-(2a(\Box)+1)(g-1)} 
 (1-y^{a(\Box)-l(\Box)}q^{h(\Box)})^{2g}\over 
(1-y^{a(\Box)-l(\Box)-1}q^{h(\Box)})
(1-y^{a(\Box)-l(\Box)+1}q^{-h(\Box)})}.
\eal
\ee

A further change of variables yields 
\be\label{eq:reforbD}
\CZ^{ref}_{\widetilde Y} (z^{-1}w,z^{-1}w^{-1},x) = 1+
\sum_{\mu\neq \emptyset} 
\CH_\mu^{g,p}(z,w) {\widetilde H}_\mu(z^2,w^2,x),
\ee
where 
\[ 
\bal 
\CH_\mu^{g,p}(z,w) = \prod_{\Box\in \mu}
{(z^{2a(\Box)}w^{2l(\Box)})^p 
(z^{2a(\Box)+1}-w^{2l(\Box)+1})^{2g}\over 
(z^{2a(\Box)+2}-w^{2l(\Box)})
(z^{2a(\Box)}-w^{2l(\Box)+2})}
\eal
\]
For $p=0$ this is the left hand side of the HLRV formula 
evaluated at formal variables 
$x=(x_0, \ldots, x_{s-1}, 0, 0,\ldots)$.  

For the remaining part of this section, let $\wY:=\wY_{\CO_C}$ 
be the product $\IA^1 \times \tS$, with $\tS={\rm tot}(K_\wC)$.
Then it easy to check
that any moduli stack of compactly supported 
Bridgeland stable pure dimension one sheaves 
on $\wY$ with fixed numerical class is isomorphic to a product 
$\IA^1 \times \CM_\beta(\tS, \gamma)$,  
where $\CM$ is a moduli stack of $\beta$-stable pure dimension one sheaves on $\tS$ with fixed numerical equivalence 
class $\gamma$. The notation used here is the 
same as in Sections \ref{orbisurfaces}, equation \eqref{eq:discrinvA}, and \ref{specstack}. Since in this particular case there is a single marked point, and the eigenvalues $\uxi$ 
are trivial, $\gamma$ will be labelled by integers $d^l\geq 1$, 
$1\leq l\leq s$, and $n\in \IZ$. According to 
Section \ref{rootsection}, the moduli stack $\CM_\beta(\tS,\gamma)$ 
is isomorphic to a moduli stack of stable strongly parabolic 
Higgs bundles on $C$. 

From a string theoretic perspective 
this chain of isomorphisms identifies parabolic Higgs 
bundles on $C$ with supersymmetric D2-D0 configurations 
on the Calabi-Yau threefold $\wY$. 
Then the HLRV formula is identified with a refined Gopakumar-Vafa 
expansion \cite{GVII,KKV,Iqbal:2007ii,CKK}
provided that one assumes a parabolic variant of the $P=W$ 
conjecture \cite{hodgechar}.  Some details are provided below 
for completeness. 

For a precise formulation of the parabolic $P=W$ conjecture, 
consider a smooth projective curve $C$ with two marked points
$p,\infty \in C$ and let 
$\gamma_p, \gamma_\infty \in \pi_1(C\setminus \{p,\infty\})$ be 
the generators associated to the marked points. 
Let $(r,e)\in \IZ_{>0}\times\IZ$ be coprime integers and let 
${\sf C}_\ulambda$ denote the $GL(r,\IC)$ conjugacy class 
of a diagonal matrix with (ordered) eigenvalues 
\[
\ulambda = (\lambda_1, \ldots, \lambda_r).
\] 
Let also $\mu=(\mu^1,\ldots, \mu^l)$ denote the partition 
of $r$ determined by the multiplicities of the above eigenvalues. 

Now let $\CC^e_\ulambda(C,p,\infty)$ be the character variety 
with conjugacy classes ${\sf C}_\ulambda$, ${\rm exp}(2e\pi\sqrt{-1}/r)$ at the marked points $p,\infty$.
According to \cite[Thm. 2.1.5]{HLRV}, for sufficiently generic $\ulambda$, 
$\CC^e_\ulambda(C,p,\infty)$ is either empty 
or 
a smooth quasi-projective variety of complex dimension 
\be\label{eq:dimcharvarA}
d_\mu = r^2(2g-2+1)-\sum_{j=1}^{l}(\mu^j)^2 +2. 
\ee
Note that $d_\mu$ is even; using the identity $r=\sum_{j=1}^{l} \mu^j$, 
\be\label{eq:dimcharvarB}
d_\mu = 2b_\mu, \qquad b_\mu = 
r^2(g-1) +1 + \sum_{\substack{1\leq j_1,j_2\leq l \\ j_1<j_2 \\}}
\mu^{j_1}\mu^{j_2}. 
\ee
Since the marked curve $(C,p,\infty)$ is fixed throughout this section,
the character variety will be denoted simply by $\CC^e_\ulambda$ in the 
following. 

Next consider the specialization of the HLRV formula \eqref{eq:HLRVformB} 
to the present case taking ${\sf x}_\infty = ( {\sf x}_{\infty, 0}, 
0, 0, \ldots)$. Since $\mu_\infty$ is the 
length one partition $(r)$, the variable ${\sf x}_{\infty,0}$ can be 
scaled off by a redefiniton of the formal variable ${\sf x}$ associated to $p$. Moreover, as observed in \cite{HLRV}, 
the mixed Poincar\'e ploynomial
$P_c(\CC^e_\ulambda; u,t)$ depends only on $\mu$ as long as a 
$e$ is coprime with $r$.  Therefore equation \eqref{eq:HLRVformB} 
yields a formula of the form 
\be\label{eq:HLRVformA}
\sum_{\mu} \CH^{g,0}_{\mu}(z,w) 
{\widetilde H}_\mu(z^2,w^2;{\sf x}) 
= {\rm exp} \bigg(\sum_{k=1}^\infty {1\over k}{w^{-kd_\mu} 
P_{c,\mu}( z^{-2k}, -(zw)^k) \over (1-z^{2k})(w^{2k}-1)}
m_\mu({\sf x}^k) 
\bigg)
\ee
where $P_{c,\mu}(u,t)= P_c(\CC^e_\ulambda; u,t)$.

By analogy with \cite{hodgechar}, the parabolic $P=W$ conjecture 
identifies the weight filtration
$W_\bullet H_{cpt}(\CC^e_\ulambda)$ 
with the perverse sheaf filtration on the compactly supported cohomology of a moduli space of stable 
strongly parabolic Higgs bundles. 
As a first step, note that Conjecture 1.2.1(ii) in  \cite{HLRV}
yields the identifications 
\[
W_{2p}H_{cpt}(\CC^e_\ulambda) = W_{2p+1} 
H_{cpt}(\CC^e_\ulambda)
\]
for all $p$, just as in the unmarked case studied in \cite{HRV}. 


Next, let $\CH_\um^e$ denote the moduli space 
of rank $r\geq 1$, degree $e$ 
stable parabolic Higgs bundles $(E,\Phi)$ 
on the marked curve $(C,p)$ with parabolic structure of 
type $\um$ at $p$. 
The Higgs field $\Phi:E\to E\otimes K_C(p)$ has nilpotent residue 
at $p$ with respect to the. Let $\mu$ be the partition of 
$r$ determined by $\um$. For primitive discrete invariants 
$(\um,e)$ and  sufficiently generic parabolic weights there are no strictly semistable objects, 
and the moduli space is a smooth quasi-projective variety 
of dimension $d_\mu$. Furthermore, $\CH_\um^e$ is 
diffeomorphic in this case with the character variety $\CC^e_\ulambda$ 
provided the eigenvalues $\lambda_i$ are related to the parabolic 
weights by 
$\lambda_i = e^{2i\pi \alpha_i}$, $1\leq i\leq r$. 
There is also a Hitchin map 
\[ 
h : \CH_\um^e \to \CB_\um 
 \]
with $\CB_\um \subset \oplus_{i=1}^r H^0(K_C(p)^{\otimes i})$
a linear subspace of dimension $b_\mu$. 
The generic fibers of $h$ are smooth abelian varieties of dimension $b_\mu$, and the total space $\CH_\um^e$ is an algebraically complete integrable system.  
By analogy with \cite{decompth,hodgechar}, this yields 
a perverse sheaf filtration $P^\bullet H(\CH_\um^e)$. The  
 parabolic $P=W$ conjecture states that
\be\label{eq:PWconj}
W_{2p}H(\CC^e_\ulambda) = P_{p}H(\CH^e_\um)
\ee
for all values of $p$.

Equation \eqref{eq:PWconj} leads to an identification of 
the HLRV formula \eqref{eq:HLRVformB} with a refined BPS expansion
in close analogy with 
\cite[Sect. 4]{BPSPW}. Very briefly, using the methods in
\cite{decompth} one can prove a
hard Lefschetz theorem for the parabolic Hitchin map and also
choose a (noncanonical) splitting of the perverse sheaf 
filtration as in 
\cite[Sect 1.4.2, 1.4.3]{hodgechar}. This yields 
an $SL(2,\IC)\times \IC^\times$ action on the cohomology 
$H(\CH_\um^e)$, which splits as a direct sum 
\be\label{eq:sltwodecomp} 
H(\CH_\mu^e) \simeq  \oplus_{p=0}^{b_\mu} 
R_{(d-p)/2}^{\oplus \mathrm{dim}(Q^{p,0})}
\ee
where $R_{j_L}$ is the irreducible $SL(2,\IC)$-representation of 
spin
$j_L\in {1\over 2}\IZ$. 
In the above formula $p$ is the perverse degree and $Q^{p,0}$ the 
primitive cohomology of perverse degree $p$. The 
cohomological degree is encoded in the $\IC^\times$ action, which 
scales the quotient $Gr^P_p H^k(\CH^e_\mu)$ with weight 
$l=k-p-b_\mu$. Then specializing ${\sf x}$ to 
 ${\sf x} = x = (x_0,\ldots, x_{s-1}, 0,
0, \ldots)$ and making 
the same change of variables 
\[ 
(z,w) = \big( (qy)^{-1/2}, (qy^{-1})^{1/2})
\] 
as in equation \eqref{eq:reforbD}
converts equation \eqref{eq:HLRVformA} into a refined 
Gopakumar-Vafa expansion. This computation is completely analogous with \cite[Sect. 4]{BPSPW}, hence the details are omitted.

\section{Recursion via wallcrossing}\label{recursion}

The recursion relation conjectured in 
\cite{wallpairs} for the Poincar\'e 
polynomial of the moduli space of Hitchin pairs admits a natural 
generalization to parabolic Higgs bundles. The derivation of this 
formula is completely analogous to loc. cit. assuming again 
all the foundational aspects of motivic Donaldson-Thomas 
theory \cite{wallcrossing}. The final result will be recorded below, 
omitting most intermediary steps. 

For simplicity it will be assumed again that the curve $C$ has only one marked point $p$. 
To fix notation, the discrete invariants 
of a parabolic rank bundle $E^\bullet$ on $C$ are the 
degree $e\in \IZ$ and the flag type $\um = (m_a)_{0\leq a\leq s-1}\in (\IZ_{\geq 0})^{\times s}$. Let 
$$|\um|= \sum_{a=0}^{s-1}m_a,\qquad 
\chi(\um,e)=e-|\um|(g-1).$$
For any weights $\ualpha=(\alpha_{a})_{0\leq a \leq s-1}$
let 
\[ 
\um \cdot \ualpha = \sum_{a=0}^{s-1} m_a\alpha_a.
\]
The parabolic slope and the parabolic $\delta$-slope are defined 
respectively by 
\[ 
\mu(\um, e,\ualpha) = {\chi(\um,e) + \um\cdot\ualpha \over |\um|},
\qquad 
\mu_\delta(\um,e,\ualpha) = {\chi(\um,e) 
+ \um\cdot \ualpha +\delta\over 
|\um|},
\]
and the ordinary slopes are given by
\[ 
\mu_\delta(\um,e) = {\chi(\um,e)+ \delta\over |\um|}, 
\qquad 
\mu(\um ,e) = {\chi(\um,e)\over |\um|}.
\]

\subsection{Generic parabolic weights}\label{generic}
The recursion formula will be derived from wallcrossing with respect 
to variations of the stability parameter $\delta$ introduced in 
Section \ref{ADHMpar}. The refined 
parabolic ADHM invariants will be denoted by $A_\delta(\um, e,\ualpha; y)$
 while the refined parabolic Higgs bundle invariants 
by $H(\um,e,\ualpha;y)$. Note that  $H(\um,e,\ualpha;y)\in 
\IQ(y)$ 
are the rational refined invariants obtained directly from the motivic integration 
map in \cite{wallcrossing}, not the integral refined invariants
${\overline H}(\um,e,\ualpha;y)\in \IZ[y,y^{-1}]$. 
The relation between the two sets of invariants for sufficiently 
generic weights  is given by the 
refined multicover formula 
\be\label{eq:refmulticover} 
H(\um,e,\ualpha;y) = \sum_{\substack{ k\geq 1,\ (\um,e) = 
k(\um',e')}} {1\over k[k]_y} {\overline H}(\um',e',\ualpha; y^k).
\ee

For fixed numerical invariants and fixed 
 parabolic weights, there are finitely many 
critical values $\delta_c \in \IR$, where strictly semistable 
objects can exist. 
Using the formalism of \cite{wallcrossing}, 
the wallcrossing formula at such a critical 
value $\delta_c\neq 0$ is
\be\label{eq:wallrecA}
\bal
& A_{\delta_{c}+}(\um,e,\ualpha; y)- A_{\delta_c-}(\um,e,\ualpha;y) = \\
& \mathop{\sum_{l\geq 2}}_{} {1\over (l-1)!}
\sum_{
\Delta_l(\delta_c, \um, e,\ualpha)}
A_{\delta_{c-}}(\um_1,e_1,\ualpha;y)
\prod_{i=2}^l [\chi(\um_i,e_i)]_{y}
H(\um_i,e_i,\ualpha;y).\\
\eal
\ee
where 
\[
\bal
& \Delta_l(\delta_c,\um,e,\ualpha) =\\
& \big\{ 
(\um_1,\ldots, \um_l) , (e_1,\ldots, e_l)
\, |\, \um_i\in (Z_{\geq 0})^{\times r},\ e_i\in \IZ,\ 
|\um_i|>0,\ 1\leq i \leq l ,\\
& \ \ (\um_1,e_1)+\cdots+(\um_l,e_l)=(\um,e),\
\mu(\um_i,e_i,\ualpha)=\mu_{\delta_c}(\um_1,e_1,\ualpha), 
\ 2\leq i\leq l \big\}\\
\eal
\]
and 
\[
[n]_y = {y^n-y^{-n} \over y-y^{-1}}
\]
for any integer $n\in \IZ$. This is the same wallcrossing formula as
\cite[Eqn. 1.3]{wallpairs}, except the Higgs invariants differ by a sign 
$(-1)^{\chi(\um,e)}$ from the used in loc. cit.  The present normalization 
is more natural in this context. 
There is a similar formula for $\delta_c=0$, including an extra 
term with $\um_1=0$ as in \cite[Eqn. 1.4]{wallpairs}. 

Applying equation 
\eqref{eq:wallrecA} 
iteratively from $\delta>>0$ to $\delta<<0$, and using the 
duality relations \eqref{eq:dualinv}, 
one obtains a wallcrossing formula of the form 
\be\label{eq:wallrecB}
\bal
& [\chi(\um,e)]_{y} H(\um,e,\ualpha;y) 
 =A_{+\infty}( \um,e;y)    -
 A_{+\infty}({\check \um},{\check e};y) \\
& +
{\sum_{l\geq 2}}{(-1)^{l-1}\over (l-1)!}
\sum_{\Delta_l^{(>)}(\um,e,\ualpha)}
A_{+\infty}(\um_1,e_1;y)
\prod_{i=2}^l [\chi(\um_i,e_i)]_{y}
H(\um_i,e_i,\ualpha;y)\\
& -\mathop{\sum_{l\geq 2}}_{}{(-1)^{l-1}\over (l-1)!}
\sum_{\Delta_l^{(\geq)}({\check \um}, {\check e},{\check \ualpha})}
A_{+\infty}(\um_1,e_1;y)
\prod_{i=2}^l [\chi(\um_i,e_i)]_{y}
H(\um_i, e_i,\ualpha;y)\\
& -\mathop{\sum_{l\geq 2}}_{} {1\over l!}
\sum_{\Delta_l^{(=)}(\um,e,\ualpha)}
\prod_{i=1}^l [\chi(\um_i,e_i)]_{y}
H(\um_i, e_i,\ualpha;y)\\
\eal
\ee
where 
\be\label{eq:sumrangeA}
\bal
& \Delta_l^{(\Diamond)}(\um,e,\ualpha) =\\
& \big\{ 
(\um_1,\ldots, \um_l) , (e_1,\ldots, e_l)
\, |\, \um_i\in (Z_{\geq 0})^{\times r},\ e_i\in \IZ,\ 
|\um_i|>0,\ 1\leq i \leq l ,\\
& \ \  (\um_1,e_1)+\cdots+(\um_l,e_l)=(\um,e),\
\mu(\um_i,e_i,\ualpha) \ \Diamond \ 
\mu(\um,e,\ualpha), \ 2\leq i\leq l\big\},
\eal
\ee
 the symbol $\Diamond$ taking values 
  $ >, \geq, =$ respectively. 
  Note that for any discrete invariants $\un$ there exists 
a lower bound $d_0\in \IZ$ such that $A_\infty(\un,d;y)=0$ 
for all $d<d_0$. This can be proven by standard bounding arguments, or, alternatively, it follows easily from the conjectural 
formula \eqref{eq:reforbC}. Therefore  the number of terms 
in the right hand side of equation \eqref{eq:wallrecB} 
is finite and bounded above by a constant independent of
the parabolic weights $\ualpha$. 

The recursion formula \eqref{eq:wallrecB} together with the 
geometric engineering conjecture \eqref{eq:reforbB} 
completely determines the parabolic refined invariants 
$H(\um,e,\ualpha; y)$. 
Using the arguments employed by Mozgovoy in \cite{ADHMrecursion},  
it will be shown below that the resulting invariants are compatible 
with those determined by the HRLV formula \eqref{eq:HLRVformA}. 

For simplicity, consider local curves on type $(0,2g-2)$ in the 
following. Using the same notation as \cite{ADHMrecursion}, the refined 
partition function \eqref{eq:orbstpairAB} will be denoted by 
$A_\infty(q,y,x)$. Hence 
\[ 
A_\infty(q,y,x) = \sum_{\um, e} A_{+\infty}(\um,e;y) q^{\chi(\um,e)} x^{\um}. 
\] 
Since there is a single marked point, the formal variable 
$x=(x_0,x_1,\ldots)$ does not carry an extra index. 

As shown in Section \ref{parPW},
 geometric engineering predicts that 
$A_\infty(q,y,x)$ is determined by equation \eqref{eq:reforbB}
\[
A_\infty(q,y,x) =1+\sum_{\mu\neq \emptyset} Z_\mu^{g,0}(q,y) 
\wH_\mu(q^{-1}y^{-1}, qy^{-1}; x).
\] 
Following \cite{ADHMrecursion}, let ${\widetilde P}_\um(q,y)$ be defined by the formula 
\be\label{eq:tildeP}
A_\infty(q,y,x) ={ \rm{exp}}\bigg[\sum_{k\geq 1} \sum_{\um} 
{x^{k\um} \over k} f(q^k,y^k){\widetilde P}_\um(q^k,y^k)\bigg]
\ee
where 
\[ 
f(q,y) = {q
\over (1-qy)(y-q)}.
\] 
Note that 
${\widetilde P}_\um (q,y)$ is related to the mixed Poincar\'e polynomial of 
the character variety $\CC_\lambda^e$ defined in Section \ref{parPW},
where $\lambda$ is the partition of $r=|\um|$ determined by $\um$. 
Using the change of variables $(z^2,w^2)=(q^{-1}y^{-1}, qy^{-1})$ in 
equation \eqref{eq:HLRVformA}, one obtains  
\[
{\widetilde P}_\um (q,y) = y^{b_\lambda+2}q^{-b_\lambda} 
P_c(\CC_\lambda^e, qy, -y^{-1})
\]
where $b_\lambda = d_\lambda/2$ is half the complex dimension 
of the character variety. 

Now let 
\[
\Omega(\um, e; y) =  y{\overline H}(\um, e, \ualpha; y)
\] 
for any discrete invariants $(\um,e)$. Assuming that
 the invariants $\Omega(\um,e;y)$ are 
independent of the degree $e\in \IZ$ for any $\um$, 
it will be shown below that 
\be\label{eq:recformula}
\Omega(\um,e;y) = {\widetilde P}_\um(1,y)
\ee
for all $(\um,e)$. The proof is entirely analogous to the proof 
of \cite[Thm. 4.6]{ADHMrecursion}, some details being presented below 
for completeness. Note that the assumption 
that $\Omega(\um,e;y)$ are 
independent of $e$ is a standard conjecture \cite{genDTI} 
for Donaldson-Thomas 
invariants of pure dimension one sheaves on Calabi-Yau threefolds. 

Following \cite{ADHMrecursion}, for any series 
\[ 
I = \sum_{(\um,e)} I(\um,e) q^{\chi(\um,e)} x^\um \
\in\ \IQ(y)[[q^{\pm 1}, x]]
\]
and any $\mu \in \IR$, let 
\[
I_{\Diamond \mu} = \sum_{\substack{
\mu(\um,e,\ualpha)  \Diamond \mu}} I(\um,e) q^{\chi(\um,e)}
x^\um 
\]
where $\Diamond \in \{=, \geq, >, \leq, <\}$. 
Furthermore, for any $\mu\in \IR$ define 
\[
C_\mu(q,y,x) =
{\rm exp} \bigg[\sum_{k\geq 1} \sum_{\mu(\um, e, \ualpha)=\mu} 
{x^{k\um}\over k(y^{2k}-1)} \Omega(\um,e;y^k) \big( (qy)^{k\chi(\um,e)} 
-(qy^{-1})^{k\chi(\um,e)}\big)\bigg]
\]
 and 
 \[ 
 C_{\Diamond \mu}(q,y,x) = \prod_{\eta \Diamond \mu} C_\eta(q,y,x)
  \] 
  Then the recursion relation \eqref{eq:wallrecB} 
  can be recast in the form 
  \be\label{eq:wallrecE} 
  C_\mu(q,y,x)= \big( {\overline A}_\infty (q,y,x)C^{-1}_{>\mu}(q,y,x)\big)_\mu 
  -({\overline A}_\infty(q^{-1},y,x) C^{-1}_{\geq -\mu}(q^{-1},y,x)\big)_{-\mu}
  \ee 
  by analogy with \cite[Remark. 4.5]{ADHMrecursion}, where 
  \[
  {\overline A}_{\infty}(q,y,x) = A_\infty(q,y,x) -1.
  \]
  
In order to prove equation \eqref{eq:recformula} it suffices to show that the 
statement of \cite[Thm 4.7]{ADHMrecursion} holds in the present 
context. Namely, it suffices to prove the identity 
\be\label{eq:wallrecF}
A_\infty(q,y,x) C_{>\mu}^{-1}(q,y,x) = A_{\infty}(q^{-1},y,x) C^{-1}_{\geq -\mu}
(q^{-1},y,x)
\ee
in $\IQ(y)[[q^{\pm 1},x]]$. The proof given in \cite[Sect 5]{ADHMrecursion} 
is based on several essential facts. 

First note that independence of $\Omega(\um,e;y)$ of degree yields 
a factorization of the form 
\[ 
\bal 
& \sum_{(\um,e)} \Omega(\um,e;y) q^{\chi(\um,e)} 
\big( y^{\chi(\um,e)}-y^{-\chi(\um, e)}\big) x^\um 
= \\
& \bigg(\sum_{\um} \Omega(\um;y) x^{\um} \bigg) 
\sum_{n \in \IZ} \big((qy)^n-(qy^{-1})^n)\big),
\eal
\] 
where $\Omega(\um;y)$ denotes the common value of $\Omega(\um,e;y)$. 
Moreover, note that the function $f(q,y)$ defined below equation \eqref{eq:tildeP}
satisfies 
\[
f(q,y) = f(q^{-1},y).
\]
These two facts imply that 
completely analogous statements to 
\cite[Lemma 5.1]{ADHMrecursion}, 
\cite[Lemma 5.4]{ADHMrecursion} and
\cite[Prop. 5.5]{ADHMrecursion} hold in the present context.

The next important observation is that 
\be\label{eq:Ainfty}
A_\infty(q,y,x) = A_\infty(q^{-1},y,x). 
\ee
In the present context, this 
 follows from equation \eqref{eq:transpformA}, which shows that 
\[
Z^{g,0}_\lambda(q^{-1},y) = Z^{g,0}_{\lambda^t}(q,y),
\]
and the standard property of MacDonald polynomials
\[ 
\wH_\lambda(t,s;x) = \wH_{\lambda^t}(s,t;x),
\] 
which yields 
\[ 
\wH_\lambda(qy^{-1}, q^{-1}y^{-1};x) = \wH_{\lambda^t}(q^{-1}y^{-1}, qy^{-1};x).
\] 
Since $f(q,y)$ is invariant under $q\mapsto q^{-1}$, equation 
\eqref{eq:Ainfty} implies that 
\[
{\widetilde P}_\um(q,y) = {\widetilde P}_\um(q^{-1},y)
\]
which is analogous to \cite[Lemma 5.7]{ADHMrecursion}. 

From this point on, the proof of identity \eqref{eq:recformula} is 
identical with the proof of \cite[Thm. 4.7]{ADHMrecursion} 
given in Section 5 of loc. cit.

\subsection{Trivial weights}\label{zero}

An alternative recursion formula may be derived along the same lines, working with trivial weights, $\alpha_a=0$ for 
all $0\leq a\leq r-1$,  rather than generic weights. 
This is usually a very degenerate limit in the theory of parabolic 
Higgs bundles. However, the theory of motivic Donaldson-Thomas 
invariants \cite{wallcrossing} works equally well for trivial 
weights. Moreover, the wallcrossing formula 
of \cite{wallcrossing} shows that the 
refined Donaldson-Thomas 
invariants $H(\um, e,\ualpha;y)$ are in fact independent of
the weights, as long as the weights are sufficiently generic. 
In fact, if one is willing to grant the refined generalization 
of \cite[Thm. 6.16]{genDTI}, even more is expected to be true. That is, the integral refined invariants ${\overline H}(\um,e,\ualpha;y)$ are expected to be independent of the weights $\ualpha$, 
for all possible values, including non-generic ones. 

Moreover, note that for coprime numerical invariants 
$(r,e)=1$, the moduli space of stable parabolic bundles 
with fixed flag type $\um$ is independent of the parabolic 
weights as long as they are sufficiently small, including 
non-generic values. 
This can be proven using a boundedness argument. 
Therefore the strong form of the conjecture 
in the previous paragraph holds at least 
for $(|\um|,e)$ coprime. 

In the following we will simply write down the $\ualpha=0$ 
version of the recursion formula \eqref{eq:wallrecB}. 
As explained above, it will yield the same results for 
the integral refined invariants as \eqref{eq:wallrecB} 
at least for $(|\um|,e)$ coprime. If one is willing to grant the 
strong form of the refined non-dependence conjecture, 
it will yield the same values even for non-coprime pairs 
$(|\um|, e)$.

Setting $\ualpha=0$, equation \eqref{eq:sumrangeA} 
specializes to 
\be\label{eq:sumrangeB}
\bal
& \Delta_l^{(\Diamond)}(\um,e,0) =\\
& \big\{ 
(\um_1,\ldots, \um_l) , (e_1,\ldots, e_l)
\, |\, \um_i\in (Z_{\geq 0})^{\times r},\ e_i\in \IZ,\ 
|\um_i|>0,\ 1\leq i \leq l ,\\
& \ \ (\um_1,e_1)+\cdots+(\um_l,e_l)=(\um,e),\
\mu(|\um_i|,e_i) \ \Diamond \ 
\mu(|\um|,e), \ 2\leq i\leq l\big\}
\eal
\ee
with $\Diamond\in\{>,\geq, =\}$. Note that the slope 
inequalities in the right hand side of equation \eqref{eq:sumrangeB} 
depend only on $|\um_i|, |\um|$, hence 
they are invariant under permutations of the entries of 
$\um_i, \um$, $1\leq i\leq l$. Moreover, the conjectural 
formula \eqref{eq:reforbC}, and the parabolic $P=W$ conjecture 
in Section \ref{parPW} imply that the invariants 
$A_\infty(\um,e;y)$, $H(\um,e;y)$ 
are also invariant under permutations of the entries of 
$\um$. Therefore they depend only on the partition $\lambda$ 
of $|\um|=r$ determined by $\um$. Abusing notation, they 
will be denoted by $A_\infty(\lambda, e;y)$, $H(\lambda,e;y)$. 
Moreover, for each partition $\lambda$ of $r\geq 1$, let 
\be\label{eq:sumrangeC} 
\bal
\Delta_l^{(\Diamond)}(\lambda,e) =\big\{ &
(\lambda_1,\ldots, \lambda_l) , (e_1,\ldots, e_l)
\, |\, \lambda_i\neq \emptyset, \ e_i\in \IZ,\ 1\leq i \leq l ,\\
& (|\lambda_1|,e_1)+\cdots+(|\lambda_l|,e_l)=(|\lambda|,e),\
\mu(|\lambda_i|,e_i) \ \Diamond \ 
\mu(|\lambda|,e), \ 2\leq i\leq l\big\},
\eal
\ee
with $\Diamond\in\{>,\geq, =\}$. 
Then a straightforward computation shows that the zero 
weight specialization of the recursion formula \eqref{eq:wallrecB} 
can be set in the form:
\be\label{eq:wallrecC} 
\bal
& [e-|\lambda|(g-1)]_y H(\lambda,e;y)
 =A_{+\infty}( \lambda,e;y)  -
 A_{+\infty}(\lambda,{\check e};y) \\
& +\bigg[
{\sum_{l\geq 2}}{(-1)^{l-1}\over (l-1)!}
\sum_{\Delta_l^{(>)}(\lambda,e)}
A_{+\infty}(\lambda_1,e_1;y)m_{\lambda_1}(x)
\prod_{i=2}^l [e_i-|\lambda_i|(g-1)]_y 
H(\lambda_i,e_i;y)m_{\lambda_i}(x)\\
& -\mathop{\sum_{l\geq 2}}_{}{(-1)^{l-1}\over (l-1)!}
\sum_{\Delta_l^{(\geq)}(\lambda, {\check e})}
A_{+\infty}(\lambda_1,e_1;y)m_{\lambda_1}(x) 
\prod_{i=2}^l [e_i-|\lambda_i|(g-1)]_y 
H(\lambda_i, e_i;y)m_{\lambda_i}(x)\\
& -\mathop{\sum_{l\geq 2}}_{} {1\over l!}
\sum_{\Delta_l^{(=)}(\lambda,e)}
\prod_{i=1}^l [e_i-|\lambda_i|(g-1)]_y 
H(\lambda_i,e_i;y)m_{\lambda_i}(x) \bigg]_\lambda\\
\eal
\ee
where $[f(x)]_\lambda$ is the coefficient of $m_\lambda(x)$ 
in the expansion of the symmetric 
function $f(x)$ in the monomial symmetric basis.

Proceeding as in the Section \ref{generic}, it is straightforward to show 
that the 
solution to the recursion relation \eqref{eq:wallrecC} is also 
in agreement with the predictions of the HLRV formula. 
This confirms the weight independence conjecture stated at the 
beginning of the current subsection.

\section{A conifold experiment}\label{conifold}

The goal of this section is to present numerical evidence for the 
geometric engineering conjecture \eqref{eq:reforbB} for refined 
parabolic invariants on a resolved conifold. Therefore the curve $C$ will 
be the projective line $\IP^1$ and the line bundles $L_1,L_2$ will be 
isomorphic to $\CO_C(-1)$. Choosing homogeneous coordinates
$[z_0,z_1]$ on $C$, the marked point $p$ will be $z_1=0$. 
Parabolic refined invariants will be computed by virtual localization, 
using the equivariant $K$-theoretic index defined by Nekrasov and Okounkov in \cite{Mvert}. 
According to \cite{Motvan} this definition agrees under certain conditions
with the motivic construction of Kontsevich and Soibelman 
\cite{wallcrossing}. The equivariant index has been also employed in 
\cite{CKK} for similar computations of refined stable pair 
invariants of toric Calabi-Yau threefolds.

 \subsection{A parabolic conifold conjecture}\label{compsection}
Since the data $(C,L_1,L_2,p)$ will be fixed throughout this section, 
the moduli space of 
asymptotically stable parabolic ADHM sheaves 
with nunerical invariants $(\um,e)$ will be denoted by 
${\mathcal PM}_\infty(\um, e)$. 
As 
explained in Section \ref{modulisect}, this moduli space is equipped with  a 
symmetric perfect
obstruction theory $\IE^\bullet$ 
and a natural ${\bf T} = \IC^\times\times \IC^\times$
action given in equation 
\eqref{eq:toractA}. 
The symmetric obstruction theory is equivariant, but not 
equivariantly 
symmetric with respect to the {\bf T}-action.  The 
the {\bf T}-fixed locus has been shown to be proper 
using the forgetful morphism \eqref{eq:forgetflags}. Since the curve $C$ is the 
projective line in this section, there is an enhanced 
${\bf G}={\bf T}\times \IC^\times$ action on the moduli space where 
the action of the third factor $\IC^\times$ is induced by the scaling action 
on $C=\IP^1$, 
\[ 
(s\times [z_0,z_1]) \mapsto [z_0,sz_1].
\]
Using again the proper morphism \eqref{eq:forgetflags} it can be 
easily shown that the ${\bf G}$-fixed locus is finite. 
Again, the perfect obstruction theory is ${\bf G}$-equivariant but not 
{\bf G}-equivariantly symmetric. However using the deformation complex 
\eqref{eq:defcpxA} it is straightforward to check that 
\be\label{eq:equivsymmA} 
\IE^\bullet = Z^{-1}Q_1Q_2\, (\IE^\bullet)^\vee[1],
\ee
where $(Q_1,Q_2,Z)$ denote the canonical generators of the representation 
ring of ${\bf G}$. In particular $\IE^\bullet$ is equivariantly symmetric with respect to the action of the subtorus ${\bf G}_0 =\{s^{-1}t_1t_2=1\}
\subset {\bf G}$. Note moreover, that in this example, one can easily prove that the moduli space of asymtotically stable parabolic ADHM sheaves is proper. 
The details are provided in Appendix \ref{compactness} for completeness. 

In this context, following \cite{Mvert}, note that the virtual canonical bundle 
 of the moduli space admits a square root $K^{1/2}$, which is equivariant with respect to the action of the double cover 
${\widetilde {\bf G}}{\buildrel 2:1\over \longto} {\bf G}$ determined by the 
commutative diagram 
\[ 
\xymatrix{ 
1 \ar[r] & {\bf G}_0 \ar[r] \ar[d]_-{1}
& {\widetilde {\bf G}}\ar[r] \ar[d] & {\IC^\times} 
\ar[r] \ar[d]^-{\zeta\mapsto \zeta^2} & 1 \\
1 \ar[r] & {\bf G}_0 \ar[r] & {\bf G} \ar[r] & \IC^\times \ar[r] & 1.\\}
\]
Then the equivariant 
index  defined  in \cite{Mvert} is the equivariant holomorphic Euler character 
of $K^{1/2}$, 
\be\label{eq:equivindexA} 
I_{\mu,e} = \chi_{\widetilde {\bf G}}(K^{1/2}). 
\ee
According to \cite{Mvert}, relation \eqref{eq:equivsymmA} 
implies that $I_{\mu,e}$ is a Laurent polynomial in the 
element ${R}=(Z^{-1}Q_1Q_2)^{1/2}$ of the representation ring 
of ${\widetilde {\bf G}}$. 

Specializing equation \eqref{eq:reforbB} to a local rational curve 
of type $(-1,-1)$ yields 
\be\label{eq:parconifoldA} 
\CZ^{ref}_{\widetilde Y} (q,y,x) = 1+
\sum_{\mu\neq\emptyset} 
Z_\mu^{0,1}(q,y) {\widetilde H}_\mu(q^{-1}y^{-1},qy^{-1},x),
\ee
where 
\[ 
\bal 
& Z_\mu^{0,1}(q,y) = (-1)^{|\mu|}\prod_{\Box\in \mu} 
{ q^{h(\Box)} y^{a(\Box)-l(\Box)+1} \over 
(1-y^{a(\Box)-l(\Box)-1}q^{h(\Box)})
(1-y^{a(\Box)-l(\Box)+1}q^{-h(\Box)})}.
\eal
\]
The expansion of the right hand side of equation \eqref{eq:parconifoldA} 
in the monomial symmetric basis can be written as 
 \be\label{eq:parconifoldB} 
\CZ^{ref}_{\widetilde Y} (q,y,x) = 1+ 
\sum_{\mu\neq\emptyset} 
{W}_\mu (q,y) (-qy)^{|\mu|} m_\mu(x) 
\ee
with 
\[
 {W}_\mu (q,y) = \sum_{e\in \IZ} W_{\mu,e}(y) q^e. 
 \]
 Then the relation between the equivariant index \eqref{eq:equivindexA} 
 and formula \eqref{eq:parconifoldA} is the following conjectural 
 identity 
 \be\label{eq:equivindexB} 
  (-1)^e I_{\um,e}(R)\big|_{R=y} =W_{\mu,e}(y)
  \ee
for any discrete invariants $(\um,e)$. In the right hand side,
$\mu$ is the partition of $r=|\um|$ determined by $\um$, and $l(\mu)$ is 
the length of $\mu$. 
This conjecture will be verified by explicit computations in 
Section  \ref{examples}.

\subsection{Virtual localization and fixed points}\label{fixedsection}
The index \eqref{eq:equivindexA} 
can be computed explicitly by virtual localization, 
using the virtual Riemann-Roch theorem 
proven in \cite{virtRR,vfdg}. 
Suppose ${\mathfrak m}$ is an isolated ${\bf G}$-fixed 
point in the moduli space, and let $\IE^\bullet_{\mathfrak m}$ be 
the restriction of the 
two term perfect obstruction complex to ${\mathfrak m}$. 
Let 
$$[\IE_{\mathfrak m}^{\bullet}]= 
V_{2}-V_{1}$$
be the virtual ${\bf G}$-representation determined by the restriction of the 
perfect obstruction complex to ${\mathfrak m}$.
For parabolic ADHM sheaves, the $K$-theory class 
$[\IE^{\bullet}_{\mathfrak m}]$ 
is determined by the deformation complex \eqref{eq:defcpxA}, 
\be\label{eq:locformB} 
\bal 
& [\IE_{\mathfrak m}^{\bullet}]=\sum_{i=0}^1\sum_{j=0}^2 
(-1)^{i+j} [H^i(C, \CC^j(\CE^\bullet)]
\eal 
\ee
where $\CE^\bullet$ is the asymptotically stable 
parabolic ADHM sheaf on $C$ corresponding to the fixed point ${\mathfrak m}$.
Since the obstruction theory is symmetric, there is an isomorphism 
$V_2\simeq V_1^\vee$ of complex vector spaces, but not of 
{\bf G}-representations. In particular ${\rm dim}(V_2)={\rm dim}(V_1)=v$.
Then note that 
\[
V_2 = R^2\, V_1^{\vee}
\qquad 
{\rm and}\qquad  
K^{1/2}_{\mathfrak m} = R^v\, {\rm det}(V^1)^{-1}
\]
in the representation ring of ${\widetilde G}$. 
The contribution of ${\mathfrak m}$ to the 
virtual $K$-theoretic localization formula is 
\be\label{eq:locformA}
\bal
& K^{1/2}_{\mathfrak m}\Lambda_{-1}
[(\IE_{\mathfrak m}^{\bullet})^{\vee}]=
R^{v} {\rm det}(V_1)^{-1} {\Lambda_{-1}(V_2^\vee)\over 
\Lambda_{-1}(V_1^\vee)}.
\eal 
\ee
The virtual representation $[(\IE_{\mathfrak m}^{\bullet})^{\vee}]$ is 
the equivariant $K$-theoretic Euler characteristic of the deformation complex 
\eqref{eq:defcpxA} for the parabolic ADHM sheaf $\CE^\bullet$ 
on $C$ corresponding to the fixed point ${\mathfrak m}$ in the 
moduli space. 

The next task is to classify the fixed loci and compute their local contribution 
to the fixed point theorem. 
For concreteness suppose $L_1=\CO_C(-\infty)$ and $L_2=\CO_C(-\infty)$ 
as ${\widetilde {\bf G}}$-equivariant line bundles on $C$, where 
$\infty \in \IP^1$ is the point $z_0=0$ (as opposed to the marked point 
$p\in C$, which is given by $z_1=0$.) Therefore the ${\widetilde {\bf G}}$-equivariant canonical line bundle will be $K_C = \CO_C(-2\infty)$.

Using \cite[Prop. 3.1]{modADHM}, 
 asymptotically stable parabolic ADHM sheaves $\CE^\bullet$
fixed by  ${\widetilde {\bf G}}$ up to isomorphism are classified as follows. 
Forgetting the parabolic structure, the data $\CE=(E,\Phi_1,\Phi_2,\phi,
\psi)$ is an asymptotically stable ADHM sheaf on $C$ with coefficient line 
bundles $(L_1,L_2(p))$. There is a one-to-one correspondence 
between such sheaves ${\CE}$ and data $(\Delta, d,k)$, where 
$\Delta\subset (\IZ_{\geq 0})^2$ is a Young diagram, and 
$d:\Delta \to \IZ$, $k:\Delta \to \IZ$ two $\IZ$-valued functions 
satisfying the inequalities:
\be\label{eq:degineqA}
0\leq k(0,0) \leq d(0,0),
\ee
\be\label{eq:degineqB} 
0\leq k(i+1,j)-k(i,j)\leq d(i+1,j)-d(i,j)-1,
\ee
for any $(i,j) \in \Delta$ such that $(i+1,j)\in \Delta$, 
and 
\be\label{eq:degineqC} 
-1\leq k(i,j+1)-k(i,j)\leq d(i+1,j)-d(i,j)-1,
\ee
for any $(i,j) \in \Delta$ such that $(i,j+1)\in \Delta$.

Given a collection $(\Delta, d, k)$ as above, the underlying vector 
bundle of $\CE$ is of the form 
\be\label{eq:fixedA}
E \simeq \bigoplus_{(i,j)\in \Delta} E(i,j), 
\qquad E(i,j)=Q_1^{-i}Q_2^{-j}Z^{-k(i,j)}\CO_C(d(i,j)\infty)
\ee
as a ${\widetilde {\bf G}}$-equivariant bundle on $C$. 
The nonzero 
components of the morphisms $(\Phi_1,\Phi_2, \phi, \psi)$ 
are 
\be\label{eq:fixedB} 
\bal 
\Phi_1(i,j) : \CO_C\big(d(i,j)\infty\big) & \to \CO_C\big((d(i+1,j)-1)\infty\big)\\
1& \mapsto z_1^{k(i+1,j)-k(i,j)} z_0^{d(i+1,j)-d(i,j)-k(i+1,j)+k(i,j)-1}\\
\eal
\ee
\be\label{eq:fixedC} 
\bal
\Phi_2(i,j) :
\CO_C\big(d(i,j)\infty\big) &  \to \CO_C\big((d(i,j+1)-1)\infty+p\big)\\
1& \mapsto z_1^{k(i,j+1)-k(i,j)+1} z_0^{d(i,j+1)-d(i,j)-k(i,j+1)+k(i,j)-1}\\
\eal
\ee
\be\label{eq:fixedD} 
\bal 
\psi : \CO_C & \to \CO_C(d(0,0)\infty) \\
1& \mapsto  z_1^{k(0,0)}z_0^{d(0,0)-k(0,0)}
\eal 
\ee
All other components are identically zero. 

In order to simplify the computations, it will be convenient to 
choose specific generators for the cohomology of equivariant 
line bundles on $C=\IP^1$ using a  standard ${\check {\rm C}}$ech
cohomology computation. Let $z=z_1/z_0$ be an affine coordinate centered 
at the marked point $p$. Then one can easily show that 
\be\label{eq:cohA} 
H^0(\CO_C(d\infty + ap)) \simeq \IC\langle z^{-a}, z^{1-a}, \ldots, z^d\rangle 
\ee
for any $a, d\in \IZ_{\geq 0}$, and 
\be\label{eq:cohB}
H^1(\CO_C(-d\infty)) \simeq \IC\langle z^{-1}, \ldots, z^{1-d}\rangle
\ee
for any $d\in \IZ_{\geq 2}$. In this basis the nontrivial components 
\eqref{eq:fixedB}--\eqref{eq:fixedD} read 
\be\label{eq:fixedDA} 
\bal 
\Phi_1(i,j)=&  z^{k(i+1,j)-k(i,j)}, \qquad
\Phi_2(i,j) = z^{k(i,j+1)-k(i,j)}, \\
& \qquad \ \ \psi(0,0) = z^{k(0,0)}. \\
\eal 
\ee

For a ${\widetilde {\bf G}}$-fixed asymptotically stable parabolic ADHM 
sheaves $\CE^\bullet$, one has to specify in addition a flag 
\be\label{eq:flagatp} 
0=E_p^s \subseteq E_p^{s-1} \subseteq \cdots \subseteq E^0_p = E_p
\ee
in the fiber at $p$ preserved by the ${\widetilde {\bf G}}$
such that: 
\medskip

$(a)$ $\Phi_1\big|_p (E_p^{a}) \subseteq E_p^a$ for any $0\leq a\leq s$, 
and 

$(b)$ ${\rm res}_p(\Phi_2) (E_p^a) \subseteq E_p^{a+1}$ 
for any $0\leq a\leq s-1$. 
\medskip

\noindent
Note that for any such flag the subspaces $E_p^a$, $0\leq a\leq s-1$ 
must be specified 
by a third function $\varphi: \Delta \to \{0,\cdots, s-1\}$ such that 
\be\label{eq:flagatpB} 
E_p^a = \bigoplus_{(i,j) \in \varphi^{-1}(a)} Q_1^{-i}Q_2^{-j}Z^{-k(i,j)}\CO_C(d(i,j)\infty)_p.
\ee

In conclusion the ${\widetilde {\bf G}}$-fixed points in the moduli space 
of asymptotically stable parabolic ADHM sheaves are in one-to-one 
correspondence with data $(\Delta, d,k,\varphi)$ satisfying 
inequalities \eqref{eq:degineqA}-\eqref{eq:degineqC} and 
the compatibility conditions $(a)$, $(b)$ above \eqref{eq:flagatp}.
A complete enumeration 
of such data for fixed numerical invariants $(\um, e)$ is fairly tedious. 
This is done in detail in Section \ref{compsection} for low rank examples. 

To conclude this subsection, note that the contribution 
of a fixed point ${\mathfrak m}=(\Delta, d,k,\varphi)$ to the 
fixed point formula is determined by the deformation complex 
\eqref{eq:defcpxA}, using 
the exact sequences \eqref{eq:parhomA}, \eqref{eq:parhomB} 
and their strongly parabolic analogues. 

Recall that for any filtered vector space  $V^\bullet$, 
${\rm PEnd}(V^\bullet)$,   
${\rm SPEnd}(V^\bullet)$ denote the linear spaces of parabolic, 
respectively strongly parabolic morphisms with respect to the flag. 
The space 
${\rm APEnd}(V^\bullet)$ was defined in Section \ref{background} as the quotient 
\[ 
{\rm APEnd}(V^\bullet)={\rm End}(V)/{\rm PEnd}(V^\bullet).
\]
Moreover, for any two ${\widetilde {\bf G}}$-equivariant bundles $E,F$ on $C$, let 
\[
\chi(E,F) = {\rm Ext}^0_C(E,F) -{\rm Ext}^1_C(E,F)
\]
in the representation ring of ${\widetilde {\bf G}}$.
Then a straightforward computation yields 
\be\label{eq:locformC} 
\bal 
V_2 - V_1 = T + P 
\eal 
\ee
where 
\be\label{eq:locformD} 
\bal 
T = &\ \chi(E,E) - Q_1\chi(E,E\otimes_C L_1) - 
Q_2\chi(E,E\otimes_C L_2)\\
&\  + Q_1Q_2\chi(E,E\otimes_C L_1\otimes_C L_2)  
 -\chi(\CO_C,E) -Q_1Q_2 \chi(E,L_1\otimes_C L_2) \\
 \eal
 \ee
 and 
 \be\label{eq:locformE}
 \bal
P = 
& (1-Q_1){\rm  APEnd}(E^\bullet_p) + 
Q_2(Q_1-1) {\rm SPEnd}(E^\bullet_p)\otimes {\CO_C}(p)_p.\\
\eal
\ee 
Using the canonical exact sequence 
\[
0\to \CO_C {\buildrel z_1\over \longto}  \CO_C(p) \to \CO_C(p)_p\to 0,
\]
one finds that $\CO_C(p)_p = Z^{-1}$ in the representation ring of 
${\widetilde {\bf G}}$. Moreover, there is a natural ${\widetilde {\bf G}}$-equivariant isomorphism 
\[ 
{\rm SPEnd}(E_p^\bullet) \simeq {\rm APEnd}(E_p^\bullet)^\vee.
\] 
Therefore equation \eqref{eq:locformE} yields 
\be\label{eq:locformF} 
\bal 
P= &\ (1-Q_1){\rm  APEnd}(E^\bullet_p) + 
Q_2(Q_1-1) Z^{-1} {\rm  APEnd}(E^\bullet_p)^{\vee}.\\
\eal
\ee

 \subsection{Experimental evidence}\label{examples}
The goal of this section is to provide some supporting evidence for the 
conjectural formula \eqref{eq:equivindexB}. 

First note that 
\[
W_{(r)}(q,y) = (-1)^{r}Z_{(r)}^{0,1}(q,y)
\] 
for length one partitions $\mu=(r)$. In this case the conjectural formula 
\eqref{eq:parconifoldA} reduces to the case without marked points discussed 
in detail in \cite{wallpairs,BPSPW}. Then identity \eqref{eq:equivindexB} 
is already verified by the computations of \cite{CKK},  both sides being in 
agreement  
with the geometric engineering predictions. 
Therefore only partitions of length 
$l\geq 2$ will be considered in the following. 
A straightforward computation yields the following expressions  
\be\label{eq:parconifoldC}
\bal 
W_{(11)}(q,y) & = {2y^{12}+2y^{14}\over y^{13}}q +{3y^{11}+3y^{15}+4y^{13}\over y^{13}}q^2 \\
& \ \ \ \ +{4y^{10}+6y^{12}+4y^{16}+6y^{14}\over y^{13}} q^3 +\cdots \\
W_{(21)}(q,y) & = {y^{13}+y^{15} \over y^{14}} q + 
{2y^{16}+2y^{12}+4y^{14}\over y^{14}} q^2 \\
& \ \ \ \ +{8y^{13}+4y^{11}+4y^{17}+8y^{15}\over y^{14}}q^3 +\cdots \\
W_{(111)}(q,y) & = {3y^{14}+3y^{16} \over y^{15}}q +{6 y^{13}+6 y^{17}+9y^{15}\over y^{15}} q^2 \\
& \ \ \ \ +{10 y^{12}+18 y^{14}+10 y^{18}+18 y^{16}\over y^{15}} q^3 +\cdots \\
\eal
\ee
A sample computation will be displayed below for 
$\mu=(1,1,1)$ and $e=1$. Employing the results of Section 
\ref{fixedsection} the fixed loci are in this case classified as follows. 
Taking into account inequalities \eqref{eq:degineqA}--\eqref{eq:degineqC}, 
there are six fixed points:
\[
\bal 
1)\ \ & E  = \CO_C \oplus Q_2^{-1}Z\CO_C + Q_2^{-2}Z^2\CO_C(\infty)\\
& \Phi_2(0,0) = z^{-1}, \qquad \Phi_2(0,1) = z^{-1}, \qquad \psi(0,0)=1,\\
& E_p^2 = E(0,2), \qquad E_p^1 = E(0,2)\oplus E(0,1).\\
\eal
\]
\[
\bal 
2.a)\ \ & E  = \CO_C \oplus Q_2^{-1}Z\CO_C + Q_2^{-2}Z^2\CO_C(\infty)\\
& \Phi_2(0,0) = z^{-1}, \qquad \Phi_2(0,1) = 1, \qquad \psi(0,0)=1,\\
& E_p^2 = E(0,2), \qquad E_p^1 = E(0,2)\oplus E(0,1).\\
\eal
\]
\[
\bal 
2.b)\ \ & E  = \CO_C \oplus Q_2^{-1}Z\CO_C + Q_2^{-2}Z^2\CO_C(\infty)\\
& \Phi_2(0,0) = z^{-1}, \qquad \Phi_2(0,1) = 1, \qquad \psi(0,0)=1,\\
& E_p^2 = E(0,1), \qquad E_p^1 = E(0,2)\oplus E(0,1).\\
\eal
\]
\[
\bal 
2.c)\ \ & E  = \CO_C \oplus Q_2^{-1}Z\CO_C + Q_2^{-2}Z^2\CO_C(\infty)\\
& \Phi_2(0,0) = z^{-1}, \qquad \Phi_2(0,1) = 1, \qquad \psi(0,0)=1,\\
& E_p^2 = E(0,1), \qquad E_p^1 = E(0,0)\oplus E(0,1).\\
\eal
\]
\[
\bal 
3.a)\ \ & E  = \CO_C \oplus Q^{-1}\CO_C(\infty) + Q_2^{-1}Z^2\CO_C\\
& \Phi_1(0,0) = 1, \qquad \Phi_2(0,0) = z^{-1}, \qquad \psi(0,0)=1,\\
& E_p^2 = E(1,0), \qquad E_p^1 = E(1,0)\oplus E(0,1).\\
\eal
\]
\[
\bal 
3.b)\ \ & E  = \CO_C \oplus Q^{-1}\CO_C(\infty) + Q_2^{-1}Z^2\CO_C\\
& \Phi_1(0,0) = 1, \qquad \Phi_2(0,0) = z^{-1}, \qquad \psi(0,0)=1,\\
& E_p^2 = E(0,1), \qquad E_p^1 = E(1,0)\oplus E(0,1).\\
\eal
\]
The underlying vector bundle $E$ is encoded in a decorated Young diagram of the form
\[ 
 \young(1,0,0)
 \]
for cases $(1)-(2.c)$, respectively 
\[
\young(0,01)
\]
for cases $(3.a)-(3.b)$. 
The expression \eqref{eq:locformD} takes the form 
\[ 
\bal 
T_1 = &\ 2+ Z^{-1}Q_2 + ZQ_2^{-1} - Z^2Q_1Q_2^2-ZQ_1Q_2^{-1} + Z^{-2}Q_1Q_2 
-Z + Z^{-3} Q_2^2\\
&\  - 2 Z^{-1} Q_1Q_2 -Z^{-2} Q_1Q_2^2 -Q_1\\
\eal
\]
for case $(1)$, 
\[
\bal 
T_2= & \ 1+ Z^{-1}Q_2 + Q_2^{-1} - ZQ_1Q_2^{-2} - Q_1 Q_2^{-1} 
- Z^{-1} Q_1 Q_2 + Z^{-2} Q_2^3 \\
& \ + Z^{-1} Q_2^2 - Q_1 - Z^{-1} Q_1 Q_2^2 \\
\eal 
\]
for cases $(2.a)$-$(2.c)$, respectively
\[
\bal 
T_3 = & \ 1 + Z^{-1} Q_1^{-1} Q_2 + Q_1^2 Q_2^{-1} - Z^{-1} Q_1^{-1} Q_2^2 
- Z^{-1} Q_1 Q_2 - Q_1^2 
\eal
\]
for cases $(3.a)$ and $(3.b)$. 

The expression \eqref{eq:locformE} specializes to, respectively, 
\[
\bal 
P_1 = & \ -2 + Z Q_1 Q_2^{-1} + 2Q_1 - ZQ_2^{-1} + 2Z^{-1}Q_1Q_2 + 
Z^{-2} Q_1 Q_2^2 - 2 Z^{-1}Q_2 - Z^2 Q_2^2 
\eal 
\]
\[
\bal
P_{2.a} = & \ -1 + Q_1 Q_2^{-1} + Z^{-1} Q_1 + Q_1 - Z^{-1} -Q_2^{-1} + Z^{-1} 
Q_1Q_2 \\
&\ + Q_1Q_2 + Z^{-1} Q_1 Q_2^2 - Z^{-1} Q_2 - Q_2 -Z^{-1} Q_2^2 \\
\eal
\]
\[
\bal 
P_{2.b}  = & \ -1 +Q_1 + Z^{-1} Q_1Q_2 + 2Q_1Q_2^{-1} + 2Z^{-1} Q_1Q_2^2 -Z^{-1} Q_2 - 2Q_2^{-1} - 2Z^{-1} Q_2^2 
\eal 
\]
\[
\bal 
P_{2.c} = & \ -1 - Z^{-1} Q_2 - Q_2^{-1} - Z Q_2^{-2} + Z^{-1} Q_1 Q_2 + Q_1 Q_2^{-1} + Z Q_1 Q_2^{-2} - Z^2 Q_2^3 \\
& \ - Z^{-1} Q_2^2 + Z^{-2} Q_1 Q_2^3 +Z^{-1} Q_1 Q_2^2 + Q_1\\
\eal 
\]
\[
\bal 
P_{3.a} = & \ -1 - Z Q_1 Q_2^{-1} + Q_1^2 + Z Q_1^2 Q_2^{-1} + Z^{-1} Q_1Q_2 - Z^{-1} Q_1^{-1} Q_2 \\
& \ - Z^{-2} Q_1^{-1} Q_2^2 + Z^{-2} Q_2^2 \\
\eal
\]
\[ 
\bal 
P_{3.b} = & \ -1 - Q_1 - 2 Z^{-1} Q_1^{-1} Q_2 + Z^{-1} Q_2 + 2 Q_1^2 + Z^{-1} Q_1 Q_2 \\
\eal
\]
Let $F_{\mathfrak m}$ denote the right hand side of equation \eqref{eq:locformA}.
Then, using the above computations, one obtains 
\[ 
\bal 
F_1=R^3 Z^{-2} Q_1^{-1} Q_2 
 {(1-Q_1^{-1})(1-Z^2Q_1^{-1}Q_2^{-1})(1-Z^3Q2^{-3})\over 
 (1-Z^{-1})(1-ZQ_2^{-1})(1-Z^{-2}Q_1^{-1}Q_2^2)}
\eal 
\] 
\[
F_{2.a} = R^3 Q_1^{-1} Q_2 {(1-Q_1^{-1}Q_2^{-1})(1-ZQ_1^{-1})(1-Z^2Q_2^{-3})\over 
(1-Q_2^{-1})(1-Z)(1-Z^{-1}Q_1^{-1}Q_2^2)}
\]
\[
F_{2.b} = R^3 Q_1^{-1} Q_2 {(1-Z Q_1^{-1}Q_2^{-2})(1-Q_1^{-1}Q_2)
(1-Z^2 Q_2^{-3})\over 
(1-Q_2)(1-ZQ_2^{-2})(1-Z^{-1}Q_1^{-1}Q_2^2)}
\] 
\[
F_{2.c} = R Z^{-1} Q_2^2 {1-Z^2 Q_1^{-1}Q_2^{-3}\over 1- Z^{-1} Q_2^2}
\]
\[
F_{3.a} = R^3 Z^2 Q_1 Q_2^{-3} {(1-Q_1^{-2}Q_2)(1-Z^2Q2^{-2})(1-Z^{-1}Q_1^{-2}Q_2)\over 
(1-Z^{-1}Q_1^{-1}Q_2)(1-Z^2Q_1Q_2^{-2})(1-ZQ_1Q2^{-2})}
\] 
\[
F_{3.b} = R^3 Z^2 Q_1 Q_2^{-3}
{(1-Q_1^{-2}Q_2)(1-Q_1^{-2})(1-ZQ_2^{-1})\over 
(1-Q_1^{-1})(1-ZQ_1Q_2^{-2})(1-ZQ_1Q_2^{-1})}
\] 
where $R=Z^{-1}Q_1Q_2$. Adding all local contributions yields 
\[
I_{(1,1,1),1} = - 3 \big( R^{1/2} + R^{-1/2}\big),
\]
confirming conjecture \eqref{eq:equivindexB} in this case. Similar computations 
confirm the conjecture for $(\mu,e) = ((1,1),1), ((1,1),2), ((1,1),3), ((2,1),1),
((2,1),2),((2,1),3),$\\
 $((1,1,1),2),((1,1,1),3)$.

\appendix

\section{Degree zero ADHM sheaves}\label{flatADHM}
This section proves a result used in the main text stating that the underlying vector 
bundle of any rank $r$, degree 0 
asymptotically flat ADHM sheaf $\CE$ must be trivial, $E\simeq \CO_C^{\oplus r}$.

If $r=1$, the claim is obvious since $\mathrm{deg}(E)=0$ and there is a nonzero 
morphism $\psi:\CO_C\to E$. 

Suppose $r\geq 2$ and $E$ is slope semistable. For any 
$(n_1,n_2)\in (\IZ_{\geq 0})^2$, 
let $E(n_1,n_2)= \Phi_1^{n_1}\Phi_2^{n_2}
\psi(\CO_C)\subset E$. 
Note that $E(n_1,n_2)$ is either the zero sheaf or isomorphic to 
$\CO_C$ since it is a locally free quotient of $\CO_C$. 
Let $$E' =\sum_{(n_1,n_2)\in \IZ^2_{\geq 0}} E(n_1,n_2)\subseteq E.$$
The asymptotic stability condition implies that $E/E'$ is a zero 
dimensional sheaf on $C$. 
By construction there there exists a finite set $\Delta \subset (\IZ_{\geq 0})^2$ and a surjective morphism 
\[
V_\Delta=\bigoplus_{(n_1,n_2)\in \Delta} E(n_1,n_2) \twoheadrightarrow
 E'.
\]
Since $E$ is semistable of degree 0, it follows that the resulting morphism $V_\Delta \to E$
 must be surjective as well, hence $E\simeq E'$. 

Now let 
\[
0=JE_0 \subset JE_1\subset \cdots \subset JE_n=E
\]
be a Jordan-H\"older filtration of $E$. Obviously, there is a 
commutative triangle of surjective morphisms
\[ 
\xymatrix{
V_\Delta \ar[r] 
\ar[dr] & E \ar[d] \\ & 
E/JE_{n-1}.\\}
\] 
This implies that there is at least one direct summand 
$E(m_1,m_2)\subset V_\Delta$ which fits into a  commutative triangle 
\[ 
\xymatrix{
 E(m_1,m_2) \ar[r] 
\ar[dr] & E \ar[d] \\ & 
E/JE_{n-1}\\}
\]
with all maps nontrivial. Moreover, the horizontal map 
must be in fact injective. 
Since $E/JE_{n-1}$  is stable of 
degree 0, and $E(m_1,m_2)\simeq \CO_C$, it follows that 
$E/JE_{n-1}\simeq \CO_C$, and the map $E(m_1,m_2) 
\to E/JE_{n-1}$ is an isomorphism. This implies that there 
is a splitting 
\[ 
E \simeq E/JE_{n-1} \oplus JE_{n-1} \simeq \CO_C \oplus 
JE_{n-1} 
\] 
By construction, $JE_{n-1}$ is degree 0 slope semistable and 
there is a surjective morphism 
\[ 
\bigoplus_{(n_1,n_2)\in \Delta \setminus \{(m_1,m_2)\}} 
E(n_1,n_2) \twoheadrightarrow JE_{n-1}. 
\] 
Repeating the above argument shows that $JE_{n-1}/JE_{n-2}\simeq \CO_C$ and there is a splitting 
\[ 
JE_{n-1} \simeq \CO_C \oplus JE_{n-2} 
\]
Proceeding recursively, one finds that $E\simeq \CO_C^{\oplus r}$ 
in a finite number of steps. 

To finish the proof, suppose $E$ is not slope semistable. Then it 
It will be shown below that this leads to a contradiction.  
By assumption, $E$ has a 
a Harder-Narasimhan filtration 
\[
0=HE_0\subset HE_1 \subset  \cdots \subset HE_l=E
\]
with $l\geq 2$. 

The first observation is that $\Phi_{j}(E_k)\subseteq E_k$ for 
all $1\leq j\leq 2$ and $1\leq k \leq l$. Suppose this fails 
for some $1\leq k \leq l-1$ and some $1\leq j \leq 2$,
and let $k$ be minimal with this property 
i.e. $\Phi_j(E_{k'})\subseteq E_{k'}$ for all $k'<k$. 
Then let $k''>k$ be minimal such that 
$\Phi_j(E_k) \subseteq E_{k''}$ for all $j\in\{1,2\}$ 
and $\Phi_j(E_k)\nsubseteq E_{k''-1}$ for at least one value of 
$j\in \{1,2\}$. Then $\Phi_j$ yields a 
nontrivial morphism ${\overline \Phi}_j : E_{k}/E_{k-1} \to E_{k''}/E_{k''-1}$ contradicting 
the defining property of the Harder-Narasimhan filtration. 

Since $\Phi_1,\Phi_2$ preserve the Harder-Narasimhan filtration,  the asymptotic
stability condition for ADHM sheaves 
implies that $\psi(\CO_X) \nsubseteq E_{l-1}$. Hence $\psi$ yields a nontrivial 
morphism $\CO_X \to E/E_{h-1}$.  Since $E/E_{l-1}$ is semistable, this implies 
$\mu(E/E_{l-1})\geq 0$, again contradicting the properties of the Harder-Narasimhan filtration
which imply that $\mu(E/E_{l-1})<\mu(E)=0$. 

In conclusion, the underlying bunde of an asymptotically stable 
degree 0 ADHM sheaf must be indeed isomorphic to 
$\CO_C^{\oplus r}$.

\section{Fermion zero modes}\label{zeromodes}

The goal of this section is to determine the bundle of fermion 
zero modes on the moduli space of supersymmetric D2-D6 configurations found in Section \ref{geomeng}. 
As  proven in Section \ref{nested},  
supersymmetry constraints 
require the Chan-Paton bundle on $r$ such D2-branes to be isomorphic to the trivial rank $r$ bundle, and 
all field configurations  to be constant. This shows that the low 
energy effective action of such a configuration is reduced to 
supersymmetric quantum mechanics.
 The detailed action of a similar system has been written in 
 \cite[Sect 2.2]{ADHMsurface} as the dimensional 
reduction of a two dimensional $(0,2)$ gauged linear sigma model. 
Analogous considerations will yield the action in the present case
by dimensional reduction of a two dimensional $(0,4)$ 
gauged linear sigma model. Omitting the details, note the resulting quantum mechanical system will have a moduli space of flat directions isomorphic to $\CN(\gamma)$, as expected. 
Using standard $(0,2)$ sigma model technology 
\cite{phases}, the bundle of fermion zero modes is isomorphic to 
the middle cohomology of a monad complex, as shown below. 
 
In absence of the orbifold point $p$, 
the D2-D6 moduli space is isomorphic to the Hilbert scheme 
of points $\CH^r$. 
For any stable ADHM data  $(A_1,A_2,I)$, 
the space of fermion zero modes is isomorphic to the middle 
cohomology group of the complex $\CF_{(A_1,A_2,I)}$ 
\be\label{eq:fermizeroA} 
0 \to H^1(End_C(E)) {\buildrel d_1\over \longto} 
\begin{array}{c}
H^1(End_C(E))^{\oplus 2} \\
\oplus \\
H^1(E)\\
\oplus \\
H^1(E^\vee) \\
\end{array}
{\buildrel d_2 \over \longto} 
H^1(End_C(E))\to 0
\ee
where 
\be\label{eq:diffs}
\bal
d_1(\alpha) & = \big( [\alpha,A_1], [\alpha,A_2], \alpha I\big) \\
d_2(\beta_1,\beta_2, \gamma, \delta) & = [\beta_1,A_2]+[A_1,\beta_2] + I \delta.\\
\eal 
\ee
Since $E\simeq \CO_C^{\oplus r}$, one can easily prove using 
Serre duality that $\CF_{(A_1,A_2,I,J)}$ is left and right exact while its middle cohomology is isomorphic to 
$\big(\CT^*_{(A_1,A_2,I,J)}\CH^r\big)^{\oplus g}$, where 
$\CT^*_{(A_1,A_2,I,J)}\CH^r$ is the fiber of the cotangent 
bundle to the Hilbert scheme $\CH^r$ at the point $(A_1,A_2)$.
Using the same argument in flat families of stable ADHM data, 
it follows that the bundle of fermion zero modes is isomorphic 
to the direct sum $\big(\CT^*\CH^r\big)^{\oplus g}$.

Now suppose there is an orbifold point, in which case the supersymmetric configurations are in one-to-one correspondence 
with stable parabolic ADHM data  $(A_1,A_2,I,J; V^\bullet)$ 
as shown in Section \ref{nested}.
The space of fermion zero modes 
will then given by the middle cohomology 
of a complex $\CF_{(A_1,A_2,I,J; V^\bullet)}$
of the form \eqref{eq:fermizeroA}, where 
$E$ is replaced with an orbi-bundle ${\tilde E}$ on ${\widetilde C}$. 
Using the correspondence described in Section \ref{orbisection}, this complex can be written in terms of parabolic data as follows. 

Recall that there is a root line bundle ${\tilde L}$ on ${\tilde 
C}$ such that ${\tilde L}^s \simeq \nu^* \CO_C(p)$, 
where $\nu:{\widetilde C}\to C$ is the natural projection.
Moreover, the canonical class of ${\widetilde C}$ is given by 
\[
K_{\widetilde C}\simeq \nu^*K_C\otimes_{\widetilde C} {\tilde L}^{(s-1)}
\simeq \nu^*K_C(p) \otimes_{\widetilde C} {\tilde L}^{-1}.
\]
Then 
Serre duality on the stack ${\widetilde C}$ yields the following isomorphisms
\[ 
\bal
H^1(End_{\widetilde C}({\tilde E})) & \simeq H^0(End_{\widetilde C}({\tilde E})\otimes_{\widetilde C} \nu^*K_C(p)\otimes_{\widetilde C} {\tilde L}^{-1})^\vee\\
H^1({\tilde E}^\vee) & \simeq H^0({\tilde E}\otimes_{\widetilde C}
\nu^*K_C(p)\otimes_{\widetilde C} {\tilde L}^{-1})^\vee.
\eal
\]
Now recall that the pushforward $E=\nu_*{\tilde E}$ is a vector 
bundle on $C$ equipped with a filtration by subsheaves $F_a = 
\nu_*({\tilde E}\otimes_{\widetilde C} {\tilde L}^{-1})$, $a\geq 1$. 
This filtration determines a flag $E_p^\bullet$ in the fiber $E_p$, 
hence a parabolic structure on $E$ at $p$. 
Moreover the higher direct images $R^k\nu_*{\tilde E}$ are trivial
and there is a one-to-one correspondence between 
morphisms ${\tilde \Phi}: {\tilde E} \to {\tilde E}$ and 
parabolic morphisms $\Phi:E^\bullet\to E^\bullet$. 
Therefore one obtains isomorphisms of the form 
\[ 
\bal
H^1(End_{\widetilde C}({\tilde E})) & \simeq H^0(SPEnd_{C}(E)\otimes_{C} \nu^*K_C(p))^\vee\\
H^1({\tilde E}^\vee) & \simeq H^0({F_1}\otimes_{C}
\nu^*K_C(p))^\vee\\
H^1({\tilde E}) & \simeq H^1(E) \simeq H^0(E^\vee \otimes_C 
K_C)
\eal
\]
Then dual complex is isomorphic to 
\be\label{eq:fermizeroC} 
\bal 
0 & \to H^0(SPEnd_C(E^\bullet)\otimes_C K_C(p)) {\buildrel d'_1\over \longto} 
\begin{array}{c}
H^0(SPEnd_C(E^\bullet)\otimes_C K_C(p))^{\oplus 2} \\
\oplus \\
H^0(E^\vee\otimes_C K_C(p))\\
\oplus \\
H^0(F_1\otimes_C K_C(p)) \\
\end{array}\\
& 
{\buildrel d'_2 \over \longto} 
H^0(SPEnd_C(E^\bullet)\otimes_C K_C(p))\to 0
\\
\eal
\ee
where $SPEnd_C(E^\bullet)$ denotes the sheaf of strongly parabolic 
endomorphisms of $E^\bullet$. 
The expressions of the differentials are formally identical 
with the ones given in 
\eqref{eq:diffs}. 

Next note that by construction there is an exact sequence 
\be\label{eq:spexseq}
0\to End_C(E) \otimes_C K_C \to SPEnd_C(E^\bullet) \otimes 
K_C(p) \to SP{\rm End}(E_p^\bullet)\otimes_C \CO_p(p)\to 0
\ee
of sheaves on $C$. Moreover, the inclusions 
\[
0\subset E(-p) \subset F_1 \subset E
\] 
yield inclusions of vector spaces 
\[ 
0\subset H^0(E\otimes_C K_C) \subseteq H^0(F_1\otimes_C K_C(p)) \subseteq H^0(E\otimes_C K_C). 
\]
However since $E\simeq \CO_C^{\oplus r}$, there is an isomorphism 
\[
H^0(E\otimes_C K_C) \simeq H^0(E\otimes_C K_C(p)).
\]
Therefore there is an isomorphism 
\be\label{eq:Fonesect}
H^0(E\otimes_C K_C) \simeq H^0(F_1\otimes_C K_C(p)).
\ee
Using the exact sequence \eqref{eq:spexseq} and isomorphism 
\eqref{eq:Fonesect} a straightforward computation shows that there is an exact sequence of complexes 
\be\label{eq:cpxexseq}
0 \to \CF_{(A_1,A_2,I)}^\vee \to 
\CF_{(A_1,A_2,I; V^\bullet)}^\vee 
\to {\mathcal D}_{(A_1,A_2; V^\bullet)} \to 0
\ee
where ${\mathcal D}_{(A_1,A_2; V^\bullet)}$ is the three term complex 
\[
0\to {\rm SPEnd}(V^\bullet) {\buildrel \delta_1\over \longto} 
 {\rm SPEnd}(V^\bullet) ^{\oplus 2} {\buildrel \delta_2\over \longto}  {\rm SPEnd}(V^\bullet) \to 0.
\]
The differentials $\delta_1,\delta_2$ are given by 
\[ 
\bal
\delta_1(f) & = \big( [f,A_1], [f,A_2]\big) \\
\delta_2(g_1,g_2) & = [g_1,A_2]+[A_1,g_2] .\\
\eal 
\]
Now note that under the current assumptions $\delta_1$ is 
injective and $\delta_2$ is surjective, hence the 
complex ${\mathcal D}_{(A_1,A_2; V^\bullet)}$ 
has trivial cohomology. 

To prove this claim, recall that $(A_1,A_2)$ is by assumption a 
cyclic commuting 
pair preserving the flag $V^\bullet$. In particular $(A_1,A_2)$is regular i.e. the subspace $f\in 
{\rm End}(\IC^r)$ such that $[f,A_1]=[f,A_2]$ is isomorphic 
to a Cartan subalgebra of ${\rm End}(\IC^r)$. On the other hand 
if $f\in {\rm SPEnd}(V^\bullet)$, it follows that $f$ is nilpotent, 
hence it must be trivial. This shows that ${\rm Ker}(\delta_1)=0$. 
Surjectivity of $\delta_2$ follows by an analogous argument for the dual 
morphism 
\[
\delta_2^\vee : {\rm SPEnd}(V^\bullet)^\vee 
\to {\rm SPEnd}(V^\bullet)^\vee \oplus 
{\rm SPEnd}(V^\bullet)^\vee
\]
The dual vector space ${\rm SPEnd}(V^\bullet)^\vee$ is isomorphic 
to a space of strongly parabolic maps on the dual vector space 
$V^\vee$ equipped with the dual flag
\[
V^\vee_{s-a} = {\rm Ker}\big(V^\vee \twoheadrightarrow (V^a)^\vee\big),
\qquad 0\leq a\leq s.
\]
That is ${\rm SPEnd}(V^\bullet)^\vee \simeq {\rm SPEnd}(V_\bullet^\vee)$. Moreover, 
\[ 
\delta_2^\vee(\xi) = \big([\xi,A_1^\vee], [A_2^\vee,\xi]\big) = 
\big([A_1,\xi^\vee]^\vee, [\xi^\vee, A_2]^\vee\big).
\]
 Then the same argument shows that ${\rm Ker}(\delta_2^\vee)=0$, hence $\delta_2$ is surjective.
 
 In conclusion, the exact sequence \eqref{eq:cpxexseq} implies that 
 the complexes $\CF_{(A_1,A_2,I; V^\bullet)}^\vee$ and 
 $\CF_{(A_1,A_2,I)}^\vee$ are quasi-isomorphic.

\section{Some basic facts on nested Hilbert schemes}\label{reducednested}
The goal of this section is to prove that the nested Hilbert scheme $\CN(\gamma)$ 
used in Section \ref{nested} is reduced and connected. 
The proof relies on an 
alternative
presentation of $\CN(\gamma)$ given in 
\cite{nested_quivers} as a moduli space of stable  
framed quiver representations. Namely, consider the moduli 
space of  stable framed quiver  representations of the form:
\be\label{eq:nestedquiverA} 
\xymatrix{ 
\IC^{r_\ell} \ar@(ur,ul)_-{A_{\ell,1}}\ar@(dl,dr)_-{A_{\ell,2}} 
\ar[rr]^{f_{\ell-1,\ell}} & & \IC^{r_{\ell-1}}
\ar[rr]^-{f_{\ell-2,\ell-1}} 
\ar@(ur,ul)_-{A_{\ell-1,1}}\ar@(dl,dr)_-{A_{\ell-2,2}}& & 
 & \cdots  & \IC^{r_0} \ar@(ur,ul)_-{A_{0,1}}
\ar@(dl,dr)_-{A_{0,2}} &&  \IC \ar[ll]_-{I}}
\ee
 with quadratic relations
 \be\label{eq:nestedquiverB} 
 [A_{0,1},A_{0,2}]=0, \quad A_{\imath, 1} f_{\imath, \imath+1} 
 - f_{\imath,\imath+1} A_{\imath, 1} =0,\quad
 A_{\imath, 2} f_{\imath, \imath+1} 
 - f_{\imath,\imath+1} A_{\imath, 2} =0.
  \ee 
The discrete invariants $r_\imath$, $0\leq \imath\leq \ell$, are given by 
\[ 
r_\imath = \sum_{\jmath = \imath}^{\ell} \gamma_\jmath.
\]
For generic King stability parameters $(\theta_\imath, \theta_\infty)\in \IR^{\ell+2}$ satisfying
\[
\theta_\infty = -\sum_{\imath=0}^{\ell} n_\imath 
\theta_\imath, \qquad \theta_\imath >0, \qquad 
 0\leq \imath\leq \ell.
\]
a representation of the above quiver is semistable if and only if 
the ADHM data $(A_{0,1}, A_{0,2}, I)$ is stable 
and the linear maps $f_{\imath, \imath+1}$ are injective for all $0\leq \imath \leq \ell-1$.
Here $\infty$ denotes the framing node corresponding 
to the tail of the arrow 
$I$ in the above diagram. 
In particular for $(\theta_\imath, \theta_\infty)$ sufficiently 
generic there are no strictly semistable objects  
and the stabilizer group of any stable framed 
representation is trivial.

Let ${\mathbb A}(\gamma)$ denote the 
linear space of all linear maps of the form
\eqref{eq:nestedquiverA}, not subject to any 
stability condition or relations. Then the subset
of stable quiver representations is an open subspace  
$U(\gamma)\subset {\mathbb A}(\gamma)$. 
Let $V(\gamma) \subset {\mathbb A}(\gamma)$ be the closed 
subscheme determined by the quadratic equations 
\eqref{eq:nestedquiverB}, and $V_U(\gamma)$ its restriction 
to $U(\gamma)$. 
Since all stabilizers are trivial, $V_U(\gamma)$ is a principal 
$G(\gamma)$ bundle over $\CN(\gamma)$, where 
$G(\gamma) = \times_{\imath=0}^\ell GL(n_\imath,\IC)$. 
Therefore in order to conclude that $\CN(\gamma)$ is 
reduced it suffices to prove that $V(\gamma)$ is reduced. 
Now recall that any ideal 
$I\in \IC[x_1,\ldots, x_N]$ generated by irreducible 
polynomials is a radical ideal. This statement can be easily proven by induction on $N$. Then it suffices to prove that all quadrics in 
equation \eqref{eq:nestedquiverB} are irreducible. A straightforward 
computation shows that any quadric in \eqref{eq:nestedquiverB} 
is of the form 
\[
\sum_{i=1}^s x_iy_i 
\]
with $s\geq 2$, which is indeed irreducible. 

In order to prove $\CN(\gamma)$ is connected, 
recall that the morphism $\rho_{\sf red}: \CN(\gamma)\to \wCH^r_{\sf red}$ constructed in 
diagram \eqref{eq:hilbdiagA} was shown there to have 
connected fibers for 
$\gamma=(1,\ldots, 1)$. This implies that $\CN(\gamma)$ is connected since $\rho_{\sf red}$ is also surjective and 
$\wCH^r_{\sf red}$ is connected. 
The above quiver moduli space  yields a natural 
 morphism $\CN(1,\ldots, 1) \to \CN(\gamma)$ 
for any ordered partition $\gamma$ of $r$. Using the 
Jordan normal for the linear maps $A_{\imath,1}$, $A_{\imath,2}$, 
$0\leq \imath\leq \ell$, it is straightforward to show that 
this morphism is surjective. Therefore 
$\CN(\gamma)$ must also 
be connected, as required in the proof 
of equation \eqref{eq:pushfwdC}.

\section{A compactness result}\label{compactness} 
This section proves that the moduli spaces of asymptotically stable parabolic ADHM 
sheaves in the example considered in Section \ref{conifold} are proper. 
In that case $C\simeq \IP^1$ and there is a single orbifold point $p$, which is 
one of the fixed points of the canonical torus action on $C$. The second fixed 
point is denoted by $\infty$. 
The orbifold 
${\widetilde Y}$ is the total space of the rank two bundle $K_{\widetilde C}\otimes_{\widetilde C}\nu^*\CO_C(\infty)\oplus \nu^*\CO_C(-\infty)$. Therefore one 
has a moduli space of asymptotically stable parabolic ADHM sheaves on $C$ 
with coefficient line bundles $\CO_C(-\infty), K_C\otimes_C\CO(-\infty)
\otimes_C\CO_C(p)$. The underlying vector bundle $E$ of any such 
ADHM sheaf $\CE$ splits as a direct sum 
\[
E\simeq\oplus_{j=1}^{l} \CO_C(e_j\infty)^{\oplus r_j}
\]
with 
\[
0\leq d_1 < d_2 < \cdots < d_l.
\]
Positivity follows from asymptotic ADHM stability, which requires $\CE$ to be generically generated by the image of the section $\psi:\CO_C \to E$ as a 
quiver sheaf. The Higgs fields $\Phi_1,\Phi_2$ have components 
\[ 
\bal
\Phi_1(j,j') : \CO_C(e_j\infty)^{\oplus r_j} & \to \CO_C((e_{j'}-1)\infty)^{\oplus r_{j'}}\\
\Phi_2(j,j') : \CO_C(e_j\infty)^{\oplus r_j} & \to \CO_C((e_{j'}-1)\infty)^{\oplus r_{j'}}\otimes_C \CO_C(p)\\
\eal
\]
For degree reasons $\Phi_1(j,j')=0$ for all $j'\leq j$, and 
$\Phi_2(j,j')=0$ for all $j'<j$. Moreover, note that the diagonal components 
$\Phi_2(j,j)$ must be constant maps. Since the residue ${\rm Res}_{p}\Phi_2$ 
must be nilpotent, it follows that the components $\Phi_2(j,j)$ must vanish as well. 
This implies that all polynomial invariants of the quiver sheaf $\CE$ are identically zero since $\phi:E\to \CO_C$ is identically zero. Since the generalized 
Hitchin map determined by the polynomial invariants is proper, it follows 
that the moduli space is proper.

\end{document}